%% file: main_file.tex
\documentclass{amsart}

\usepackage{amsfonts}
\usepackage{float}
\usepackage{amstext}
\usepackage{amsmath}
\usepackage{amssymb}
\usepackage{color} 
\usepackage{hyperref}
\usepackage{graphicx}
\usepackage{amsthm}

\newtheorem{theorem}{Theorem}

\newtheorem{lemma}[theorem]{Lemma}

\newtheorem{proposition}[theorem]{Proposition}

\newcommand{\RR}{\mathbb{R}}

\newcommand{\EE}{\mathbb{E}}
\newcommand{\PP}{\mathbb{P}}
\newcommand{\TT}{\mathbf{T}}

\DeclareMathAlphabet{\itbf}{OML}{cmm}{b}{it}

\def\bx{{{\itbf x}}}

\def\bX{{{\itbf X}}}

\def\bW{{{\itbf W}}}
\def\balpha{{\boldsymbol{\alpha}}}

\title{A risk analysis for a system stabilized by a central agent}

\author{Josselin Garnier}
\address{Laboratoire de Probabilit\'{e}s et Mod\`{e}les Al\'{e}atoires \& Laboratoire Jacques-Louis Lions, Universit\'{e} Paris Diderot}
\email{garnier@math.univ-paris-diderot.fr}

\author{George Papanicolaou}
\address{Department of Mathematics, Stanford University}
\email{papanicolaou@stanford.edu}

\author{Tzu-Wei Yang}
\address{School of Mathematics, University of Minnesota}
\email{yangx953@umn.edu}

\date{}

\begin{document}
\maketitle

\begin{abstract}
We formulate and analyze a multi-agent model for the evolution 
of individual and systemic risk in which the local
agents interact with each other through a central agent
who, in turn, is influenced by the mean field of the local agents.
The central agent is stabilized by a bistable potential, the only
stabilizing force in the system. The local agents derive their
stability only from the central agent. In the mean field limit
of a large number of local agents we show that the systemic
risk decreases
when the strength of the interaction of the local agents
with the central agent increases.
This means that the probability of transition from one of the two
stable quasi-equilibria to the other one decreases.
We also show that the systemic risk increases when the strength of the
interaction of the central agent with the mean field of the local
agents increases. Following the financial interpretation of such models
and their behavior given in our previous paper 
(Garnier, Papanicolaou and Yang, SIAM J. Fin. Math. 4, 2013, 151-184), 
we may interpret the results of this paper in the following way. From the point of
view of systemic risk, and while keeping the perceived risk of the
local agents approximately constant, it is better to strengthen
the interaction of the local agents with the central agent than the
other way around.
\end{abstract}


\input{introduction.tex}
\input{formulation.tex}
\input{zero_h.tex}

\input{empirical_measure.tex}
\input{control.tex}
\input{numerics.tex}
\input{summary.tex}

\appendix

\input{proof_zero_h.tex}
\input{proof_empirical_measure.tex}
\input{proof_control.tex}

\bibliographystyle{plain}
\bibliography{reference}

\end{document}

%% file: introduction.tex
\section{Introduction}

In recent years, interacting particle systems have been extensively used to model financial systemic risk for complex, inter-connected systems. 
An interacting particle system with binary risk variables is considered in \cite{DaiPra2009} and the law of large numbers, central 
limit theorem and large deviation principle are derived for this model. 
An interacting particle system of diffusion processes is used in \cite{Fouque2013}  to model the interbank lending system.  In \cite{Carmona2013}, a model simplified from the one in 
\cite{Fouque2013} is considered, in which each agent can control the lending flow rate and optimizes the individual objective function, and thus the system can 
be put in the framework of mean field games.  In \cite{Ichiba2013}, the authors use interacting Bessel-like diffusion processes to model systemic risk and 
establish a large deviation principle. 
In \cite{Garnier2013, Garnier2013a}, we consider an interacting particle system with a bistable potential and we use the large deviation principle to explain that the overall systemic risk may increase while individual risks are decreased. 
The large deviation principle in \cite{Garnier2013, Garnier2013a} is solved numerically in \cite{Lauriere2014}.  
In \cite{Bo2015}, the authors consider interacting jump-diffusion processes modeling interbank lending and borrowing 
and prove the weak law of large numbers (LLN) of the empirical measure as the number of individuals goes to infinity, and define systemic indicators based on the  
LLN result.  In \cite{Giesecke2013, Spiliopoulos2014, Giesecke2015, Spiliopoulos2015}, the authors model large portfolios and default clustering and derive
 the law of large numbers, fluctuation analysis and large deviations.

In our previous work \cite{Garnier2013}, we used an interacting agent-based, mean-field model to show that individual risk may 
not affect systemic risk in an obvious way. That is, each agent may have relatively low individual risk by diversification through 
risk-sharing while the overall, systemic risk  is increased as a result of diversification.
We considered the following model that was studied extensively before by
\cite{Dawson1983, Dawson1987, Gartner1988, Dawson1989}:
\begin{equation}
	\label{eq:SDE for x_j, old model}
	dx_j (t)= -h V'(x_j(t))  dt - \theta (x_j(t) - \bar{x}_N(t)) dt + \sigma dW^j_t, \quad \quad j=1,\ldots,N,
\end{equation}
where $x_j(t)$ represents a risk variable for agent $j$ at time $t$ and $N$ is the number of agents. The potential $V(x)=\frac{1}{4}x^4-\frac{1}{2}x^2$ is  
taken to be bistable with two stable states $\pm 1$, and the constant $h>0$ quantifies intrinsic stability for each agent. We define $-1$ as the 
normal state of an agent and $+1$ as the failed state. The empirical mean $\bar{x}_N(t):=\frac{1}{N}\sum_{j=1}^{N}x_j(t)$ is the 
mean risk, and the constant $\theta$ is positive so that $x_j$ tends to stay close to $\bar{x}_N$. 
The standard Brownian motions $\{W^j_t\}_{j=1}^{N}$ are independent and model external risk factors, with 
$\sigma>0$ their strength.

It was shown in \cite{Dawson1983} that the empirical measure $U_N(t,dx):=\frac{1}{N}\sum_{j=1}^{N}\delta_{x_j(t)}(dx)$ converges 
weakly in probability to $u(t,dx)=u(t,x)dx$, the weak solution of the nonlinear Fokker-Planck equation:
\begin{equation*}
	\frac{\partial}{\partial t}u
	= h\frac{\partial}{\partial x}\left[V'(x)u\right] 
	- \theta\frac{\partial}{\partial x}\left\{\left[\int_{-\infty}^\infty yu(t,dy)-x\right]u\right\}
	+ \frac{1}{2}\sigma^2\frac{\partial^2}{\partial x^2}u,
	\end{equation*}
starting from $u(0,dx)=\lim_{N\to\infty}U_N(0,dx)$ (provided the weak limit exists).
Given $h$ and $\theta$, for sufficiently small $\sigma$,
 $u(t,x)$ has two equilibria $u^e_{\pm\xi_b}(x):=\lim_{t\to\infty}u(t,x)$,
where $\bar{x}_N(t)$ converges to either $\xi_b>0$ or $-\xi_b$ as $t\to\infty$, depending on the initial condition. 
Thus we define $u^e_{-\xi_b}$ as the normal state of the system and $u^e_{+\xi_b}$ as the failed state of the system.

Given that $N$ is large but finite, and $U_N(0,dx)\approx u^e_{-\xi_b}(x)dx$, we showed 
\cite[Theorem 6.2 and Corollary 6.4]{Garnier2013} that by using the large deviation principle in \cite{Dawson1987} and assuming 
that $h$ is small, the systemic risk, defined as the probability of the transition of $U_N(t,dx)$ 
from $u^e_{-\xi_b}(x)dx$ at time $0$ to 
$u^e_{+\xi_b}(x)dx$ at some time $t\leq T<\infty$ has the following exponentially small but nonzero value:
\begin{equation}
\PP\left(U_N(0,dx)\approx u^e_{-\xi_b}(x)dx,~ U_N(t,dx)\approx u^e_{+\xi_b}(x)dx ~~ t\leq T<\infty\right) 
\overset{\substack{N\gg 1\\h\ll 1}}{\approx}
 \exp\left(-N\frac{2\xi_b^2}{\sigma^2 T}\right),
 	\label{eq:systemic risk, old model}
\end{equation}
where
\begin{equation*}
	\xi_b = \sqrt{1 - 3\frac{\sigma^2}{2\theta}} 
	\left( 1 + h\frac{6}{\sigma^2}\left(\frac{\sigma^2}{2\theta}\right)^2
	\frac{1 - 2(\sigma^2/2\theta)}{1 - 3(\sigma^2/2\theta)}\right) + O(h^2).\notag
\end{equation*}
Fluctuation analysis on (\ref{eq:SDE for x_j, old model}) \cite[Lemma 6.5]{Garnier2013}, shows that the risk of each agent has the form
$x_j(t)=-1+z_j(t)$ and $\lim_{t\to\infty}\mathbf{Var}z_j(t)\lesssim\frac{\sigma^2}{2\theta}$.  Thus, the quantity 
$\frac{\sigma^2}{2\theta}$ can be considered as the individual risk for each agent.

We then see that if the strength of the external risk $\sigma^2$ is increased, 
either because the agents are more risk-prone or 
because the economic environment is 
more uncertain, then the agents can  increase $\theta$, the risk-diversification parameter, so that that their individual risk is still 
low. However, from the analysis of the systemic risk (\ref{eq:systemic risk, old model}) we see that the systemic risk is
increased when $\sigma^2$ increases even if the individual risk ${\sigma^2}/({2\theta})$ is very low: there is a systemic level 
effect of $\sigma^2$ that cannot be observed by the agents and it tends to destabilize the system. 

In this paper, we extend the previous model (\ref{eq:SDE for x_j, old model}) by introducing a central agent with 
the risk variable $x_0^{(N)}(t)$. The model we study in this paper is given by
\begin{align}
	\label{eq:mn1a}
	&dx_0^{(N)} = -h_0 V_0'(x_0^{(N)}) dt - \theta_0 (x_0^{(N)} - \bar{x}_N) dt + \frac{\sigma_0}{\sqrt{N}} dW^0_t,\quad \quad 
	\bar{x}_N=\frac{1}{N}\sum_{j=1}^N x_j,\\
	\label{eq:mn1b}
	&dx_j = -h V'(x_j)  dt - \theta \big(x_j - x_0^{(N)} \big) dt + \sigma dW^j_t, \quad \quad j=1,\ldots,N .
\end{align}
Here $V_0(x)$ and $V(x)$ are potentials with two stable states and in this paper we again assume that 
$V_0(x)=V(x)=\frac{1}{4}x^4-\frac{1}{2}x^2$ with the stable states $\pm 1$. The parameters $h_0, h\geq 0$ are the strengths of intrinsic 
stability of the central and local agents, respectively.  The parameters $\theta_0, \theta \geq 0$ determine the strength of the mean-field interactions. 
The central agent $x_0^{(N)}$ is 
intrinsically stable when $h_0>0$ and may be destabilized through a mean field interaction with the local agents where $\theta_0>0$. 
Depending on whether $h>0$ or $h=0$, the 
local agents $\{x_j\}_{j=1}^N$ are or are not intrinsically stable. They may be stabilized  through their interaction with 
the central agent $x_0^{(N)}$. The independent, standard Brownian motions $\{W_t^j\}_{j=0}^N$ model the external risk for the central and 
local agents.  We note that the normalization factor $1/\sqrt{N}$ in (\ref{eq:mn1a}) makes $x_0^{(N)}$ and $\bar{x}_N$ have external risks of comparable 
size for $N$ large, and we will assume that $\sigma_0<\sigma$ or $\sigma_0=0$ since we want the 
central agent to operate with less risk than the local agents.

In the regime of no cooperation,  $\theta_0=\theta=0$, the central agent and the local agents are independent of each 
other and Kramers' large deviation law states that when $\sigma_0$ and $\sigma$ are small, the probabilities of transition from 
one stable state to the other within the time interval $[0,T]$ are proportional to $T \exp (-2 h_0  V_0(0) / \sigma_0^2 )$ and 
$T \exp (-2 h  V(0) / \sigma^2 )$, for the central and local agents, respectively. 
We want to analyze stabilization effects in the cooperative regime $\theta_0, \theta>0$. 

In this paper, we will assume that the intrinsic stability of the local agents, $h$, is exactly zero, while we only 
assume that $h$ is small in \cite{Garnier2013}. Because of this simplifying assumption, instead of considering the pair 
$(x_0^{(N)}(t), \frac{1}{N} \sum_{j=1}^N \delta_{x_j(t)}(dx))$ as a scalar and a measure-valued process, we can simply consider 
$(x_0^{(N)}(t),\bar{x}_N(t))$ as a two-dimensional process and get results that are more detailed than it was possible in the setup of \cite{Garnier2013}. 
First, we compute numerically the minimizing path for the associated large deviation problem, and we 
are able to explore how the various parameters affect the agents' fluctuations and the systemic risk. We also recover 
the main result in \cite{Garnier2013}, that is, that the systemic risk is increased, with the local risks kept fixed, if we 
increase $\sigma^2$ and $\theta$ with the ratio $\sigma^2/\theta$ fixed. Another result is that because we assume that 
$0=h<h_0$ and $\sigma_0<\sigma$, the central agent is more stable than the
empirical mean of the local agents. In this setting, we find that 
$\theta_0$ and $\theta$ tend to play opposite roles: higher $\theta_0$ increases the systemic risk as we force the stable term $x_0^{(N)}$ 
to be close to the relatively unstable term $\bar{x}$, but on the other hand, increasing $\theta$ lowers the systemic risk as 
$\bar{x}$ tends to be close to $x_0^{(N)}$. This is the main result of this paper.
The third result here, for a case not considered in the previous paper, concerns
the introduction of optimal controls for 
the local agents. We use optimal control theory and find that the use of controls amounts to replacing $\theta$  
by an effective one that is larger, and thus it reduces the systemic risk.

This paper is organized as follows. In Section \ref{sec:mean-field limit} we state the mean field limit of the pair 
$(x_0^{(N)}(t), \frac{1}{N} \sum_{j=1}^N \delta_{x_j(t)}(dx))$ as $N\to\infty$.  We then discuss the equilibria of the limit
Fokker-Planck equation. In Section \ref{sec:zero h} 
we analyze the special case where $h$ is exactly zero. In this case, explicit solutions of the fluctuation analysis can be 
obtained, and we have a large deviations principle for $(x_0^{(N)}(t), \bar{x}_N(t))$ using the Freidlin-Wentzell theory. In Section 
\ref{sec:empirical measure} we give the formal large deviation principle for the empirical measure 
$(x_0^{(N)}(t), \frac{1}{N} \sum_{j=1}^N \delta_{x_j(t)}(dx))$ that is necessary when $h>0$. We do not use this
general formulation but we do show that the large deviation 
problems for $(x_0^{(N)}(t), \bar{x}_N(t))$ and $(x_0^{(N)}(t), \frac{1}{N} \sum_{j=1}^N \delta_{x_j(t)}(dx))$ are the same if $h=0$. In Section 
\ref{sec:optimal control} we formulate a control problem for the local agents in (\ref{eq:mn1b}) and use optimal control theory to 
analyze the effect of the control on the system. Finally, in Section \ref{sec:numerics}  we present results of extensive numerical
simulations. The technical  details of the proofs are in the appendices.

%% file: formulation.tex
\section{The mean field limit of a large number of local agents}
\label{sec:mean-field limit}
 
We begin by recalling the main results of mean field limit theory as they apply to
problem (\ref{eq:mn1a}),(\ref{eq:mn1b}), in the next section, and then discuss the 
equilibrium solutions of the limit, non-linear Fokker-Planck equation.

\subsection{The non-linear Fokker-Planck equation}

The stochastic model (\ref{eq:mn1a}),(\ref{eq:mn1b}) is a simple extension of the model in \cite{Dawson1983, Gartner1988} (see also 
\cite{Tanaka1984, Sznitman1991, Meleard1996, Kurtz1999}). We let $M_1(\RR)$ denote the space of 
probability measures endowed with the metric of the weak convergence, and $C([0,T],M_1(\RR))$ the space of continuous $M_1(\RR)$-valued processes 
in the time interval $[0,T]$ endowed with the maximum distance in $[0,T]$. In the limit $N \to \infty$, the pair 
$(x_0^{(N)}(t), \frac{1}{N} \sum_{j=1}^N \delta_{x_j(t)}(dx))$ converges in $(\RR,M_1(\RR))$ to $(y_0(t), p(t,x)dx)$ in probability, the 
weak solution of the nonlinear Fokker-Planck equation and ordinary differential equation
\begin{align}
	\label{eq:y0}
	&\frac{d}{dt}y_0 =-h_0 V_0'(y_0) -\theta_0 \Big( y_0 - \int xp(t,x) dx \Big), \\
	\label{eq:p}
	&\frac{\partial}{\partial t}  p(t,x) = h \frac{\partial}{\partial x} [V'(x) p(t,x)] + \theta \frac{\partial}{\partial x} [( x- y_0(t)) p(t,x)] 
	+ \frac{\sigma^2}{2} \frac{\partial^2}{\partial x^2} p(t,x) ,
\end{align}
with the initial condition $y_0(0)=\lim_{N \to \infty} x_0^{(N)}(0)$ and $p(0,dx)=\lim_{N\to\infty}\frac{1}{N}\sum_{j=1}^N\delta_{x_j(0)}(dx)$, given that 
the limits exist. 
Equivalently, we can characterize the pair  $(y_0(t), p(t,x)dx)$ by noting that $p(t,x)$ is the
transition  probability density of 
the process $X_t$, the solution of 
\begin{align*}
	&\frac{d}{dt}y_0 =  -h_0 V_0'(y_0)   - \theta_0 (y_0 -\EE X_t),\\
	&dX_t = - h V'(X_t)  dt - \theta ( X_t - y_0 ) dt + \sigma dW_t,
\end{align*}
where $W_t$ is a standard Brownian motion. In addition, if $h=0$ and $\bar{y}(t):=\EE(X_t)$, then $(y_0(t),\bar{y}(t))$ satisfies
\begin{align}
	\label{eq:mean-field of zero h}
	&\frac{d}{dt}y_0 =  -h_0 V_0'(y_0)   - \theta_0 ( y_0 - \bar{y} ),\\
	&\frac{d}{dt}\bar{y} = - \theta ( \bar{y} - y_0 ).
\end{align}

\subsection{Equilibrium states}

Given the existence of a stationary state $(y_0^e, p^e(x;y_0^e)):=(\lim_{t\to\infty} y_0(t), \lim_{t\to\infty} p(t,x))$, it satisfies 
\begin{equation}
	\label{eq:pe}
	p^e(x;y_0^e) = \frac{1}{Z(y_0^e)} \exp \Big( - \frac{2 h V(x) +\theta(x-y_0^e)^2 }{\sigma^2} \Big),
\end{equation}
which is obtained from  (\ref{eq:p}), and satisfies the consistency equation 
\begin{equation}
	\label{eq:ce}
	\int x  p^e(x;y_0^e) dx = y_0^e +\frac{h_0}{\theta_0} V_0'(y_0^e),
\end{equation}
obtained from (\ref{eq:y0}). If $h=0$, then $p^e(x;y_0^e)$ is a Gaussian density function, given by (\ref{eq:pe}), with mean $y_0^e$ and 
(\ref{eq:ce}) implies $V_0'(y_0^e)=0$. Therefore $y_0^e=\pm 1$. The equilibrium states for the system are determined by the 
equilibrium states of the central agent. Indeed, if the central agent takes the equilibrium value $y_0^e=-1$, then
the individual agents take a Gaussian distribution with mean $-1$ and variance $\sigma^2/(2\theta)$:
\begin{equation}
	\label{eq:statpdf0}
	p^e(x) = \frac{1}{ \sqrt{ \pi \frac{ \sigma^2}{\theta}} } 
	\exp \Big( -\frac{\theta (x+1)^2}{ \sigma^2} \Big) .
\end{equation}

When $h$ is positive but small, we let $y_0^{e0}=\pm 1$ and therefore $V_0'(y_0^{e0})=0$ with $V_0''(y_0^{e0})>0$. It is then 
possible to find an equilibrium state $y_0^{e-}$, resp. $y_0^{e+}$,  close to $y_0^{e0}=-1$, resp. $y_0^{e0}=1$, and we have
\[
	y_0^e = y_0^{e0} + h y_0^{e1} + o(h) ,
\]
with
\[
	y_0^{e1} = -\frac{\theta_0}{h_0 \theta V_0''(y_0^{e1})} \frac{ \int e^{-\theta x^2/\sigma^2} V'(y_0^{e0}+x) dx }
	{\int e^{-\theta x^2/\sigma^2} dx }.
\]
If $V_0(x)=V(x) =\frac{1}{4} x^4 -\frac{1}{2} x^2 $ then $y_0^{e0}=\pm 1$ and 
$y_0^{e1}=\mp\frac{3\theta_0 \sigma^2}{4h_0 \theta^2}$. This result shows that the positions of the equilibrium states of the 
central agent will be shifted when the individual agents have their own stabilization potential. The states $y_0^{e-}$ and 
$y_0^{e+}$ are the two equilibrium states of the central agent, and $y_0^{e-} +(h_0/\theta_0) V_0'(y_0^{e-})$ and 
$y_0^{e+} +(h_0/\theta_0) V_0'(y_0^{e+})$ are the two associated equilibrium means of the individual agents.

%% file: zero_h.tex
\section{The case of no intrinsic stabilization for the local agents ($h=0$)}
\label{sec:zero h}

In this section we consider the special case where the individual agents have no intrinsic stability, i.e., $h=0$. In this case, 
(\ref{eq:mn1b}) is linear so instead of considering the empirical distribution $\frac{1}{N} \sum_{j=1}^N \delta_{x_j(t)}(dx)$, we can 
focus on the empirical mean $\bar{x}_N(t)=\frac{1}{N}\sum_{j=1}^{N}x_j(t)$.  The pair $(x_0^{(N)}(t),\bar{x}_N(t))$ satisfies the joint SDEs:
\begin{align}
	\label{eq:dynamics of zero h}
	&dx_0^{(N)} =  -h_0 V_0'(x_0^{(N)})  dt - \theta_0 (x_0^{(N)} - \bar{x}_N) dt + \frac{\sigma_0}{\sqrt{N}} dW^0_t,\\
	&d\bar{x}_N = - \theta (\bar{x}_N - x_0^{(N)}) dt + \frac{\sigma}{\sqrt{N}} d\bar{W}_t^{(N)},\notag
\end{align}
where $\bar{W}_t^{(N)} = \frac{1}{\sqrt{N}} \sum_{j=1}^N W^j_t$ is a standard Brownian motion independent of $W^0_t$. The mean-field limit, 
$(y_0(t), \bar{y}(t)):=\lim_{N\to\infty}(x_0^{(N)}(t), \bar{x}_N(t))$,
satisfies (\ref{eq:mean-field of zero h}) with the equilibria 
$y_0^e:=\lim_{t\to\infty}y_0(t)=\pm 1$ and $\bar{y}^e:=\lim_{t\to\infty}\bar{y}(t)=\pm 1$ depending on the initial condition 
$(y_0(0), \bar{y}(0))$.

\subsection{Fluctuation analysis in the case $h=0$}

Here we analyse the fluctuations of $(x_0^{(N)}(t),\bar{x}_N(t))$ centred at $(y_0(t),\bar{y}(t))$ when $N$ is large. To simplify, we assume 
that $y_0(0)=y_0^e=-1$ and $\bar{y}(0)=\bar{y}^e=-1$, and thus $y_0(t)\equiv y_0^e=-1$ and $\bar{y}(0)\equiv\bar{y}^e=-1$. 
Define $z_0^{(N)}=\sqrt{N}(x_0^{(N)}-y_0^e)$ and $\bar{z}_N=\sqrt{N}(\bar{x}_N-\bar{y}^e)$. As $N\to\infty$, $(z_0^{(N)}, \bar{z}_N)$ 
converges in distribution to the process $(z_0, \bar{z})$ where 
\begin{align}
	\label{eq:dynamics of the fluctuations with zero h}
	& dz_0 = -h_0 V_0''(y_0^e) z_0 dt - \theta_0(z_0 - \bar{z}) dt + \sigma_0 dW^0_t,\\
	& d\bar{z} = -\theta (\bar{z} - z_0) dt + \sigma d\bar{W}_t ,  \notag
\end{align}
where $\bar{W}_t$ is a standard Brownian motion independent of $W^0_t$.
This means that, 
when $N$ is large, $x_0^{(N)}(t)\approx y_0^e+\frac{1}{\sqrt{N}}z_0$ and $\bar{x}(t)\approx \bar{y}^e+\frac{1}{\sqrt{N}}\bar{z}$ in distribution. 
Because $y_0^e=\bar{y}^e=-1$ is the normal state, $z_0$ and $\bar{z}$ are regarded as the central risks (as opposed to the large 
deviations that will be discussed in the next section) of $x_0^{(N)}$ and $\bar{x}_N$, respectively. We note that 
(\ref{eq:dynamics of the fluctuations with zero h}) is a system of linear differential equations and thus the explicit solution is:
\begin{equation*}
	\begin{pmatrix} z_0(t) \\ \bar{z}(t) \end{pmatrix}
	 = e^{t{\bf A}} \begin{pmatrix} z_0(0)\\ \bar{z}(0) \end{pmatrix}
	 + \int_0^t e^{(t-s){\bf A}} \begin{pmatrix} \sigma_0 dW^0_s \\ \sigma d\bar{W}_s \end{pmatrix},\quad
	{\bf A} = 
	\begin{pmatrix} 
		- h_0 V_0''(y_0^e)  - \theta_0 & \theta_0 \\ 
		\theta & -\theta 
	\end{pmatrix}.	 
\end{equation*}
Therefore $(z_0(t),\bar{z}(t))$ is a Gaussian process with
\begin{equation}
	\label{eq:expectations of fluctuations}
	\EE\begin{pmatrix} z_0(t) \\ \bar{z}(t) \end{pmatrix} = e^{t{\bf A}}\begin{pmatrix} z_0(0)\\ \bar{z}(0) \end{pmatrix},
\end{equation}
\begin{equation}
	\label{eq:covariance matrix for fluctuations}
	\begin{pmatrix} 
		\mathbf{Var}z_0(t) & \mathbf{Cov}(z_0(t), \bar{z}(t)) \\ 
		\mathbf{Cov}(z_0(t), \bar{z}(t)) & \mathbf{Var}\bar{z}(t) 
	\end{pmatrix}
	= \int_0^t e^{(t-s){\bf A}} \begin{pmatrix} \sigma_0^2 & 0 \\ 0 & \sigma^2 \end{pmatrix} e^{(t-s){\bf A}^\mathbf{T}} ds.
\end{equation}

We want to analyse the impact of the various parameters on $(z_0(t), \bar{z}(t))$, in particular, for the case that $t\to\infty$ and 
$\sigma,\theta\to\infty$ with a fixed ratio $\alpha:=\sigma^2/\theta<\infty$. To do this, we use the eigen-decomposition 
of ${\bf A}$ to compute (\ref{eq:covariance matrix for fluctuations})  and obtain the following.
\begin{proposition}
	\label{prop:limits of fluctuations}
	If $h_0$, $\theta_0$ and $\theta$ are positive, then $\lim_{t\to\infty}\EE z_0(t)=\lim_{t\to\infty}\EE\bar{z}(t)=0$. In addition, 
	the variances and covariance of the fluctuations $z_0(t)$ and $\bar{z}(t)$ have the following limits as $t\to\infty$ and 
	$\sigma,\theta\to\infty$ with a fixed ratio $\alpha=\sigma^2/\theta<\infty$:
	\begin{equation}
		\label{eq:limit of Var delta y_0, large theta}
		\lim_{\substack{\sigma,\theta\to\infty\\ \sigma^2/\theta=\alpha}} \lim_{t\to\infty}
		\mathbf{Var}z_0(t)=\frac{\sigma_0^2}{2h_0 V_0''(y_0^e)},
	\end{equation}
	\begin{equation}
		\label{eq:limit of Var delta y_bar, large theta}
		\lim_{\substack{\sigma,\theta\to\infty\\ \sigma^2/\theta=\alpha}} \lim_{t\to\infty}
		\mathbf{Var}\bar{z}(t)=\frac{\sigma_0^2}{2h_0 V_0''(y_0^e)} + \frac{\sigma^2}{2\theta}, 
	\end{equation}
	\begin{equation}
		\label{eq:limit of Cov(delta y_0, delta y_bar), large theta}
		\lim_{\substack{\sigma,\theta\to\infty\\ \sigma^2/\theta=\alpha}} \lim_{t\to\infty}
		\mathbf{Cov}(z_0(t), \bar{z}(t))=\frac{\sigma_0^2}{2h_0 V_0''(y_0^e)}.
	\end{equation}
	This means that after the limits are applied, $z_0=Z_1$ and $\bar{z}=Z_1+Z_2$, where $Z_1$ and $Z_2$ are two independent 
	Gaussian random variables with mean $0$ and variances $\frac{\sigma_0^2}{2h_0 V_0''(y_0^e)}$ and $\frac{\sigma^2}{2\theta}$, 
	respectively.
\end{proposition}
\begin{proof}
	This involves basic computations given in Appendix \ref{pf:limits of fluctuations}.
\end{proof}

We see that the variances and the covariance of the limits of $z_0$ and $\bar{z}$ increase with increasing $\sigma_0$ or 
decreasing $h_0$. We also note that these three statistics blow up as $\sigma_0\to\infty$ even if 
$\sigma_0^2/\theta_0$ is finite and small. This is because when $h$ is exactly zero, $\bar{x}_N$ cannot serve as a stabilizing 
term and $x_0^{(N)}$ cannot diversify its risk to $\bar{x}_N$ by increasing $\theta_0$.

\subsection{Large deviations}
\label{sec:large deviations}

\subsubsection{A general large deviation principle}

From the mean field and fluctuation analysis we see that if $N$ is large and $x_0^{(N)}(0)=x_j(0)=-1$ for all $j=1,\ldots,N$, then one 
can expect that $(x_0^{(N)}(t),\bar{x}_N(t))\approx(y_0^e,\bar{y}^e)=(-1,-1)$
for all $t$. However, as long as $N$ is finite, $x_0^{(N)}(t)$ and 
$\bar{x}_N(t)$ are stochastic processes and therefore the event that the overall system has a transition in a finite time interval has a 
small but nonzero probability. Mathematically speaking, we consider the event of the continuous paths 
$(x_0^{(N)}(t),\bar{x}_N(t))\in C([0,T],\RR^2)$ starting from $(y_0^{e-},\bar{y}^{e-}):=(-1,-1)$ 
at time $0$ to ending around 
$(y_0^{e+},\bar{y}^{e+}):=(1,1)$ at time $T$:
\begin{multline}
	\label{eq:event of system crash}
	\mathcal{A}_\delta 
	= \big\{(x_0(t),\bar{x}(t))_{t \in [0,T]} \in C([0,T],\RR^2) : \\
	(x_0(0),\bar{x}(0))=(-1,-1), \|(x_0(T),\bar{x}(T))-(1,1)\|\leq\delta\big\},
\end{multline}
where $\|\cdot\|$ is the standard Euclidean norm in $\RR^2$.

The Freidlin-Wentzell theory \cite[Section 5.6]{Dembo2010}
 says that, for $N$ large, $\PP((x_0^{(N)},\bar{x}_N)\in\mathcal{A}_\delta)$ satisfies the following large deviation principle:
\begin{align*}
	-\inf_{\bx\in\mathring{\mathcal{A}_\delta}} I(\bx) 
	&\leq \liminf_{N\to\infty}\frac{1}{N}\log\PP\big((x_0^{(N)},\bar{x}_N)\in\mathcal{A}_\delta\big)\\
	&\leq \limsup_{N\to\infty}\frac{1}{N}\log\PP\big((x_0^{(N)},\bar{x}_N)\in\mathcal{A}_\delta \big)
	\leq -\inf_{\bx\in\bar{\mathcal{A}_\delta}} I(\bx),
\end{align*}
where $\mathring{\mathcal{A}_\delta}$ and $\bar{\mathcal{A}_\delta}$ are the interior and closure of $\mathcal{A}_\delta$ under 
the standard $C([0,T],\RR^2)$-topology, respectively, and
$I(\bx)$ is the rate function for the exponential 
decay of the probability that will be specified later. By using a similar argument as in \cite[Lemma 5.2]{Garnier2013}, we can show that for 
any $\epsilon>0$, there exists sufficiently small $\delta>0$ such that 
\begin{align*}
	-\inf_{\bx\in\mathcal{A}} I(\bx) 
	&\leq \liminf_{N\to\infty}\frac{1}{N}\log\PP\big((x_0^{(N)},\bar{x}_N)\in\mathcal{A}_\delta\big)\\
	&\leq \limsup_{N\to\infty}\frac{1}{N}\log\PP\big((x_0^{(N)},\bar{x}_N)\in\mathcal{A}_\delta \big) 
	\leq -\inf_{\bx\in\mathcal{A}} I(\bx) + \epsilon,
\end{align*}
where 
\begin{multline}
	\mathcal{A} = \big\{(x_0(t),\bar{x}(t))_{t \in [0,T]}\in C([0,T],\RR^2):   \\
	(x_0(0),\bar{x}(0))=(-1,-1), (x_0(T),\bar{x}(T))=(1,1)\big\}.
\end{multline}
In other words, for large $N$ and small $\delta$,
\begin{equation}
	\label{eq:systemic risk in LD}
	\PP\big((x_0^{(N)},\bar{x}_N)\in\mathcal{A}_\delta\big) \approx \exp\left(-N\inf_{\bx\in\mathcal{A}} I(\bx)\right),
\end{equation}
and we define this probability as the systemic risk of the overall system. We will discuss the rate 
function $I(\bx)$ separately for the cases that $\sigma_0=0$ and $\sigma_0>0$ in the following sections. We will next compute the minimum of the rate function $\inf_{\bx\in\mathcal{A}} I(\bx)$ to obtain the 
systemic risk in (\ref{eq:systemic risk in LD}).

The minimizer $\bx^*=\arg\min_{\bx\in\mathcal{A}_\delta} I(\bx)$ is \textbf{the most probable path} for 
\textbf{the rare event} $\mathcal{A}_\delta$ in the sense that the mass of the conditional probability  
$\PP(\cdot|\mathcal{A}_\delta)$ is concentrated around $\bx^*$ exponentially fast as $N\to\infty$ . Indeed, if $\bx^*$ exists 
and is unique, then for any open neighbourhood $\mathbf{N}(\bx^*)$ containing $\bx^*$,
\begin{multline}
	\label{eq:conditional probability of the most probable path}
	\PP((x_0^{(N)},\bar{x}_N)\in \mathbf{N}(\bx^*)|(x_0^{(N)},\bar{x}_N)\in \mathcal{A}_\delta) \\
	= 1 - \PP((x_0^{(N)},\bar{x}_N)\notin \mathbf{N}(\bx^*)|(x_0^{(N)},\bar{x}_N)\in \mathcal{A}_\delta)\\
	= 1 - \frac{\PP((x_0^{(N)},\bar{x}_N)\in \mathbf{N}^C(\bx^*)\cap\mathcal{A}_\delta)}{\PP((x_0^{(N)},\bar{x}_N)\in\mathcal{A}_\delta)}\\
	\gtrsim 1 - \frac{\exp(-N\inf_{\bx\in\mathbf{N}^C(\bx^*)\cap\mathcal{A}_\delta}I(\bx))}
	{\exp(-N\inf_{\bx\in\mathcal{A}_\delta}I(\bx))}
	\overset{N\to\infty}{\to} 1	,
\end{multline}
by using the fact that $\bx^*$ is unique and $\mathcal{A}_\delta$ is closed.

\subsubsection{Degenerate case}

We first consider the degenerate case where $\sigma_0=0$ and $\sigma>0$. Then (\ref{eq:dynamics of zero h}) becomes 
\begin{align*}
	&\frac{d}{dt}x_0^{(N)} =  -h_0 V_0'(x_0^{(N)}) - \theta_0 (x_0^{(N)} - \bar{x}_N),\\
	&d\bar{x}_N = - \theta (\bar{x}_N - x_0^{(N)}) dt + \frac{\sigma}{\sqrt{N}} d\bar{W}_t^{(N)}.
\end{align*}
 The rate function $I(\bx)$ in (\ref{eq:systemic risk in LD}) is of the form
\begin{equation}
	\label{eq:rate function, degenerate case}
	I(\bx) = I(x_0, \bar{x}) = \frac{1}{2\sigma^2} \int_0^T \left( \dot{\bar{x}}(t) + \theta(\bar{x}(t) -x_0(t)  \right)^2 dt,
\end{equation}
if $(\bar{x}(t))_{t \in [0,T]}$ is absolutely continuous in time and $\dot{x}_0 = -h_0 V_0'(x_0)  - \theta_0(x_0-\bar{x})$ and $I(x_0, \bar{x})=+\infty$ otherwise.
Here the dot stands for a time derivative.
By (\ref{eq:systemic risk in LD}), in order to compute the systemic risk, we need to solve the optimization problem:
\begin{equation}
	\label{eq:optimization of I in terms of x_bar, degenerate case}
	\inf_{\bar{x}(t)} \frac{1}{2\sigma^2} \int_0^T \left( \dot{\bar{x}}(t) + \theta(\bar{x}(t) - x_0(t)  \right)^2 dt,
\end{equation}
with the constraints that $(\bar{x}(t))_{t \in [0,T]}$ is absolutely continuous in time, $\dot{x}_0 = -h_0 V_0'(x_0)  - \theta_0(x_0-\bar{x})$, 
$x_0(0)=\bar{x}(0)=-1$ and $x_0(T)=\bar{x}(T)=1$. By using  
$\bar{x}=\frac{1}{\theta_0}\dot{x}_0+\frac{h_0}{\theta_0}V'(x_0)+x_0$, the constrained optimization problem is equivalent to 
\begin{equation}
	\label{eq:optimization of I in terms of x_0, degenerate case}
	\inf_{x_0}\frac{1}{2\sigma^2}\int_0^T\left[\frac{1}{\theta_0}\ddot{x}_0+\frac{h_0}{\theta_0}V''_0(x_0)\dot{x}_0
 	+(1+\frac{\theta}{\theta_0})\dot{x}_0+\frac{\theta h_0}{\theta_0}V'_0(x_0)\right]^2 dt ,
\end{equation}
with the boundary conditions $x_0(0)=-1$, $x_0(T)=1$ and $\dot{x}_0(0)=\dot{x}_0(T)=0$. 
From basic calculus of variations, 
the minimizer $x_0$ satisfies a fourth-order boundary 
	value problem that we describe in the fllowing proposition.
	
\begin{proposition}
	\label{prop:BVP for x0, degenerate case}
	The minimizer $(x_0,\bar{x})$ of $\inf_{(x_0,\bar{x})\in\mathcal{A}} I(x_0,\bar{x})$ of the rate function (\ref{eq:rate function, degenerate case}) satisfies the following boundary 
	value problem
	\begin{align}
		\label{eq:BVP for x0, degenerate case}
		&\frac{d^4}{dt^4}x_0 - (\theta_0+\theta)^2\frac{d^2}{dt^2}x_0 
		+ h_0\Bigg[V''''_0(x_0)\left(\frac{d}{dt}x_0\right)^3+3V'''_0(x_0)\left(\frac{d}{dt}x_0\right)\left(\frac{d^2}{dt^2}x_0\right)\\
		&\quad - \theta_0V'''_0(x_0)\left(\frac{d}{dt}x_0\right)^2-2\theta_0V''_0(x_0)\left(\frac{d^2}{dt^2}x_0\right)\Bigg]\notag\\
		&\quad + h_0^2V''_0(x_0)\left[-V'''_0(x_0)\left(\frac{d}{dt}x_0\right)^2
		-V''_0(x_0)\left(\frac{d^2}{dt^2}x_0\right)+\theta^2V'_0(x_0)\right] = 0 , \notag
	\end{align}
	with $x_0(0)=-1$, $x_0(T)=1$, $\frac{d}{dt}x_0(0)=\frac{d}{dt}x_0(T)=0$, and
$$
\bar{x}(t) = \frac{1}{\theta_0}\frac{d}{dt}{x}_0(t)+\frac{h_0}{\theta_0}V'(x_0(t))+x_0(t)
.
$$	
\end{proposition}
\begin{proof}
	See Appendix \ref{pf:BVP for x0, degenerate case}.
\end{proof}

If $h_0=0$, we can solve $x_0$ and $\bar{x}$ explicitly. The boundary value problem 
(\ref{eq:BVP for x0, degenerate case}) is then
\begin{equation}
	\label{eq:BVP for x0, degenerate case, h0=0}
	\frac{d^4}{dt^4} x_0 - (\theta_0+\theta)^2 \frac{d^2}{dt^2} x_0 = 0, 
\end{equation}
with the boundary conditions $x_0(0)=-1$, $\frac{d}{dt} x_0(0)=0$, $x_0(T)=1$ and $\frac{d}{dt} x_0 (T)=0$. The associated 
minimizer $\bar{x}$ is $\bar{x}(t)=x_0(t)+\frac{1}{\theta_0}\frac{d}{dt}x_0(t)$. The solution of 
(\ref{eq:BVP for x0, degenerate case, h0=0}) is  
\begin{align}
	& x_0 (t)  = \frac{ (1+e^{-(\theta_0+\theta)T}) (2t-T) + \frac{2}{(\theta_0+\theta)}e^{-(\theta_0+\theta) t} 
	- \frac{2}{(\theta_0+\theta)}e^{-(\theta_0+\theta) (T-t)}}
	{T (1+e^{- (\theta_0+\theta)T}) +\frac{2}{(\theta_0+\theta)}( e^{- (\theta_0+\theta)T}-1)},\\
	& \bar{x}(t) = x_0(t) + \frac{2}{\theta_0}\frac{(1+e^{- (\theta_0+\theta)T})-e^{-(\theta_0+\theta) t} 
	-e^{-(\theta_0+\theta) (T-t)}}{T (1+e^{- (\theta_0+\theta)T}) +\frac{2}{(\theta_0+\theta)}( e^{- (\theta_0+\theta)T}-1)}.
\end{align}

These are the most probable paths followed by the two processes to realize
 the rare event asociated with the systemic risk. Note that 
$\bar{x}(t)$ is ahead of $x_0(t)$, which means that the individual agents drive the transition. 
We also obtain the following proposition.
\begin{proposition}
	If $h_0=h=0$, then the probability of transition is
	\begin{equation}
		\PP\big((x_0^{(N)},\bar{x}_N)\in\mathcal{A}_\delta\big) \approx \exp \left( - \frac{2 N (\theta_0+\theta)^2}{\sigma^2 \theta_0^2 }
		\frac{1 + e^{-(\theta_0+\theta) T}}{ T  (1 + e^{-(\theta_0+\theta) T}) - \frac{2}
		{\theta_0+\theta}(1 - e^{-(\theta_0+\theta) T})}\right)  .
	\end{equation}
\end{proposition}
For large $T$ (i.e. $(\theta_0 +\theta)T \gg 1$), the most probable paths are
\begin{equation}
	x_0(t) \approx \bar{x}(t) \approx -1+\frac{2t}{T}  ,
\end{equation}
and  the probability of transition  is
\begin{equation}
	\PP\big( (x_0^{(N)},\bar{x}_N)\in\mathcal{A}_\delta\big) 
	\approx \exp \left( - \frac{2 N}{\sigma^2 T} \frac{ (\theta_0 + \theta)^2}{\theta_0^2} \right).
\end{equation}
This shows that stability increases with $\theta$ and decreases with $\theta_0$. This is because when $\sigma_0=0$ and 
$\sigma>0$, $x_0$ is a stabilizing term while $\bar{x}$ is a destabilizing term. When $\theta$ increases, $\bar{x}$ (unstable) is 
forced to be close to $x_0$ (stable), and therefore the systemic risk is reduced. On the other hand, the systemic risk is higher if 
$\theta_0$ increases, as we make $x_0$ stay close to $\bar{x}$.


\subsubsection{Non-degenerate case}

We next consider the non-degenerate case where $\sigma_0$ and $\sigma$ are positive. 
In this case,  the rate 
function $I(\bx)$ in (\ref{eq:systemic risk in LD}) has the form
\begin{equation}
	\label{eq:rate function, nondegenerate case}
	I(\bx) = I(x_0, \bar{x}) 
	= \frac{1}{2\sigma_0^2} \int_0^T ( \dot{x}_0 + h_0 V_0'(x_0) + \theta_0(x_0 - \bar{x}) )^2 dt
	+ \frac{1}{2\sigma^2} \int_0^T ( \dot{\bar{x}} + \theta(\bar{x} -x_0))^2 dt,
\end{equation}
if $(x_0(t))_{t\in [0,T]}$ and $(\bar{x}(t))_{t\in [0,T]}$ are absolutely continuous in time and $I(x_0, \bar{x})=+\infty$ otherwise. Again by the calculus of 
variations, the minimizer $(x_0,\bar{x})$ of $\inf_{(x_0,\bar{x})\in\mathcal{A}} I(x_0,\bar{x})$ satisfies a system of second-order 
ordinary differential equations.
\begin{proposition}
	\label{prop:BVP for x0 and x_bar, nondegenerate case}
	The minimizer $(x_0,\bar{x})$ of $\inf_{(x_0,\bar{x})\in\mathcal{A}} I(x_0,\bar{x})$ of the rate function 
	(\ref{eq:rate function, nondegenerate case}) satisfies the following system of second order boundary value problems
	\begin{align}
		\label{eq:BVP for x0 and x_bar, nondegenerate case}
		\frac{d^2}{dt^2}x_0 
		&= \frac{1}{\sigma^2}(\sigma^2\theta_0-\sigma_0^2\theta)\frac{d}{dt}\bar{x}
		+ \frac{1}{\sigma^2}(\sigma^2\theta_0^2+\sigma_0^2\theta^2)(x_0-\bar{x})\\
		&\quad + h_0\theta_0\left[V'_0(x_0)+V''_0(x_0)(x_0-\bar{x})\right]+h_0^2V'_0(x_0)V''_0(x_0) \notag \\
		\frac{d^2}{dt^2}\bar{x}
		&= \frac{1}{\sigma_{0}^{2}}(\sigma_{0}^{2}\theta-\sigma^{2}\theta_{0})\frac{d}{dt}x_0
		+ \frac{1}{\sigma_{0}^{2}}(\sigma_{0}^{2}\theta^{2}+\sigma^{2}\theta_{0}^{2})(\bar{x}-x_0)
		- h_0\frac{\sigma^2\theta_0}{\sigma_0^2}V'_0(x_0) , \notag
	\end{align}
	with $x_0(0)=\bar{x}(0)=-1$ and $x_0(T)=\bar{x}(T)=1$.
\end{proposition}
\begin{proof}
	The proof is essentially the same as the proof of Proposition \ref{prop:BVP for x0, degenerate case} in Appendix 
	\ref{pf:BVP for x0, degenerate case} and thus is omitted.
\end{proof}

Although (\ref{eq:BVP for x0 and x_bar, nondegenerate case}) is solvable when $h_0=0$, the explicit solution is very 
complicated even for zero $h_0$. Therefore we compute the transition probability by using the fact that $(x_0(T),\bar{x}(T))$ are 
jointly Gaussian random variables and obtain the exponential rate of the decay of the probability.

\begin{proposition}
	\label{prop:exponential rate of the transition probability, nondegenerate case}
	If $h_0=h=0$ and $x_0(0)=\bar{x}(0)=-1$, then the probability of  transition has the following exponential rate of decay:
	\begin{equation}
		\label{eq:exponential rate of the transition probability, nondegenerate case}
		\PP\big( (x_0^{(N)},\bar{x}_N)\in\mathcal{A}_\delta\big) 
		\approx \exp \Big(- N \frac{2(\theta_0+\theta)^2}{T(\theta^2\sigma_0^2+\theta_0^2\sigma^2)} \Big),
	\end{equation}
	for large $T$.
\end{proposition}
\begin{proof}
	See Appendix \ref{pf:exponential rate of the transition probability, nondegenerate case}.
\end{proof}

\subsubsection{The case that $h_0>0$}

Most of the large deviation analysis in this section is 
about the case  $h_0=0$ in order to have explicit results. Although it 
is also possible to consider the case that $0<h_0\ll 1$ and use the small $h_0$ analysis, we will solve the large deviation problems 
numerically as the associated boundary value problems (\ref{prop:BVP for x0, degenerate case}) and 
(\ref{eq:BVP for x0 and x_bar, nondegenerate case}) can be solved easily by standard numerical methods. The details of the 
numerical analysis are presented in Section \ref{sec:numerics}.

%% file: empirical_measure.tex
\section{Formal large deviations for the empirical measures}
\label{sec:empirical measure}

In this section, we extend 
the large deviations formulation from the space of real-valued processes $(x_0^{(N)}(t), \bar{x}_N(t))_{t \in [0,T]}$ to the space of 
probability-measure-valued processes $(x_0^{(N)}(t), U_N(t,dx))_{t \in [0,T]}$, where $U_N(t,dx):=\frac{1}{N}\sum_{j=1}^{N}\delta_{x_j(t)}(dx)$. The 
reason we consider a more general and complicated space is that there is no closed equation for $\bar{x}_N$ when $h>0$, because 
(\ref{eq:mn1b}) is not linear for non-zero $h$. In addition, we obtain more information by considering the more general space 
even for $h=0$ and we show that when $h=0$ the generalized problem is (at least formally) equivalent to the problem we considered in the previous section.

We also note that there are no existing large deviation results for $(x_0^{(N)}(t), U_N(t,dx))_{t \in [0,T]}$ satisfying (\ref{eq:mn1a}) and (\ref{eq:mn1b}) 
even if $h=0$; the current most general large deviation principle for weakly interacting particle systems is \cite{Budhiraja2012}, 
but unfortunately our model still cannot be covered. Thus the results in this section are formal.

Motivated by \cite{Dawson1987}, the (formal) rate function for $(x_0^{(N)}(t), U_N(t,dx))_{t \in [0,T]}$ satisfying (\ref{eq:mn1a}) and (\ref{eq:mn1b}) is 
\begin{multline*}
	\mathcal{J}\big( (x_0(t),\phi(t,dx))_{t \in [0,T]} \big) = \frac{1}{2\sigma_0^2} \int_0^T ( \dot{x}_0 + h_0 V_0'(x_0) + \theta_0(x_0 - \bar{x}) )^2 dt\\
	+ \frac{1}{2\sigma^2} \int_0^T \sup_{f(x): \langle \phi, (f'(x))^2 \rangle \neq 0}
	\frac{\langle \phi_t - h\frac{\partial}{\partial x}[V'(x)\phi] - \frac{1}{2}\sigma^2 \phi_{xx} 
	-\theta\frac{\partial}{\partial x}[(x-x_0(t))\phi], f(x)\rangle^2}
	{\langle \phi, (f'(x))^2 \rangle}dt ,
\end{multline*}
for $\sigma_0>0$ and for $\sigma_0=0$,
\begin{multline*}
	\mathcal{J}\big( (x_0(t),\phi(t,dx))_{t \in [0,T]}\big)\\
	= \frac{1}{2\sigma^2} \int_0^T \sup_{f(x): \langle \phi, (f'(x))^2 \rangle \neq 0}
	\frac{\langle \phi_t - h\frac{\partial}{\partial x}[V'(x)\phi] - \frac{1}{2}\sigma^2 \phi_{xx} 
	-\theta\frac{\partial}{\partial x}[(x-x_0(t))\phi], f(x)\rangle^2}
	{\langle \phi, (f'(x))^2 \rangle}dt,
\end{multline*}
if $\dot{x}_0 + h_0 V_0'(x_0) + \theta_0(x_0 - \bar{x})=0$ or $\mathcal{J}\big( (x_0(t),\phi(t,dx))_{t \in [0,T]}\big)=\infty$ otherwise. 
Here $f$ is in the Schwartz space, $\langle \phi, f(x)\rangle=\int f(x)\phi(t,dx)$, and the partial derivatives 
($\frac{\partial}{\partial t}$, $\frac{\partial}{\partial x}$, $\frac{\partial^2}{\partial x^2}$) are defined in the weak sense.

By the contraction principle \cite[Theorem 4.2.1]{Dembo2010}, if the large deviation principle for $(x_0^{(N)}(t), U_N(t,dx))_{t \in [0,T]}$ exists, then 
by using the projection $x_0^{(N)}(t)\mapsto x_0^{(N)}(t)$ and $U_N(t,dx)\mapsto \bar{x}_N(t) =\langle U_N(t,dx), x \rangle$, the large 
deviation principle for $(x_0^{(N)}(t), \bar{x}_N(t))_{t \in [0,T]}$ also exists with rate function
\begin{equation}
	\label{cont0}
	\mathcal{I}\big( (x_0(t), \bar{x}(t))_{t \in [0,T]}\big) = \inf_{ \phi(t,dx): \langle \phi(t,dx), x\rangle =\bar{x}(t)
	\forall t \in [0,T]} \mathcal{J} \big( (x_0(t), \phi(t,dx))_{t \in [0,T]}\big).
\end{equation}
The following result shows that when $h=0$, for either $\sigma_0=0$ or $\sigma_0>0$, 
$\mathcal{I}\big( (x_0(t), \bar{x}(t))_{t \in [0,T]}\big)=I\big( (x_0(t), \bar{x}(t))_{t \in [0,T]}\big)$ in (\ref{eq:rate function, degenerate case}) or 
(\ref{eq:rate function, nondegenerate case}), respectively.

\begin{proposition}
	\label{prop:gaussianpath}
	If $h=0$, then the infimum in (\ref{cont0}) is reached for and only for the path of Gaussian density functions
	\begin{equation}
		\label{eq:gaussianpath}
		\bar{p}(t,x) = \frac{1}{\sqrt{2\pi \frac{\sigma^2}{2\theta}} } 
		\exp \left( -\frac{ (x-\bar{x}(t))^2}{2\frac{\sigma^2}{2\theta}}\right).
	\end{equation}
	In addition, $\mathcal{I}\big( (x_0(t), \bar{x}(t))_{t \in [0,T]}\big)=I
	\big( (x_0(t), \bar{x}(t))_{t \in [0,T]}\big)$ in (\ref{eq:rate function, degenerate case})  for $\sigma_0=0$ and  
	in (\ref{eq:rate function, nondegenerate case}) for $\sigma_0>0$.
\end{proposition}
\begin{proof}
	See Appendix \ref{pf:gaussianpath}.
\end{proof}
In other words, when $h=0$, we can simply consider the large deviation problem for $(x_0(t), \bar{x}_N(t))$ in Section 
\ref{sec:zero h} instead of $(x_0(t), X_N(t,dx))$ in a complicated space. 

However, if $h>0$, then it is necessary to consider $(x_0^{(N)}(t), U_N(t,dx))_{t \in [0,T]}$ with rate function $\mathcal{J}\big( (x_0(t),\phi(t,dx))_{t \in [0,T]}\big)$ as 
now the large deviations for $(x_0^{(N)}(t), \bar{x}_N(t,dx))_{t \in [0,T]}$ cannot be obtained by the Freidlin-Wentzell theory. Motivated from 
Proposition \ref{prop:gaussianpath} and \cite[Section 7]{Garnier2013}, we know that because for $h=0$, the most probable path for 
the empirical measure $U_N(t,dx)$ is the Gaussian probability measure $\bar{p}(t,x)dx$, it is reasonable to assume that for 
$0<h\ll 1$, the most probable $U_N(t,dx)$ is a Gaussian probability measure plus higher order corrections in $h$. In addition, as 
the base case ($h=0$) is Gaussian, we parametrize the most probable path of the density $\phi(t,x)$ by the Hermite expansion: 
$\phi=p+hq^{(1)}+h^2q^{(2)}+\cdots$, where
\begin{align*}
	&p(t,x) = \frac{1}{\sqrt{2\pi \frac{\sigma^2}{2\theta}} } \exp \left( -\frac{ (x-\mu(t))^2}{2\frac{\sigma^2}{2\theta}}\right),\quad
	\mu(t) = \langle \phi(t,x)dx, x \rangle,\\
	&q^{(1)}(t,x) = \sum_{n=2}^\infty \beta_n(t) \frac{\partial^n}{\partial x^n}p(t,x),\quad
	q^{(2)}(t,x) = \sum_{n=2}^\infty \gamma_n(t) \frac{\partial^n}{\partial x^n}p(t,x).
\end{align*}
Then
\begin{equation*}
	\min_{x_0,\phi}\mathcal{J}\big( (x_0(t),\phi(t,dx))_{t \in [0,T]}\big)
	=\min_{x_0,\mu,\beta_n,\gamma_n}\mathcal{J}\big( (x_0(t),\mu(t),\beta_n(t),\gamma_n(t))_{t \in [0,T]}\big) + o(h^2),
\end{equation*}
and we can solve the associated variational problems for $x_0(t)$, $\mu(t)$, $\beta_n(t)$ and $\gamma_n(t)$ as in
\cite[Section 7]{Garnier2013}. This task is not carried out in this paper.

%% file: control.tex
\section{Optimal control of the central agent}
\label{sec:optimal control}

In this section, we consider an optimal control problem by introducing a
 control term $\alpha_j(t)$ into (\ref{eq:mn1b}).
In order to be able to address the problem in a manageable way 
and to discuss the role of the parameters, we will write it as a 
linear-quadratic-Gaussian control problem as in \cite{Carmona2013}. 
We let $h=0$ and define $X_0^{(N)}(t)=x_0^{(N)}(t)-y_0^e=x_0^{(N)}(t)+1$ and 
$X_j(t)=x_j(t)-\bar{y}^e=x_j(t)+1$. By assuming that $X_0^{(N)}(t)$ is small so that 
$h_0 V_0'(x_0^{(N)}(t)) =  h_0 V_0'(y_0^e+X_0^{(N)}(t)) \approx H_0 X_0^{(N)}(t)$ with $H_0\geq 0$, 
we have
\begin{align}
	\label{eq:mn1a with control}
	&dX_0^{(N)} = -H_0 X_0^{(N)} dt - \theta_0(X_0^{(N)} - \bar{X}_N) dt + \frac{\sigma_0}{\sqrt{N}} dW^0_t,\quad 
	\bar{X}_N=\frac{1}{N}\sum_{j=1}^N X_j,\\
	\label{eq:mn1b with control}
	&dX_j = - \theta (X_j - X_0^{(N)}) dt + \sigma dW^j_t + \alpha_j dt, \quad \quad j=1,\ldots,N .
\end{align}
The optimal controls $\alpha_j$ are adapted to the 
past $\{(X_j(s))_{j=0,\ldots,N}, 0\leq s \leq t\}$ and such that the following cost function is minimized:
\begin{equation}
	\label{eq:cost function for optimal control}
	J(\alpha_1, \ldots, \alpha_N) = \frac{1}{2} \sum_{j=1}^N \EE \left[ \int_0^T \alpha_j^2(t) + \theta_c^2 (X_0^{(N)}(t)-X_j(t))^2 dt\right].
\end{equation}
This cost function means that the optimal controls try to make $X_j$ close to $X_0^{(N)}$ with a quadratic cost. We can regard the term 
$-\theta(X_j-X_0^{(N)})$ as a passive feedback while $\alpha_j$ is the active feedback from the central agent. A possible control (but 
not optimal as we will see) is to take the active feedback $\alpha_j=-\tilde{\theta}_c(X_j-X_0^{(N)})$ for some
well chosen $\tilde{\theta}_c$. The goal of this 
section is to study the form of feedback that the optimal control produces and whether it is different from the passive feedback 
$-\theta(X_j-X_0^{(N)})$. By using standard theory, we have the following optimal control $\alpha_j(t)$ for $(X_0^{(N)}(t), \bar{X}_N(t))$.
\begin{proposition}
	\label{prop:optimal control}
	The optimal control $\alpha_j(t)$ that minimizes $J$ in (\ref{eq:cost function for optimal control}) where $(X_0^{(N)}(t), \bar{X}_N(t))_{t \in [0,T]}$ 
	satisfies (\ref{eq:mn1a with control}) and (\ref{eq:mn1b with control}) is 
	\begin{equation}
		\label{eq:optimal control}
		\alpha_j(t)= -\theta_c \left(b(t)X_0^{(N)}(t) + d(t)X_j(t) + e(t)\bar{X}_N(t)\right),\quad j=1,\ldots, N,
	\end{equation}
	where $(a(t), b(t), d(t), e(t))_{t \in [0,T]}$ is the solution of the following Riccati equations:
	\begin{align}
		\label{eq:Riccati equation}
		\dot{a}(t) &= 2(\theta_0+H_0) a(t) - 2 \theta b(t) + \theta_c b^2(t) -\theta_c,\\
		\dot{b}(t) &= (\theta_0+H_0+\theta) b(t) -\theta d(t) -\theta_0 a(t) 
		+ \theta_c b(t)d(t) +\theta_c -\theta e(t) +\theta_c b(t)e(t),\notag\\
		\dot{d}(t) &= 2 \theta d(t) +\theta_c d^2(t) -\theta_c,\notag\\
		\dot{e}(t) &= -2\theta_0 b(t) +2 \theta e(t) + \theta_c (2d(t)e(t) +e^2(t)),\notag
	\end{align}
	with the terminal conditions $(a(T), b(T), d(T), e(T))=(0, 0, 0, 0)$.
\end{proposition}
\begin{proof}
	See Appendix \ref{pf:optimal control}.
\end{proof}

When $T\to\infty$ we have 
\begin{equation}
	\label{eq:stationary optimal control}
	\alpha_j(t)= -\theta_c \left(b_\infty X_0^{(N)}(t) + d_\infty X_j(t) + e_\infty \bar{X}_N(t)\right),
\end{equation}
where the parameters $(a_\infty, b_\infty, d_\infty, e_\infty)$ satisfy the algebraic Riccati equations:
\begin{align}
	\label{eq:algebraic Riccati equation}
	0 &= 2(\theta_0+H_0) a_\infty - 2\theta b_\infty + \theta_c b_\infty^2 -\theta_c,\\
	0 &= (\theta_0+H_0+\theta) b_\infty - \theta d_\infty - \theta_0 a_\infty 
	+ \theta_c b_\infty d_\infty + \theta_c - \theta e + \theta_c b_\infty e_\infty,\notag\\
	0 &= \theta d_\infty + \theta_c d_\infty^2 - \theta_c,\notag\\
	0 &= -2\theta_0 b_\infty + 2 \theta e_\infty + \theta_c (2d_\infty e_\infty + e_\infty^2).\notag
\end{align}
In these conditions $(X_0^{(N)}, \bar{X}_N)$ satisfies the SDE:
\begin{align*}
	&dX_0^{(N)} = -H_0 X_0^{(N)} dt - \theta_0 (X_0^{(N)} - \bar{X}_N)dt + \frac{\sigma_0}{\sqrt{N}} dW^0_t,\\
	&d\bar{X}_N=-\theta(\bar{X}_N-X_0^{(N)})dt+\frac{\sigma}{\sqrt{N}}d\bar{W}_t^{(N)}
	-\theta_c \left(b_\infty X_0^{(N)}+(d_\infty+e_\infty) \bar{X}_N\right)dt,
\end{align*} 
where $\bar{W}^{(N)}_t = \frac{1}{\sqrt{N}}\sum_{j=1}^N W^j(t)$ is a standard Brownian motion.

In order to obtain the optimal control (\ref{eq:stationary optimal control}), we need to have the coefficients 
$(b_\infty,d_\infty,e_\infty)$ that cannot be obtained analytically, in general, and must be computed numerically. However, we are 
able to find approximate solutions in certain regimes. We note that from (\ref{eq:algebraic Riccati equation}), 
$d_\infty = (-\theta+\sqrt{\theta^2+\theta_c^2})/\theta_c$, and we consider the following cases:
\begin{enumerate}
	\item If $\theta_0=0$ and $H_0=0$, then we find $b_\infty = -d_\infty$ and $e_\infty=0$, so that we obtain the system
	\begin{align*}
		&dX_0^{(N)} = - H_0 X_0^{(N)} dt + \frac{\sigma_0}{\sqrt{N}} dW^0_t,\\
		&d\bar{X}_N =  \frac{\sigma}{\sqrt{N}} d\bar{W}_t^{(N)} - \sqrt{\theta^2+\theta_c^2} (\bar{X}_N-X_0^{(N)}) dt,  
	\end{align*} 
	which shows that the passive control $-\theta(X_j-X_0^{(N)})$ and the optimal control $\alpha_j$ combine in a quadratic way to form 
	the feedback $- \sqrt{\theta^2+\theta_c^2} ( \bar{X}_N-X_0^{(N)})$.

	\item If $0<\theta_0\ll 1$ and $H_0=0$, then we find 
	$b_\infty = -d_\infty + \theta_0 d_\infty/\sqrt{\theta^2+\theta_c^2} +o(\theta_0)$ and 
	$e_\infty=- \theta_0  d_\infty/\sqrt{\theta^2+\theta_c^2} + o(\theta_0)$, so that we obtain the system
	\begin{align*}
		&dX_0^{(N)} = -\theta_0 (X_0^{(N)} - \bar{X}_N)  dt + \frac{\sigma_0}{\sqrt{N}} dW^0_t,\\
		&d\bar{X}_N = \frac{\sigma}{\sqrt{N}} d\bar{W}_t^{(N)}
		- \left(\sqrt{\theta^2+\theta_c^2} - \theta_0\frac{\sqrt{\theta^2 +\theta_c^2}-\theta}{\sqrt{\theta^2+\theta_c^2}}\right)
		(\bar{X}_N-X_0^{(N)}) dt ,  
	\end{align*} 
	which shows that the optimal control chooses to reduce the feedback, probably because $X_0^{(N)}$ is destabilized by $\theta_0$.

	\item If $0<\theta_0\ll 1$ and $0<H_0\ll 1$, then we find 
	$b_\infty = -d_\infty + (H_0+\theta_0) d_\infty/\sqrt{\theta^2+\theta_c^2} + o(\theta_0,H_0)$ and 
	$e_\infty=- \theta_0  d_\infty/\sqrt{\theta^2+\theta_c^2} + o(\theta_0,H_0)$, so that we obtain the system
	\begin{align*}
		dX_0^{(N)} &= -H_0 X_0^{(N)} dt - \theta_0(X_0^{(N)}-\bar{X}_N)dt + \frac{\sigma_0}{\sqrt{N}}dW^0_t,\\
		d\bar{X}_N 
		&= \frac{\sigma}{\sqrt{N}}d\bar{W}_t^{(N)}
		-\left(\sqrt{\theta^2+\theta_c^2}
		-(\theta_0+H_0)\frac{\sqrt{\theta^2+\theta_c^2}-\theta}{\sqrt{\theta^2+\theta_c^2}} \right)(\bar{X}_N-X_0^{(N)})dt\\
		&\quad - H_0\frac{\sqrt{\theta^2 +\theta_c^2}-\theta}{\sqrt{\theta^2+\theta_c^2}}\bar{X}_N dt,  
	\end{align*} 
	which shows that  the optimal control chooses to reduce the feedback but it also controls $ \bar{X}_N$ directly.
\end{enumerate}

%% file: numerics.tex
\section{Numerical results}
\label{sec:numerics}

\subsection{Numerical results of fluctuations}
\label{sec:numerics of fluctuations}

In this subsection we compare the analytical fluctuation results 
(\ref{eq:limit of Var delta y_0, large theta}-\ref{eq:limit of Cov(delta y_0, delta y_bar), large theta}) with the fluctuations obtained from the 
numerical simulations of $(x_0^{(N)}(t),\bar{x}_N(t))$ in (\ref{eq:dynamics of zero h}). We use the Euler scheme to discretize 
(\ref{eq:dynamics of zero h}):
\begin{align}
	\label{eq:Euler for zero h}
	& x_0^{(N)}(n+1)=\frac{\sigma_0}{\sqrt{N}}\Delta W^0_{n+1}-h_0 V_0'(x_0^{(N)}(n))\Delta t - \theta_0 (x_0^{(N)}(n) - \bar{x}_N(n))\Delta t ,\\
	&\bar{x}_N(n+1) = \frac{\sigma}{\sqrt{N}} \Delta\bar{W}_{n+1} - \theta (\bar{x}_N(n) - x_0^{(N)}(n)) \Delta t,\notag
\end{align}
with $x_0^{(N)}(0)=\bar{x}_N(0)=-1$ and $\{\Delta W^0_{n+1}\}_n$, $\{\Delta \bar{W}_{n+1}\}_n$ i.i.d. Gaussian random variables with mean $0$ 
and variance $\Delta t$. We simulate (\ref{eq:Euler for zero h}) up to time $T$ and we take $T$ large enough so that $(x_0^{(N)}(t),\bar{x}_N(t))$ 
is in equilibrium after $T/10$. Therefore, $\mathbf{Var}(\lim_{t\to\infty}x_0^{(N)}(t))$, $\mathbf{Var}(\lim_{t\to\infty}\bar{x}_N(t))$ and 
$\mathbf{Cov}(\lim_{t\to\infty}x_0^{(N)}(t), \lim_{t\to\infty}\bar{x}_N(t))$ are approximately the sample variances and sample covariance of 
$\{x_0^{(N)}(n): T/10\leq n\Delta t \leq T\}$ and $\{\bar{x}_N(n): T/10\leq n\Delta t \leq T\}$, respectively. 

\begin{figure}
	\centering
	\includegraphics[width=0.49\linewidth]{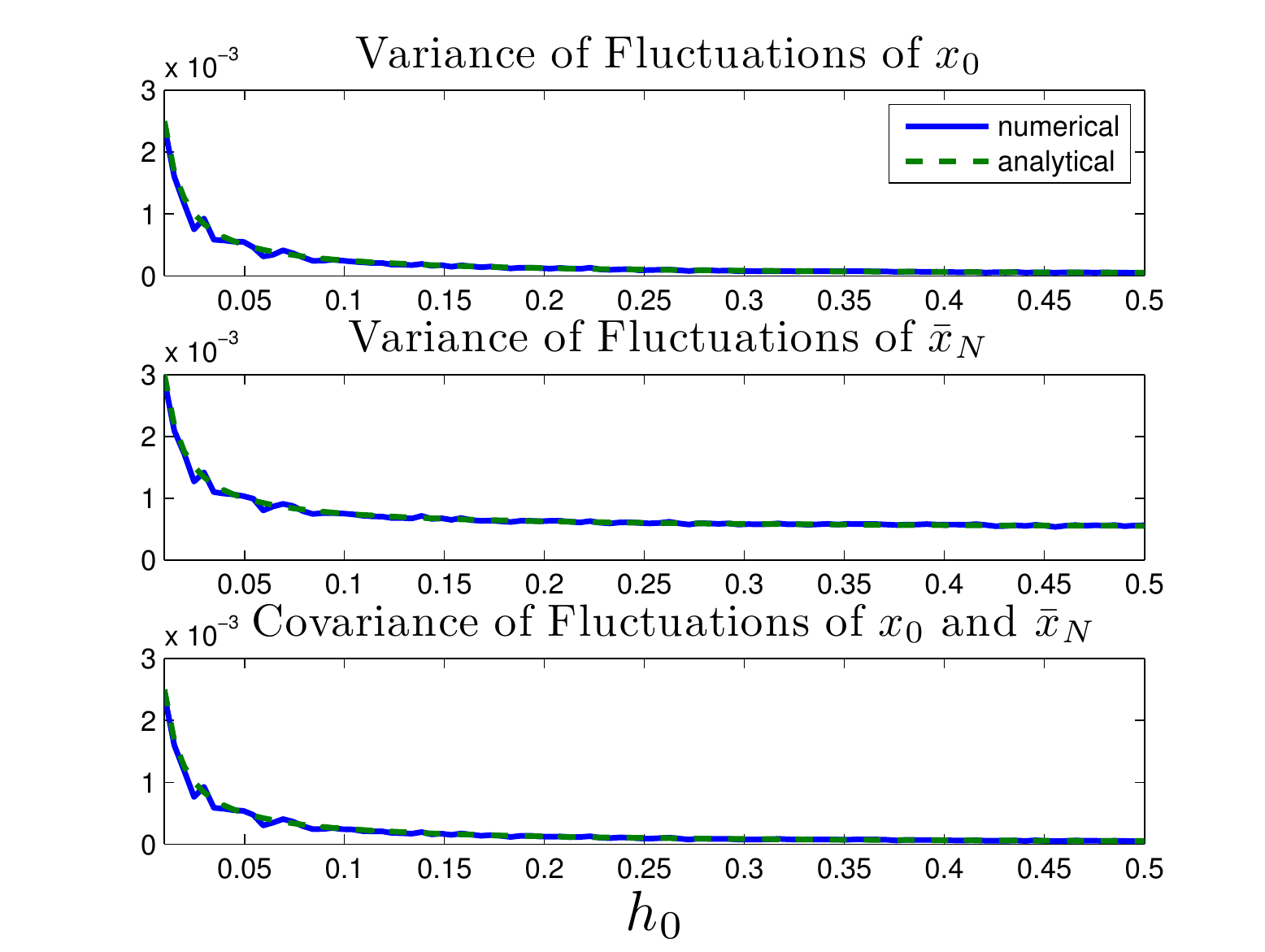}
	\includegraphics[width=0.49\linewidth]{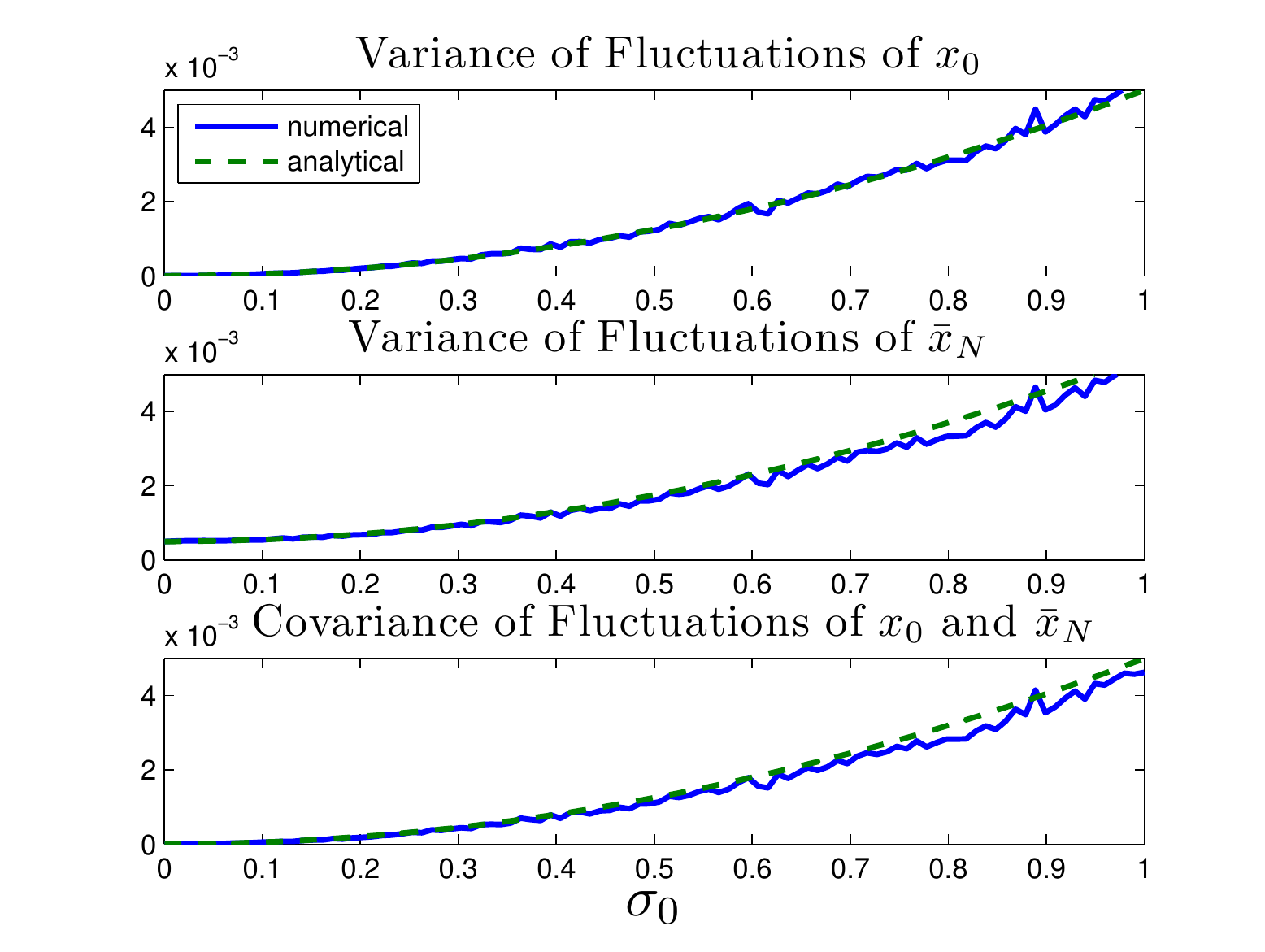}
	\caption{We compare the analytical
formulas for variances and covariances with direct numerical simulations. On the left the horizontal axis is $h_0$ and on
the right $\sigma_0$.
	\label{fig:fluctuations wrt h0 and sigma0}
}
\end{figure}

\begin{figure}
	\centering
	\includegraphics[width=0.49\linewidth]{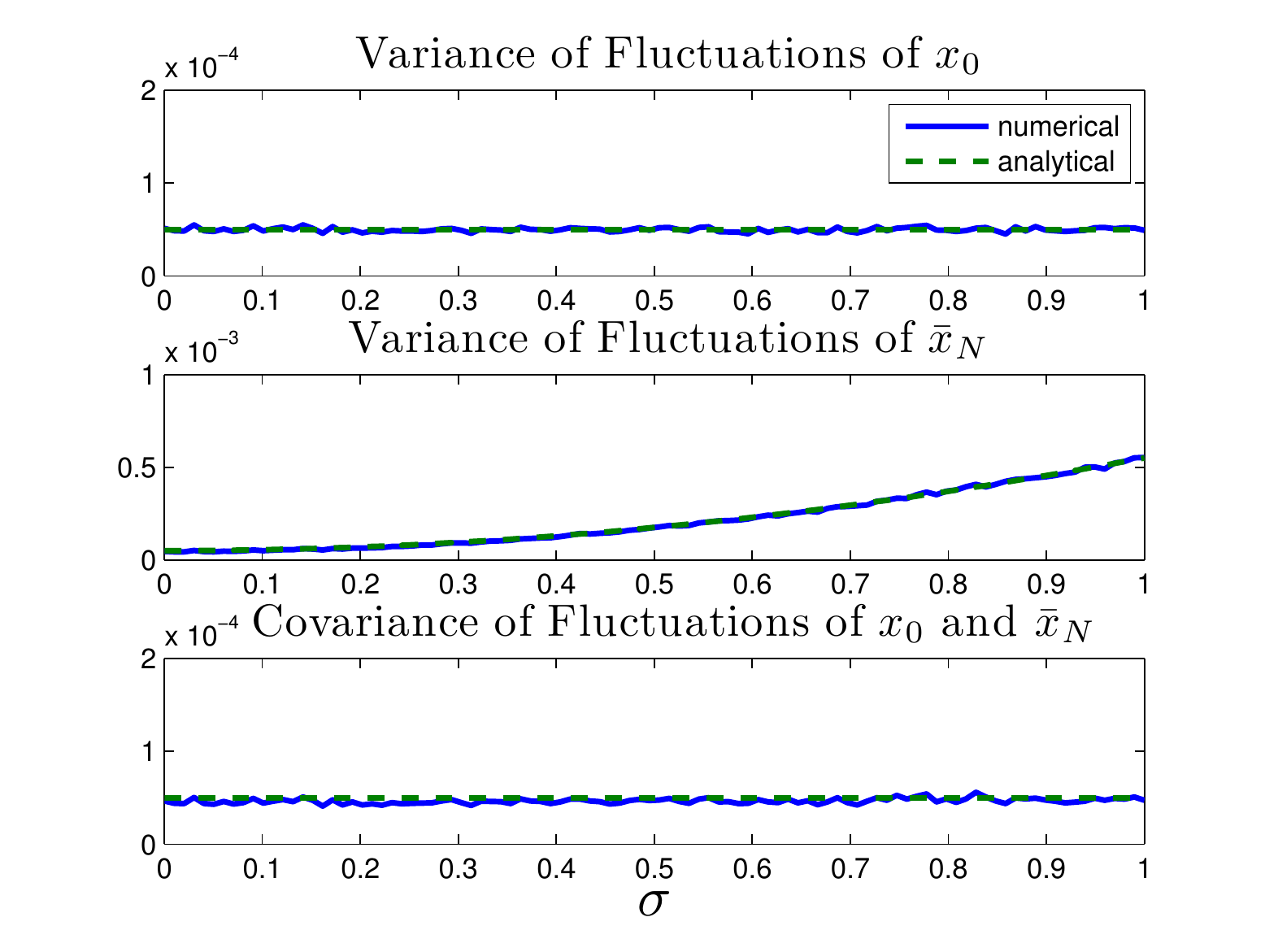}
	\includegraphics[width=0.49\linewidth]{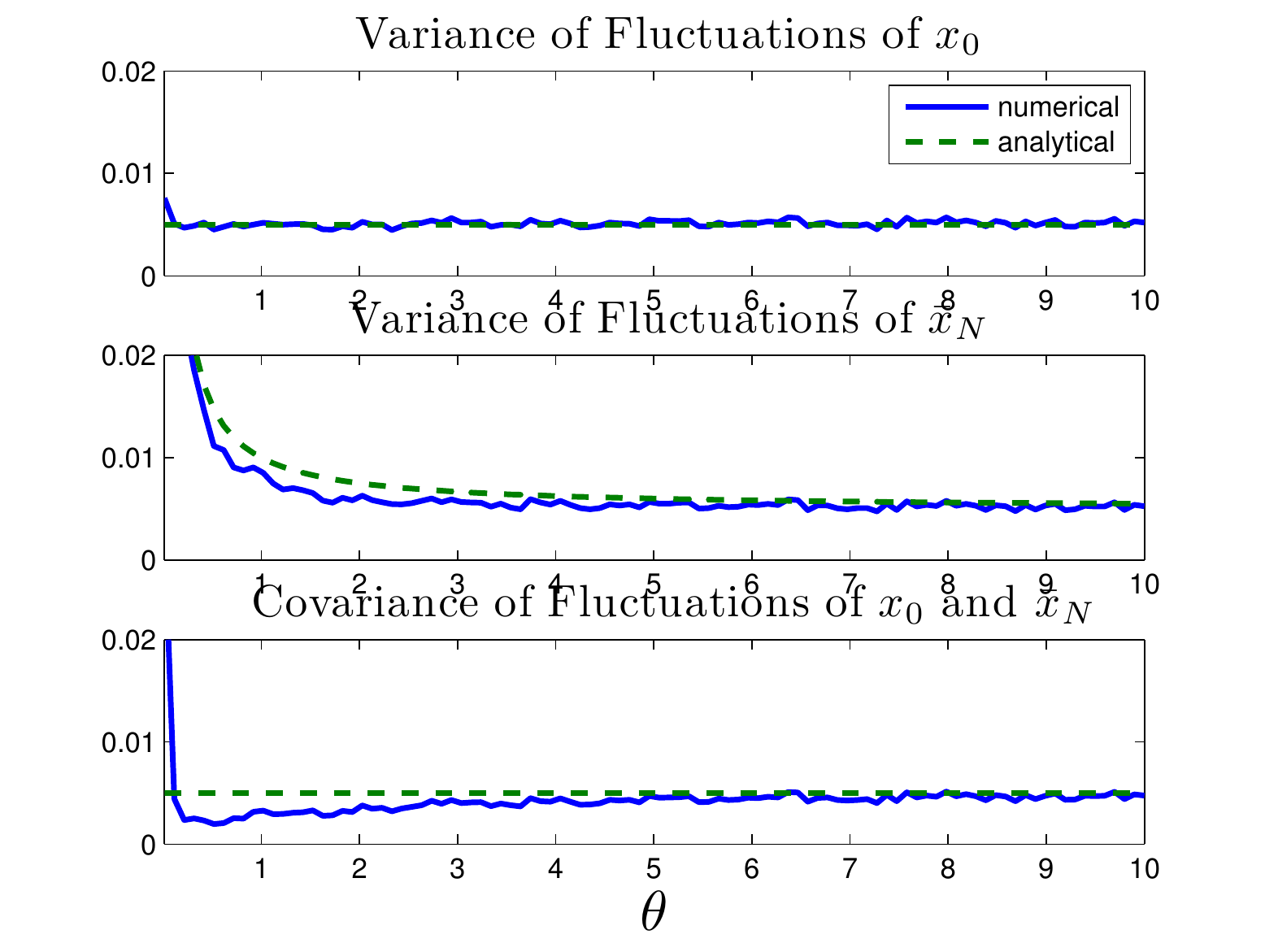}
	\caption{Same and in Figure \ref{fig:fluctuations wrt h0 and sigma0} except that the horizontal axis on
the left is $\sigma$ and on the right $\theta$.\label{fig:fluctuations wrt sigma and theta}}
\end{figure}

For each simulation, we vary one parameter for $100$ different values equally distributed in the region of interest, and use the values in Table 
\ref{tab:parameters of fluctuations} for the other parameters. The results are shown in Figures \ref{fig:fluctuations wrt h0 and sigma0} and 
\ref{fig:fluctuations wrt sigma and theta}. In Figure \ref{fig:fluctuations wrt h0 and sigma0} we compare the analytical formulas 
(\ref{eq:limit of Var delta y_0, large theta}-\ref{eq:limit of Cov(delta y_0, delta y_bar), large theta}) with the sample variances and sample 
covariances from the direct numerical simulations for $100$ different $h_0$ and $\sigma_0$ uniformly distributed in the region of interest. In 
Figure \ref{fig:fluctuations wrt sigma and theta} we compare the analytical formulas 
(\ref{eq:limit of Var delta y_0, large theta}-\ref{eq:limit of Cov(delta y_0, delta y_bar), large theta}) with the sample variances and sample 
covariances from the direct numerical simulations for $100$ different $\sigma$ and $\theta$ uniformly distributed in the region of interest. We see 
that there is good agreement between the analytical formulas and the simulations and thus 
(\ref{eq:limit of Var delta y_0, large theta}-\ref{eq:limit of Cov(delta y_0, delta y_bar), large theta}) indeed capture the fluctuations of the 
equilibrium of $(x_0^{(N)}(t),\bar{x}_N(t))$.

\begin{table}
	\begin{center}
		\begin{tabular}{|c|c|c|c|c|c|c|c|}
			\hline 
			$N$&   $T$   & $\Delta t$& $h_0$& $\sigma_0$& $\theta_0$& $\sigma$& $\theta$\\ 
			\hline 
			$100$& $10^3$& $10^{-3}$ & $0.5$& $0.1$     & $0.1$     & $1.0$     & $10$\\ 
			\hline
		\end{tabular}
	\end{center}
	\caption{\label{tab:parameters of fluctuations}
	The typical values of parameters used in Sec \ref{sec:numerics of fluctuations}. For each simulation, we vary one parameter and the other 
	parameters are fixed at the values in the table.}
\end{table}

\subsection{Numerical results of large deviations}

In this subsection, we compute the most probable paths $(x_0,\bar{x})$, defined in Section \ref{sec:large deviations}, 
by numerically solving the associated boundary value problems 
(\ref{eq:BVP for x0, degenerate case}) and (\ref{eq:BVP for x0 and x_bar, nondegenerate case}) for $\sigma_0=0$ and 
$\sigma_0>0$, respectively. We use the boundary value problem solver \texttt{bvp4c} in MATLAB to solve 
these problems.
The details of the algorithm can be 
found in \cite{Shampine2003}.

For the non-singular cases, for $h_0$ small, we use $x_0(t)\equiv-1$ or $x_0(t)=(2t/T)-1$ for 
(\ref{eq:BVP for x0, degenerate case}), and $x_0(t)=\bar{x}(t)\equiv-1$ or $x_0(t)=\bar{x}(t)=(2t/T)-1$ for 
(\ref{eq:BVP for x0 and x_bar, nondegenerate case}), depending on which one gives better results. We found that 
\texttt{bvp4c} sometimes did not give an accurate solution even for the non-singular cases. The numerical solutions failed to 
pass their internal accuracy check of the MATLAB routine. The reason for this is not clear. 
However, this issue can be 
bypassed  by iterating \texttt{bvp4c} 
several times. More precisely, we use the inaccurate solution as a new initial guess and use \texttt{bvp4c} to solve the 
same boundary value problem again to obtain a new solution and so on. After several iterations, \texttt{bvp4c} finds the 
correct solution that passes its accuracy check.

For the nearly-singular case, when $h_0$ is large, the method just described fails to find the correct solutions even 
with several iterations. To get past this issue,  we use  as initial guesses
solutions of the less singular cases obtained by the 
above technique. For example, we use the solution of the problem with $h_0=1$ as an initial guess to 
solve the problem with $h_0=2$, and so on. Eventually we can solve some quite singular problems, for example, with $h_0=10$.

\subsubsection{Impact of $h_0$}

\begin{figure}
	\centering
	
	\includegraphics[width=0.49\textwidth]{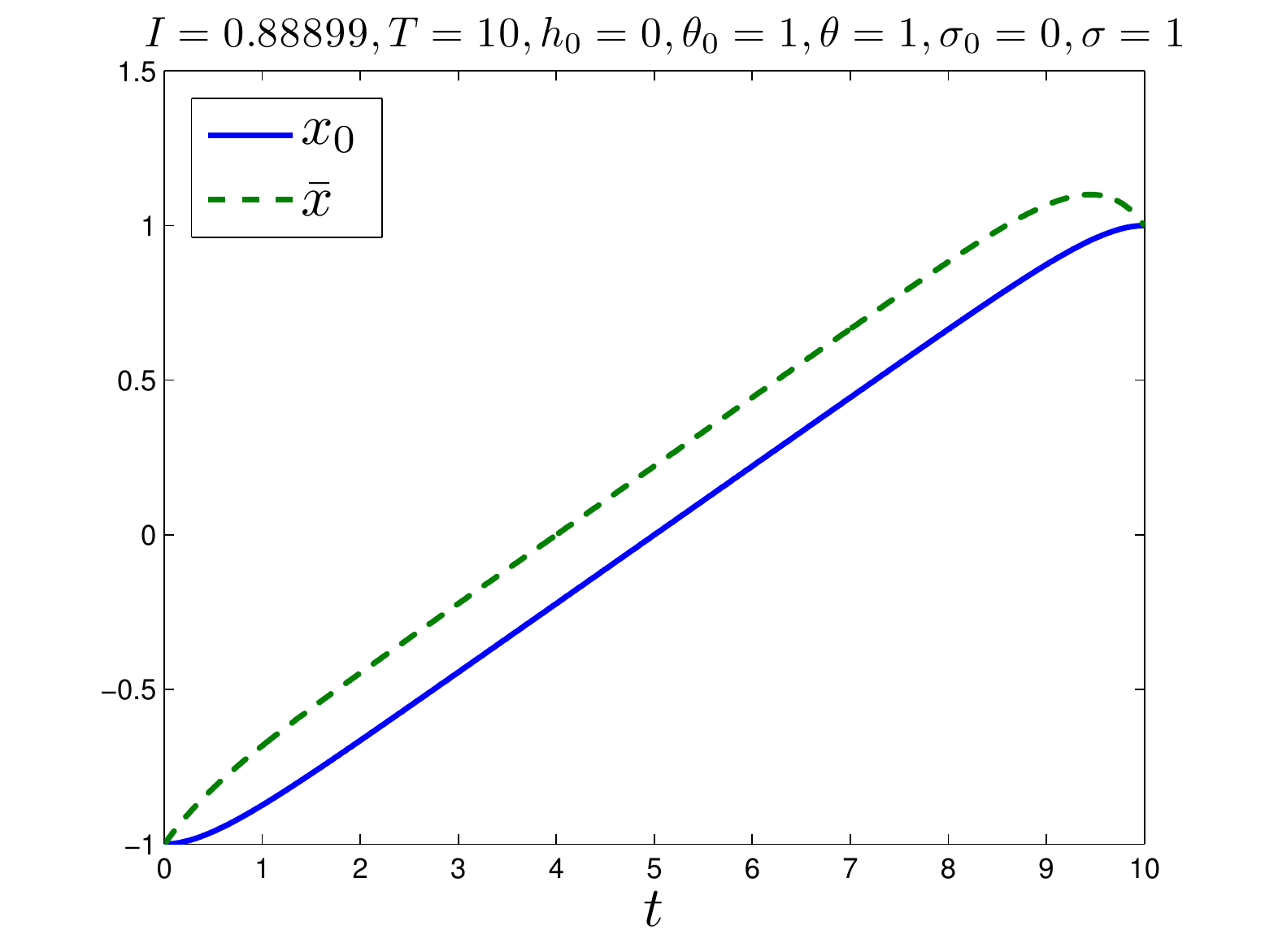}
	\includegraphics[width=0.49\textwidth]{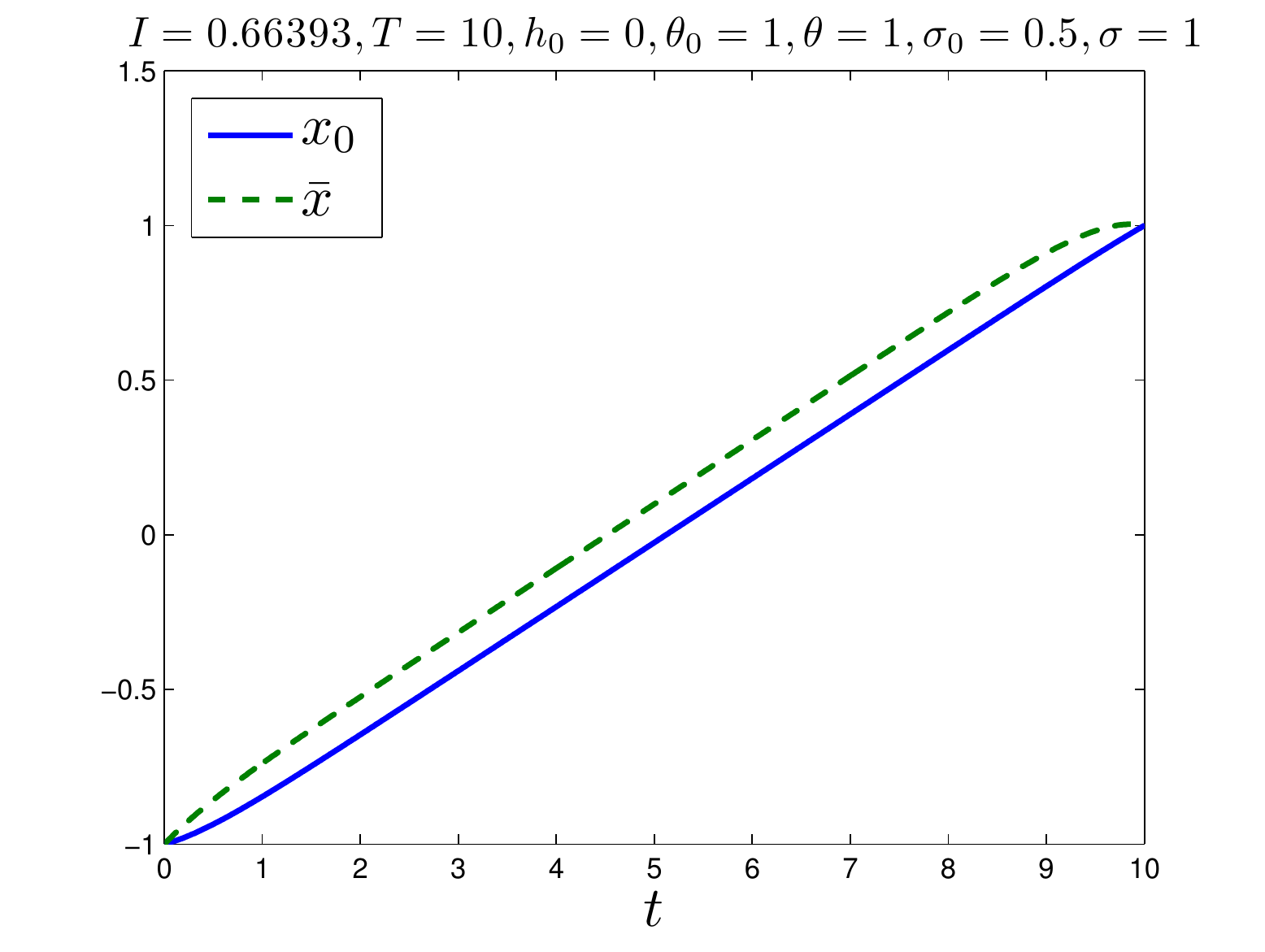}
	
	\includegraphics[width=0.49\textwidth]{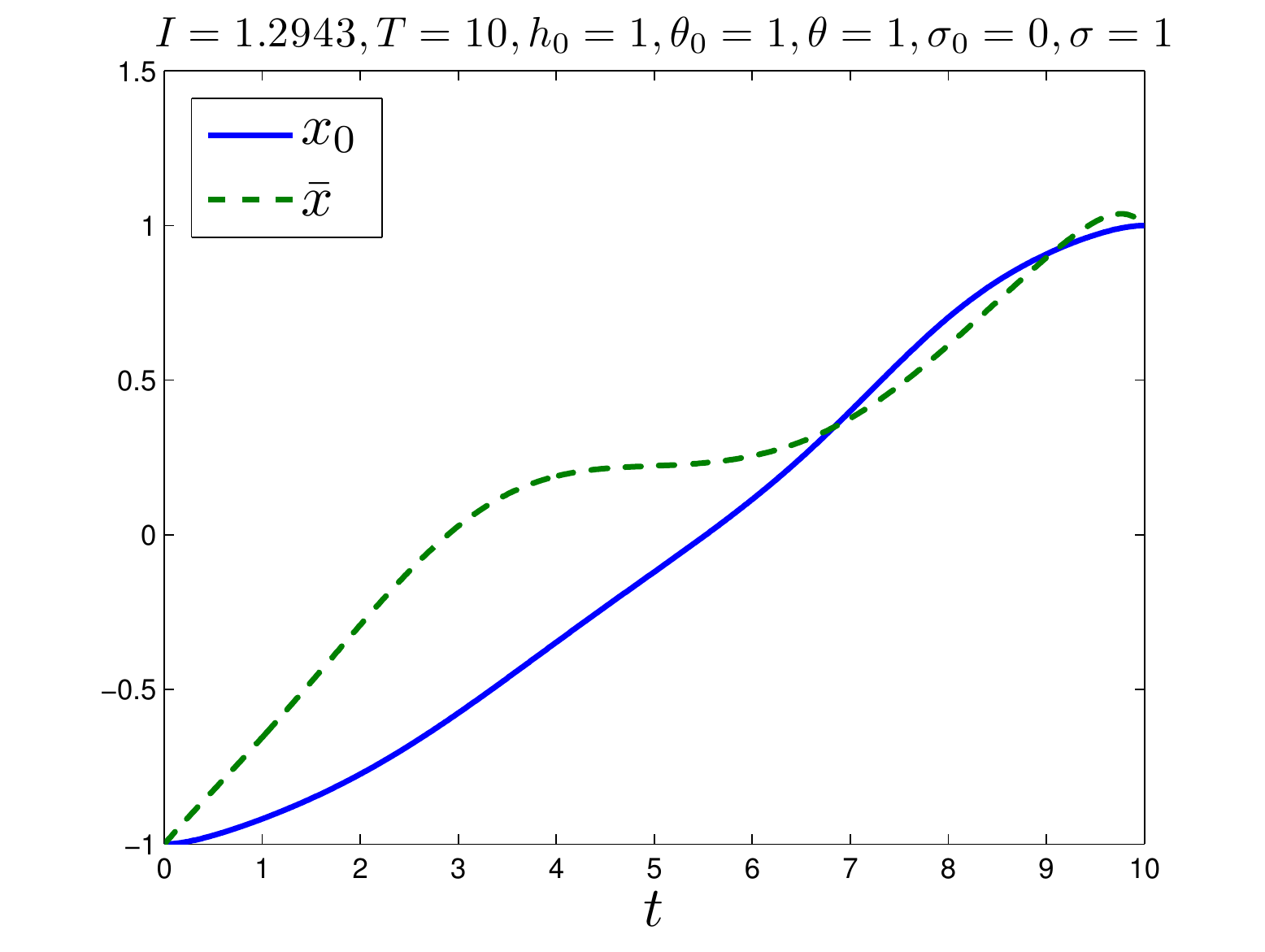}
	\includegraphics[width=0.49\textwidth]{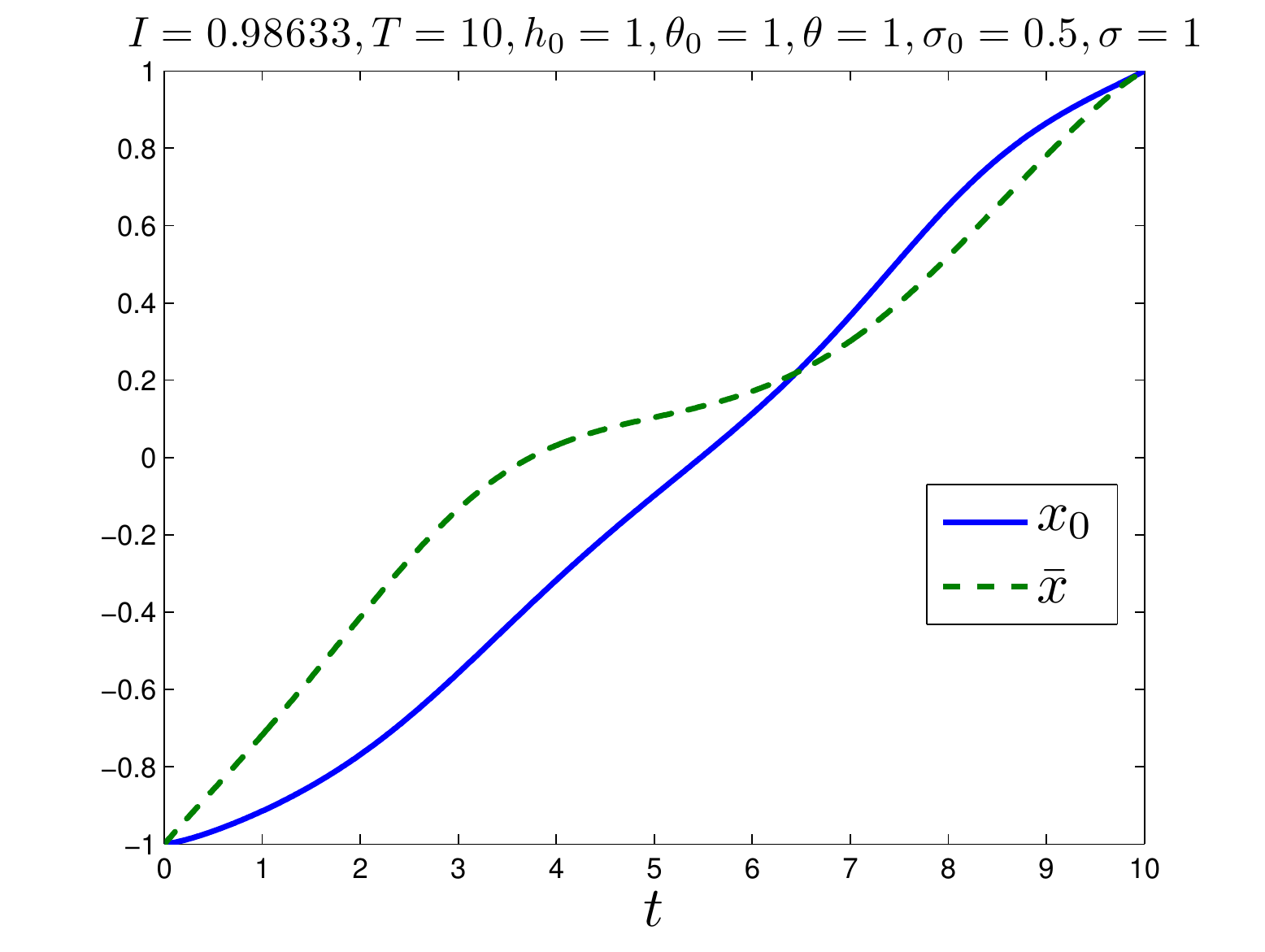}
	
	\includegraphics[width=0.49\textwidth]{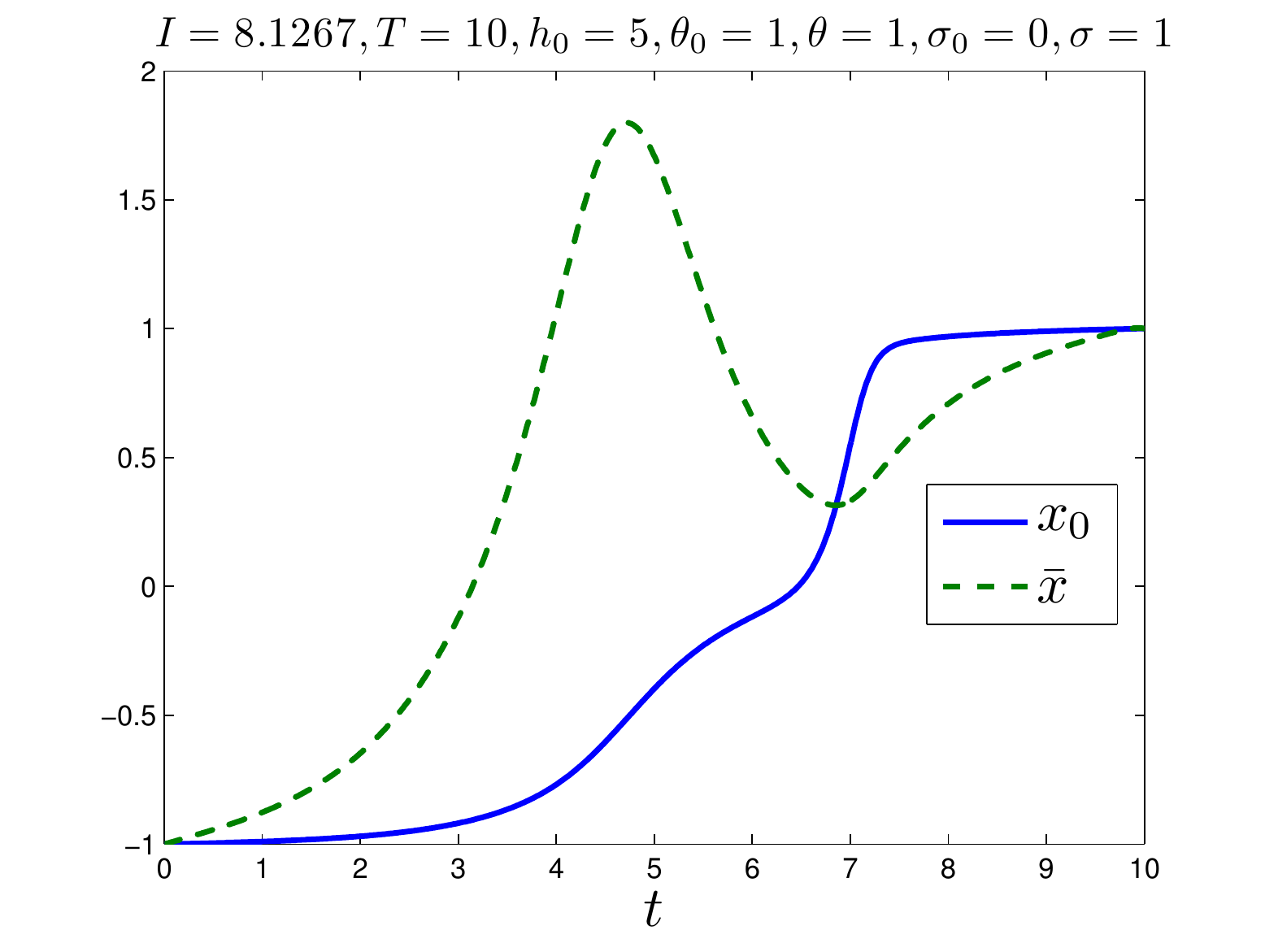}
	\includegraphics[width=0.49\textwidth]{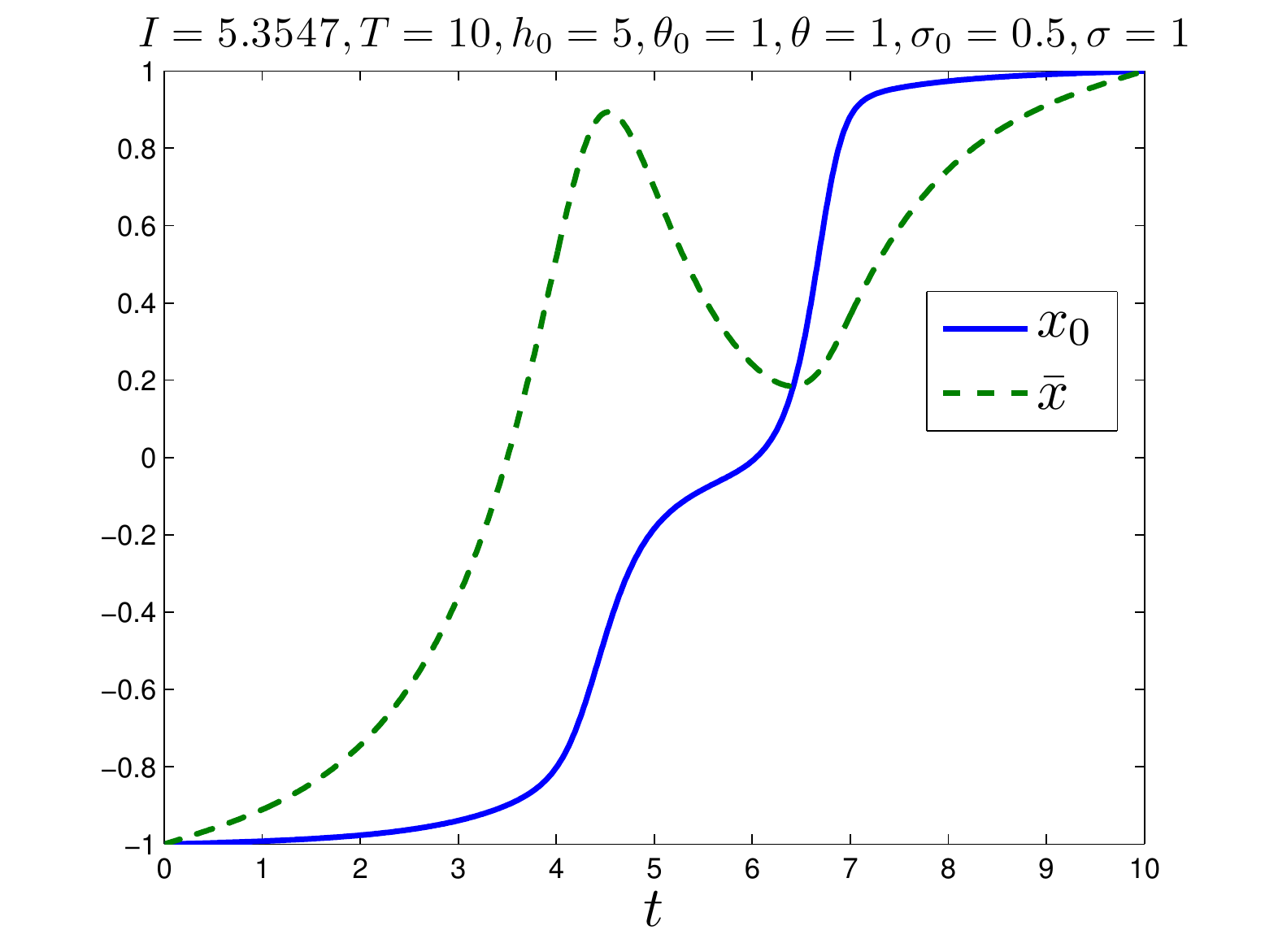}	
	
	\includegraphics[width=0.49\textwidth]{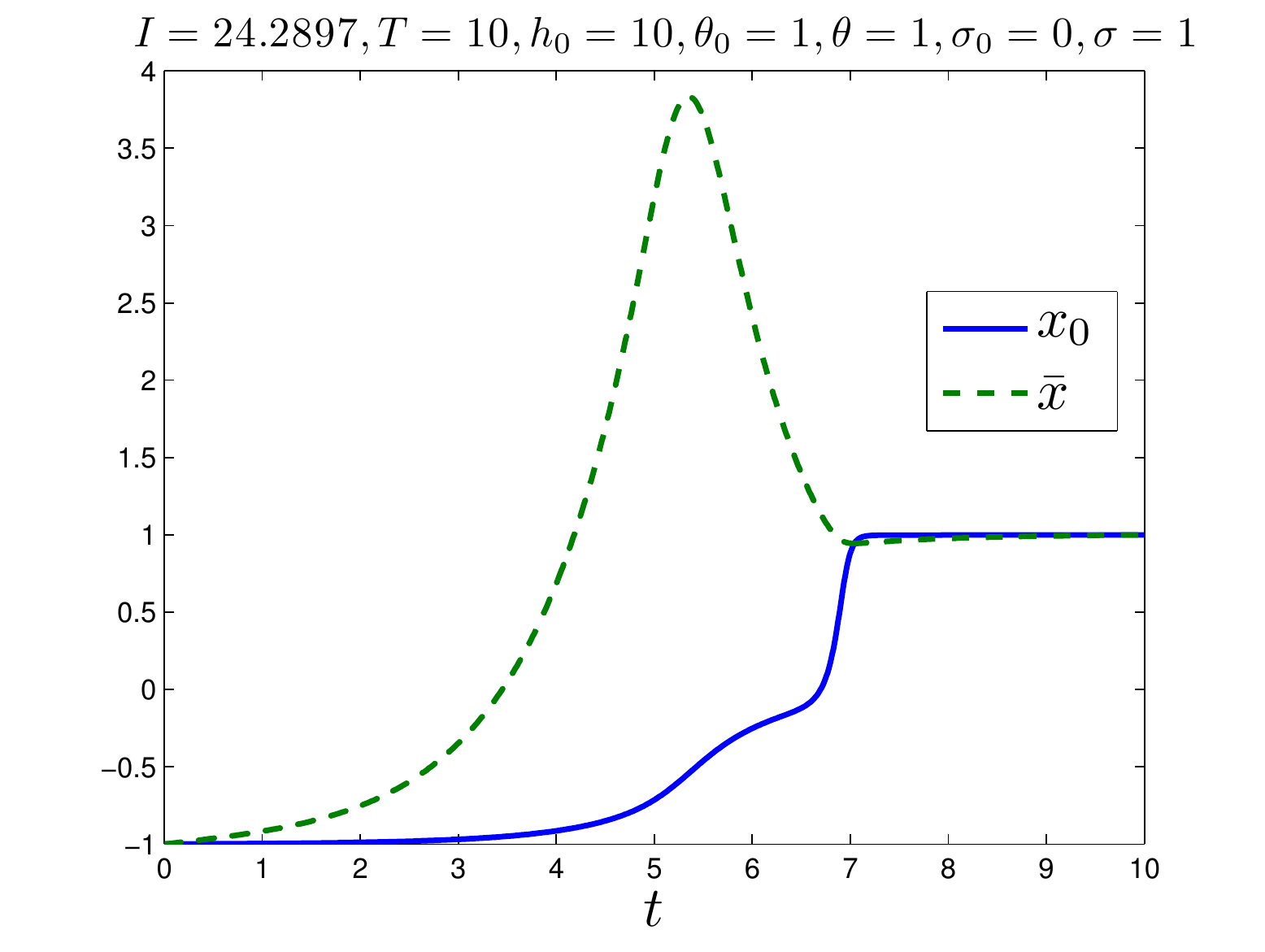}
	\includegraphics[width=0.49\textwidth]{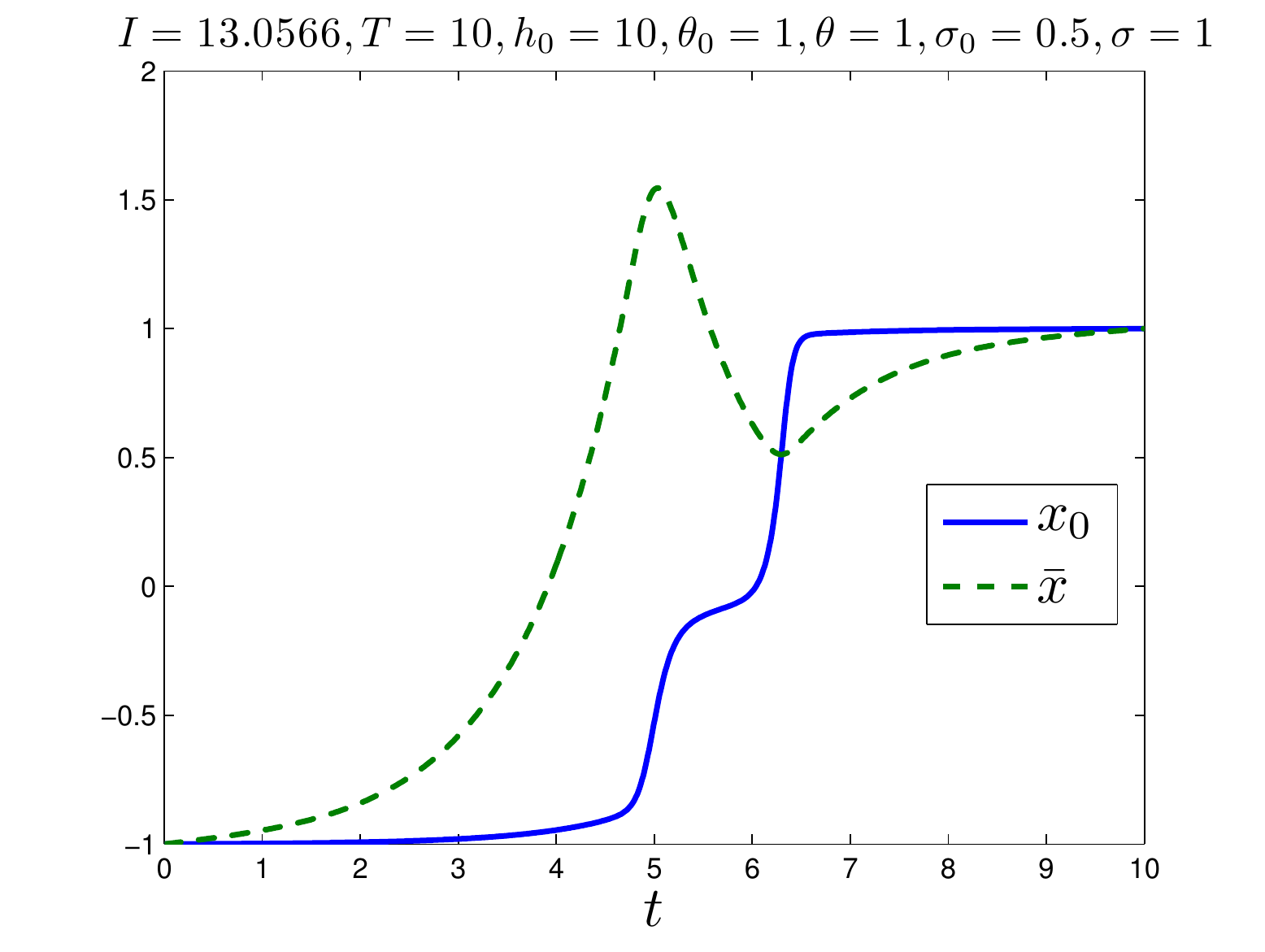}
	
	\caption{The most probable paths $(x_0,\bar{x})=\arg\min_\mathcal{A}I$ for $h_0=0, 1, 5, 10$. We let $T=10$, 	
	$\theta_0=1$, $\theta=1$ and $\sigma=1$. The left column is the case $\sigma_0=0$ and the right column is the case $\sigma_0=0.5$.
	\label{fig:x0 and xbar for different h0}}
\end{figure}

In Figure \ref{fig:x0 and xbar for different h0}  
we plot the most probable paths $(x_0,\bar{x})$ as functions of time, for $h_0$ from $0$ to $10$. On the 
left all the plots are with $\sigma_0=0$ and on the right $\sigma_0=0.5$. We note that when $h_0=0$, $(x_0,\bar{x})$ is 
smooth and in fact it is approximately linear, while $(x_0,\bar{x})$ is quite curved for $h_0=10$. We see 
that when $x_0(t)\leq 0$,  the destabilization of the system is driven by $\bar{x}(t)$. 
Indeed, $\bar{x}$ has higher external risk ($\sigma=1$) than $x_0(t)$ does ($\sigma_0=0$ or $\sigma_0=0.5$) and has no 
intrinsic stability ($h=0$), and therefore in the most probable path 
$\bar{x}(t)$ destabilizes $x_0(t)$. Nevertheless, once 
$x_0(t)>0$, the system transition is driven by $x_0(t)$ because the double-well potential forces $x_0$ to go to the failed state $1$, 
and $\bar{x}(t)$ is driven by $x_0(t)$. This effect is strengthened when $h_0$ is large because the double-well potential plays 
a more important role in that case.

\begin{figure}
	\centering
	
	\includegraphics[width=0.49\linewidth]{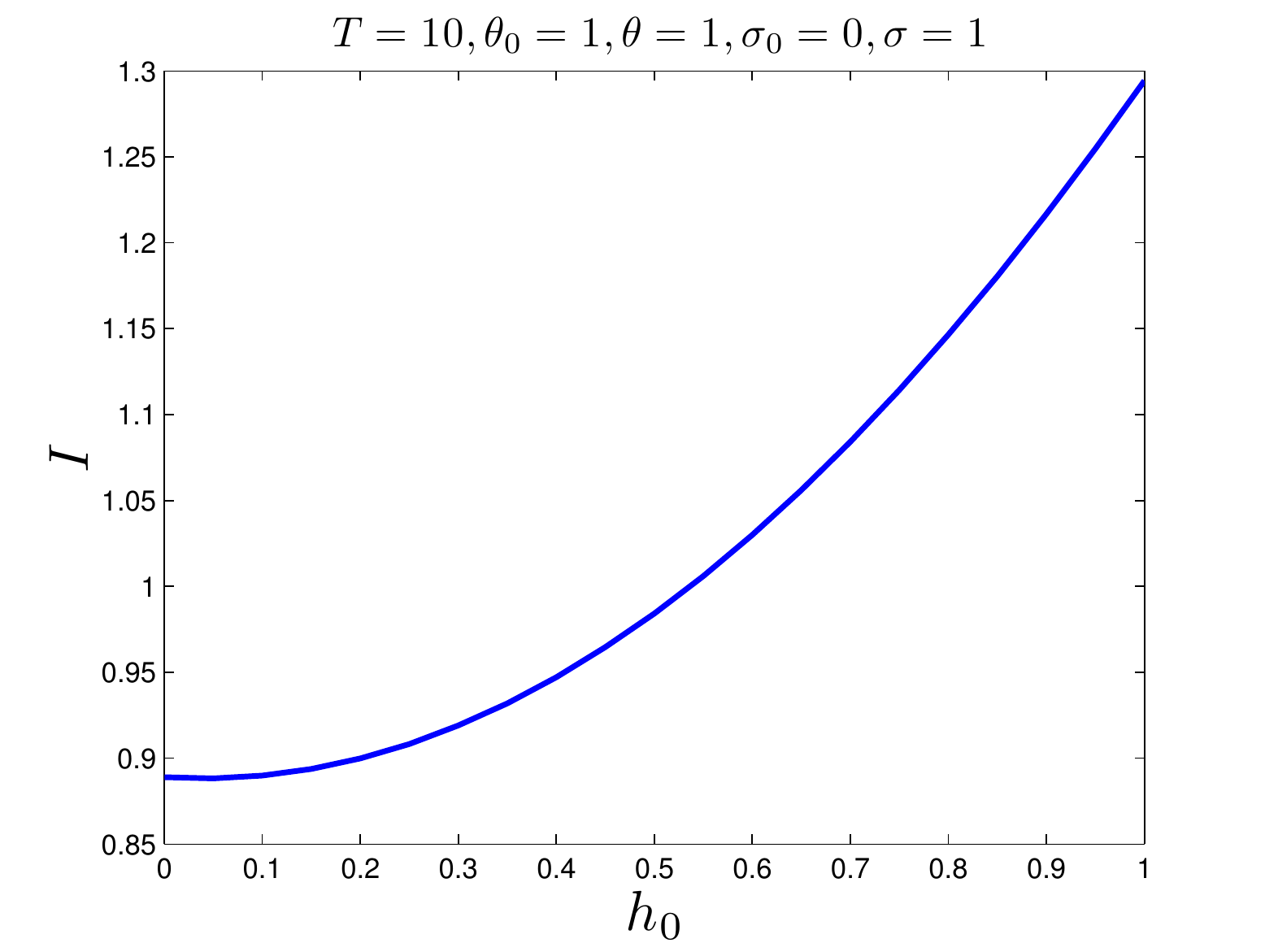}
	\includegraphics[width=0.49\linewidth]{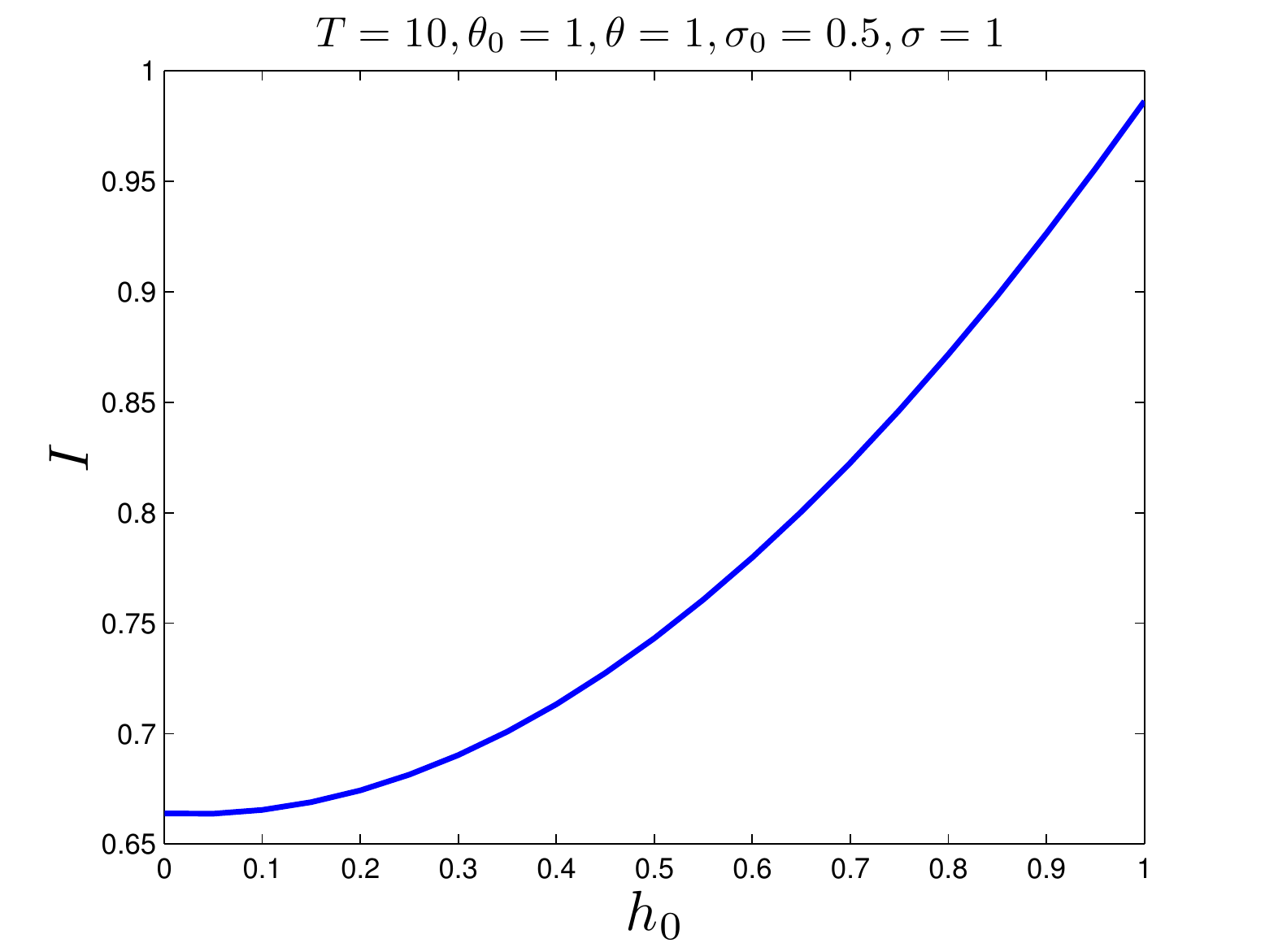}
	
	\includegraphics[width=0.49\linewidth]{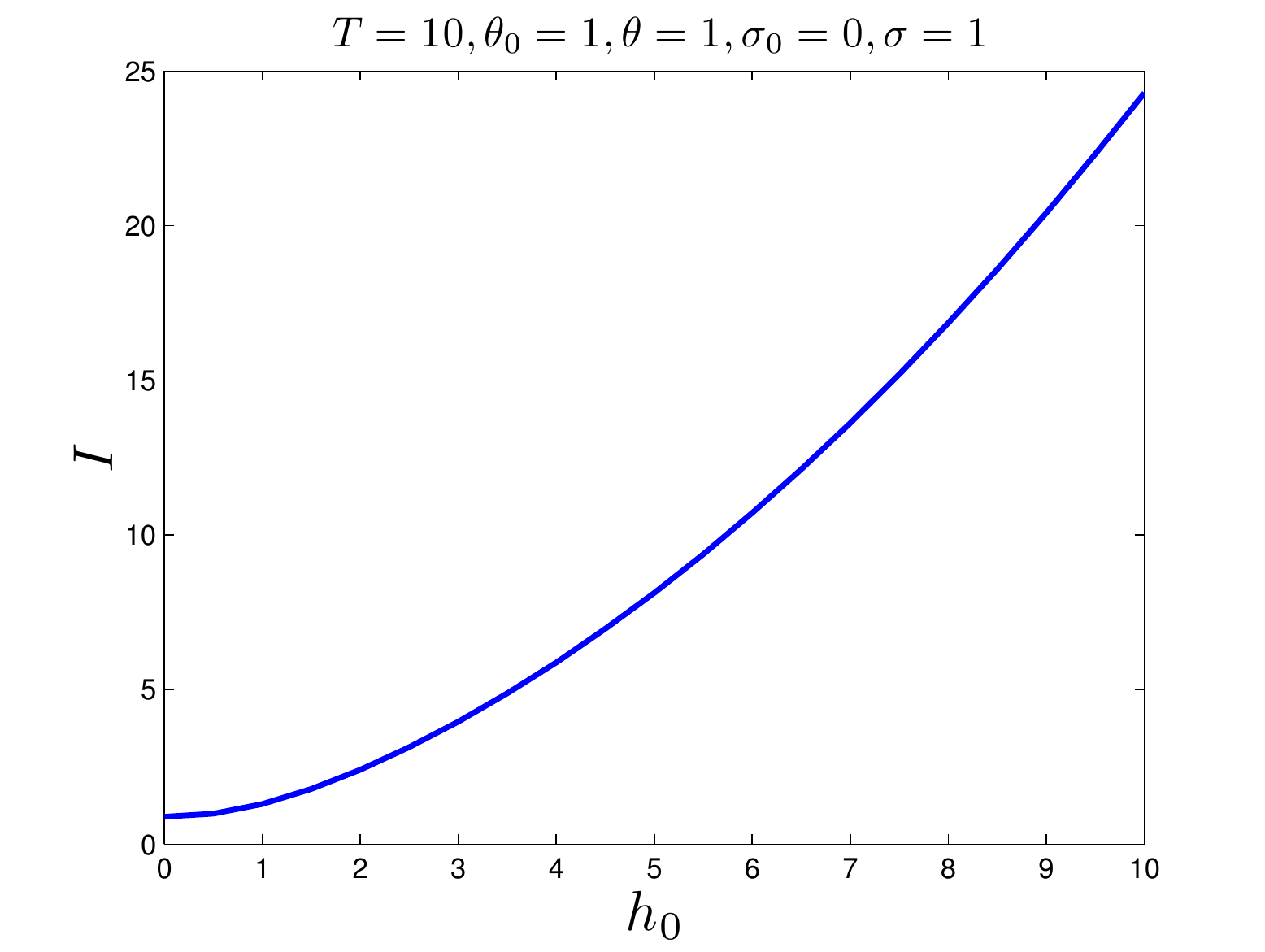}
	\includegraphics[width=0.49\linewidth]{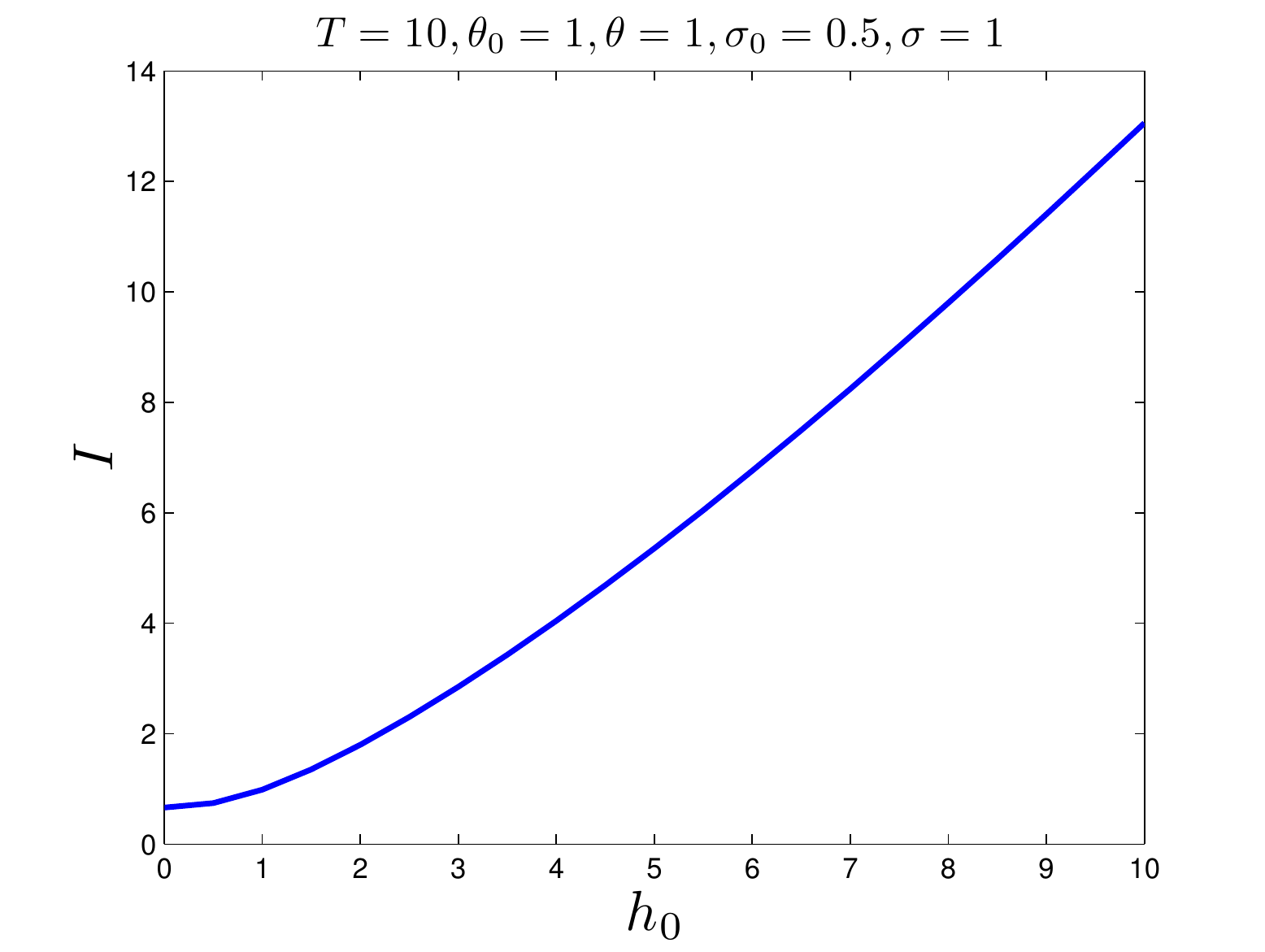}
	
	\caption{The infimum of $I$ over $\mathcal{A}$: $\inf_\mathcal{A}I$ for $h_0=0, 0.1, 0.2, \ldots, 1$ and for
	$h_0=0, 1, 2, \ldots, 10$. We let $T=10$, $\theta_0=1$, $\theta=1$ and $\sigma=1$. The left column is the case
	$\sigma_0=0$ and the right column is the case $\sigma_0=0.5$.
	\label{fig:I to different h0}}
\end{figure}

In Figure \ref{fig:I to different h0}  we plot the values of $\inf_{\bx\in\mathcal{A}}I(\bx)$ for different $h_0$. We see that 
$\inf_{\bx\in\mathcal{A}}I(\bx)$ is an increasing function of $h_0$. This is expected because the system is more stable if it has 
more intrinsic stability ($h_0$). We also see in Figure \ref{fig:I to different h0} that $\inf_{\bx\in\mathcal{A}}I(\bx)$ has quadratic 
behavior with respect to $h_0$ for small $h_0$ and linear behavior for large $h_0$.

\subsubsection{Comparison between small fluctuations and large deviations}

Here we compare the small fluctuations of $(x_0^{(N)},\bar{x}_N)$ described by the processes 
$z_0$ and $\bar{z}$ in (\ref{eq:dynamics of the fluctuations with zero h}) and the large 
deviations  of $(x_0^{(N)},\bar{x}_N)$ described by the infimum of the rate function $\inf_{\bx\in\mathcal{A}}I(\bx)$. For the characterization of the small fluctuations, we compute 
$\lim_{t\to\infty} \mathbf{Var}z_0(t)$ in (\ref{eq:limit of Var delta y_0}) and $\lim_{t\to\infty} \mathbf{Var}\bar{z}(t)$ in 
(\ref{eq:limit of Var delta y_bar}). For  the characterization of the large deviations, we compute $I(x_0,\bar{x})$ in 
(\ref{eq:rate function, degenerate case}) for $\sigma_0=0$ where $(x_0,\bar{x})$ is the solution of 
(\ref{eq:BVP for x0, degenerate case}) and compute $I(x_0,\bar{x})$ in 
(\ref{eq:rate function, nondegenerate case}) for $\sigma_0=0.5$ where $(x_0,\bar{x})$ is the solution of 
(\ref{eq:BVP for x0 and x_bar, nondegenerate case}).
The goal is to visualize the fact that the systemic risk
characterized by $\inf_{\bx\in\mathcal{A}}I(\bx)$ may vary significantly even though
the individual risk measured by $\lim_{t\to\infty} \mathbf{Var}\bar{z}(t)$ is kept at a fixed level.

\begin{figure}
	\centering
	\includegraphics[width=0.49\linewidth]{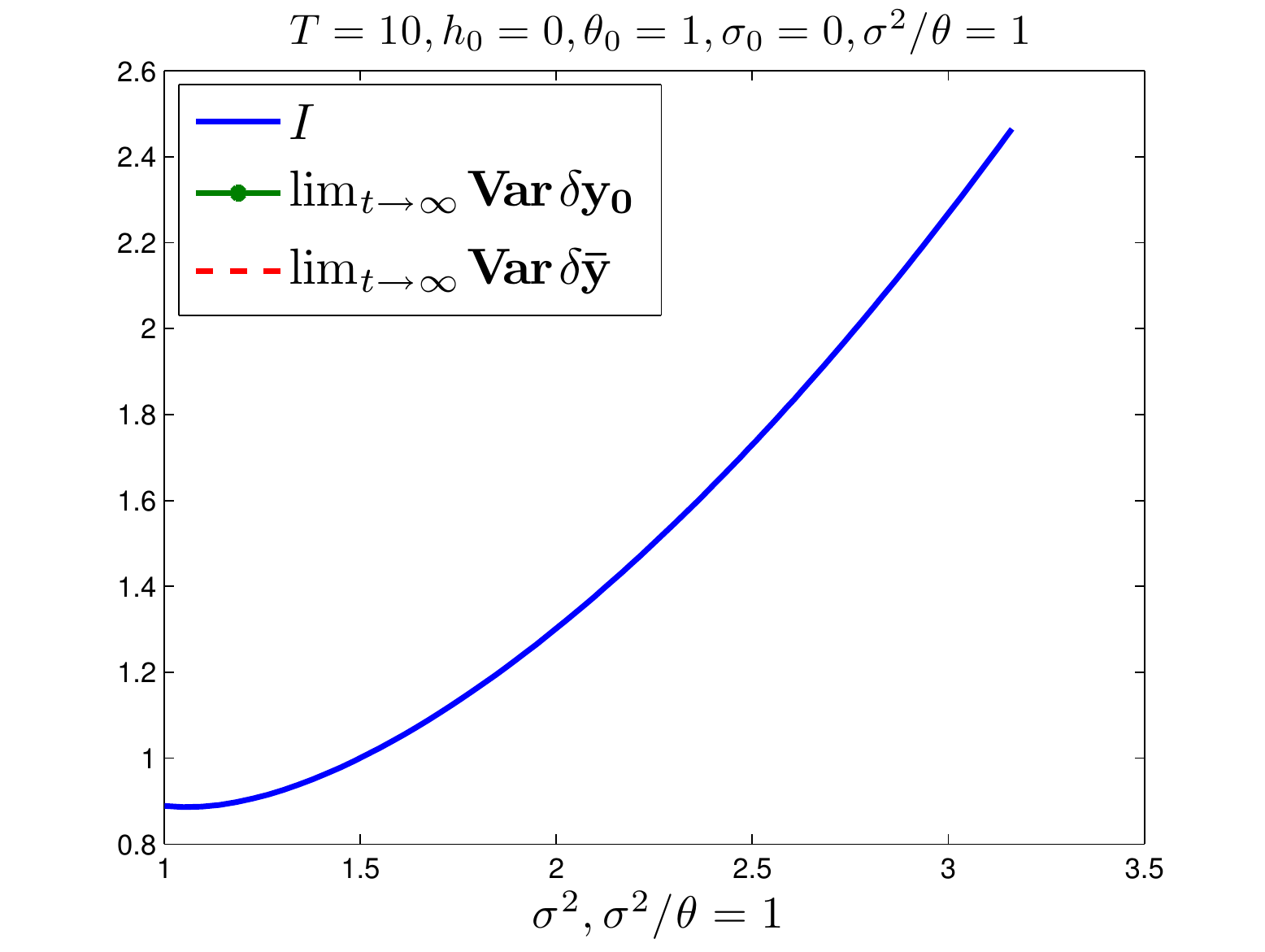}
	\includegraphics[width=0.49\linewidth]{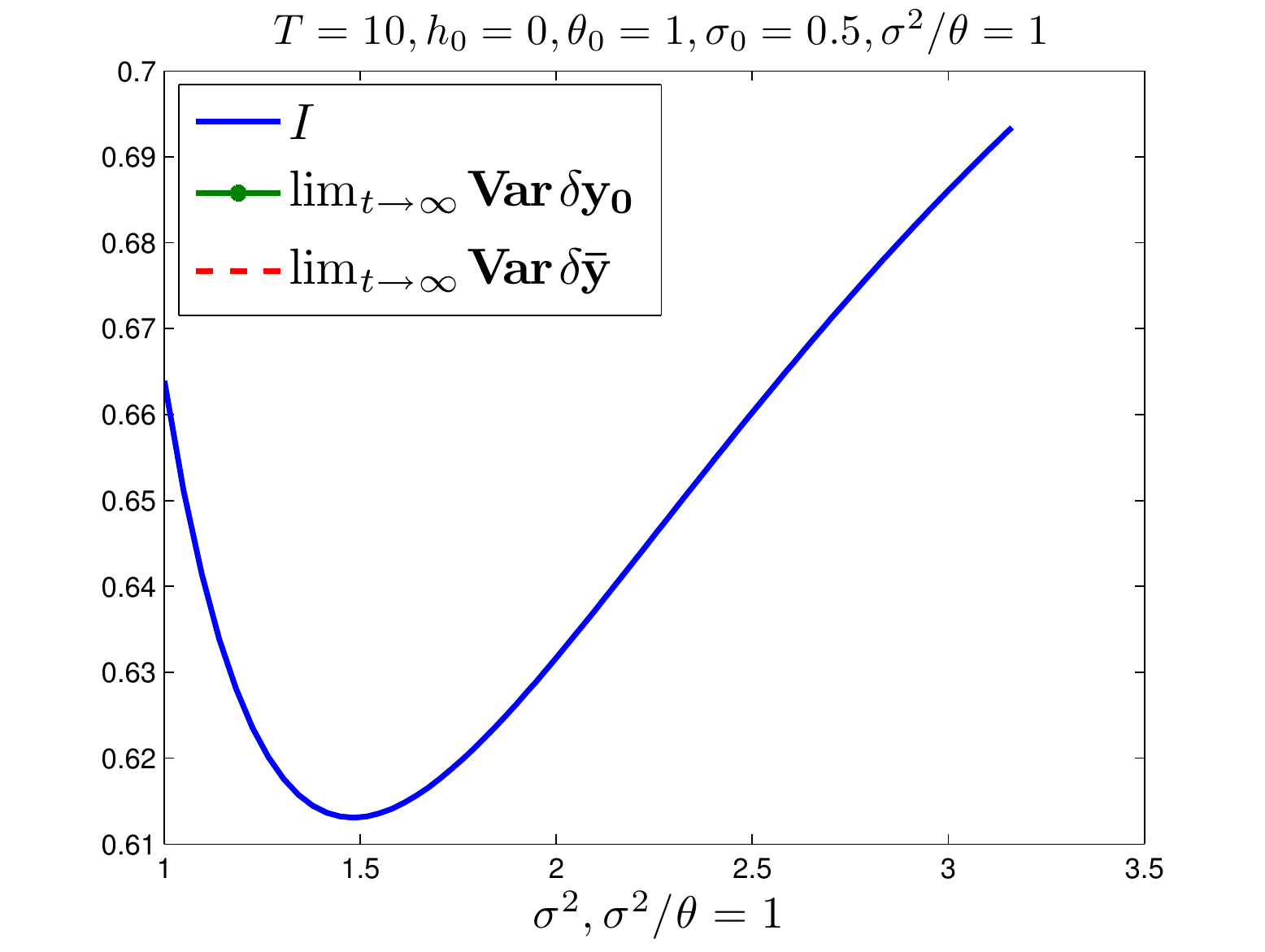}
	
	\includegraphics[width=0.49\linewidth]{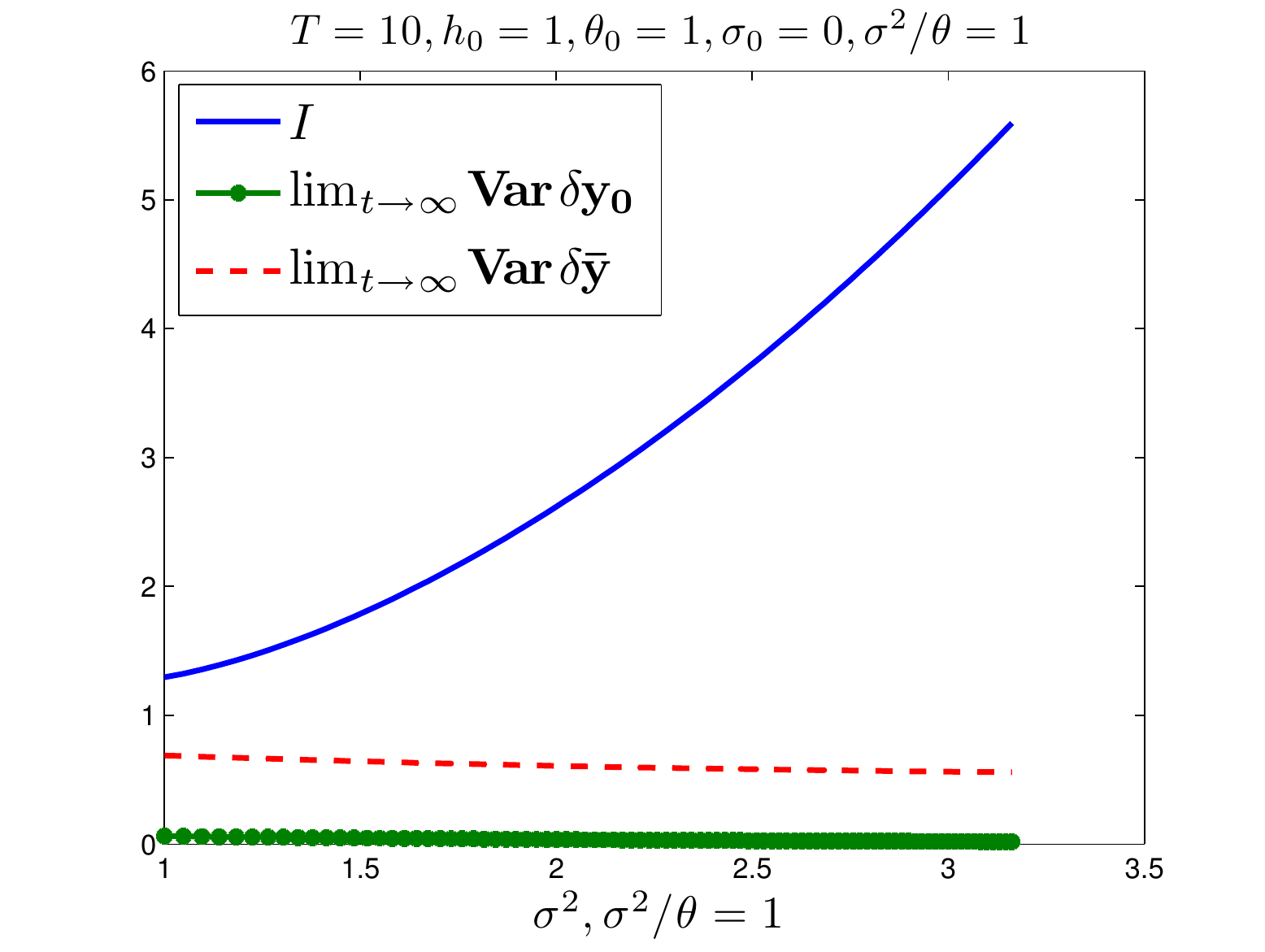}
	\includegraphics[width=0.49\linewidth]{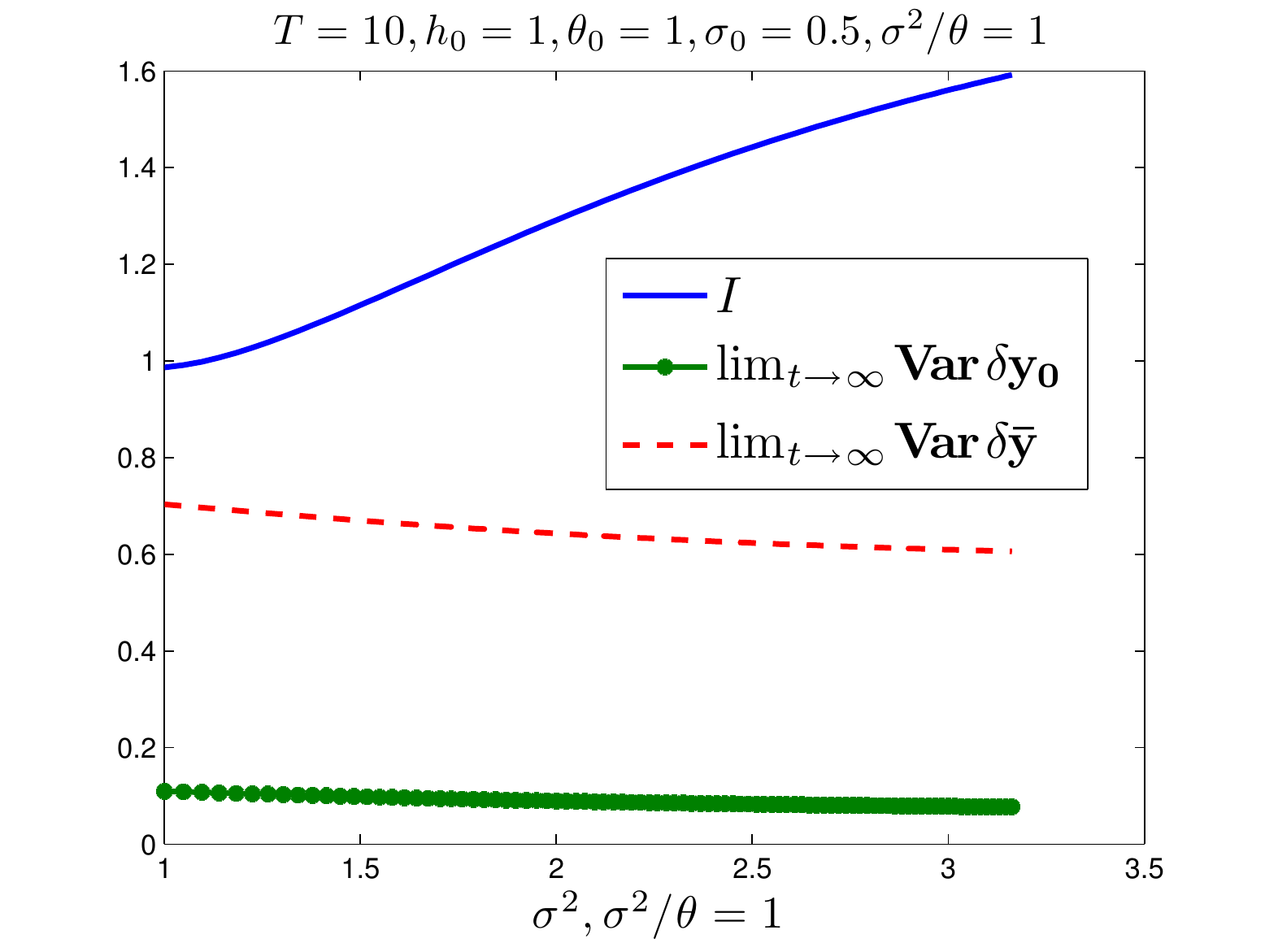}
	
	\includegraphics[width=0.49\linewidth]{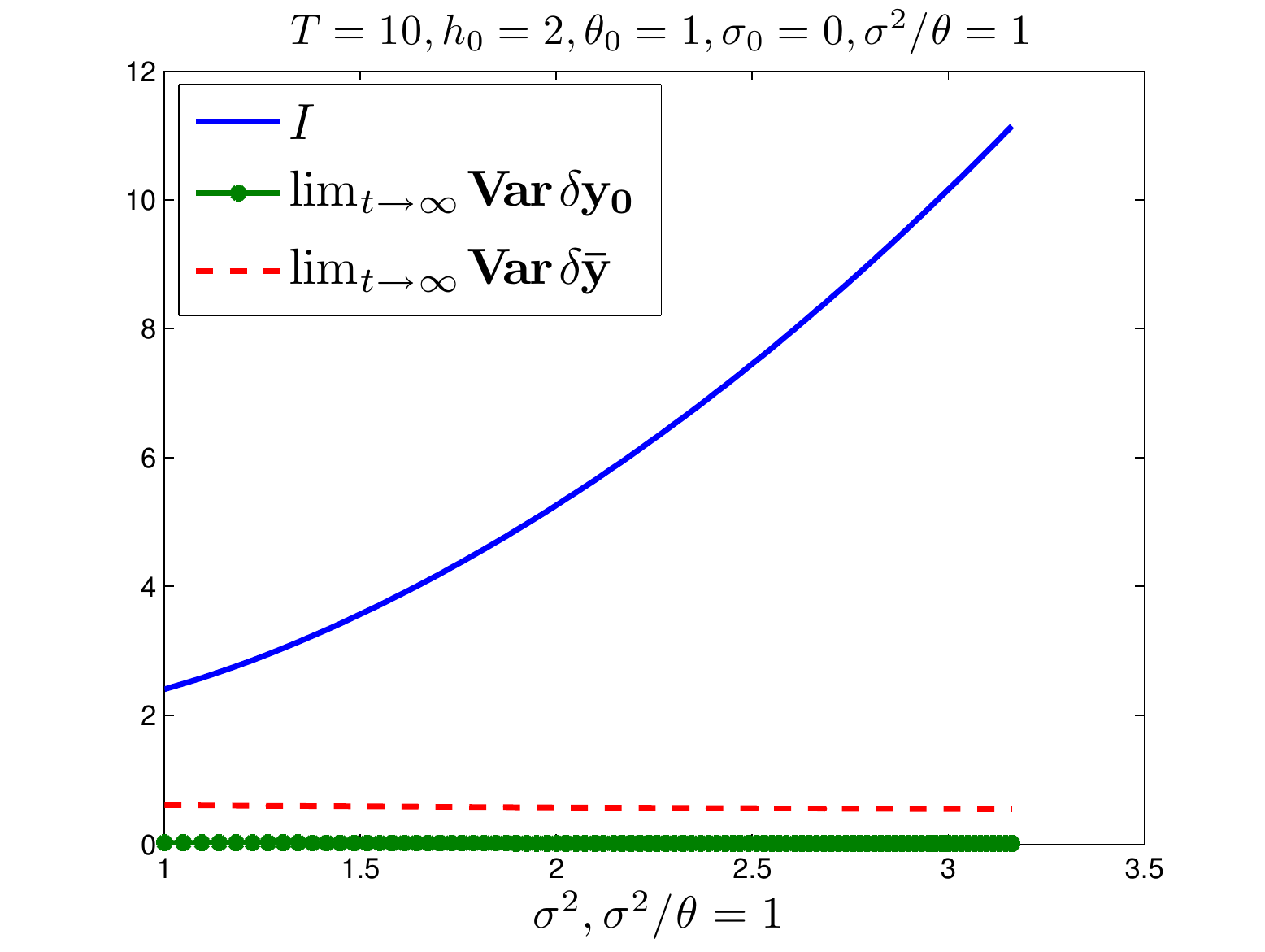}
	\includegraphics[width=0.49\linewidth]{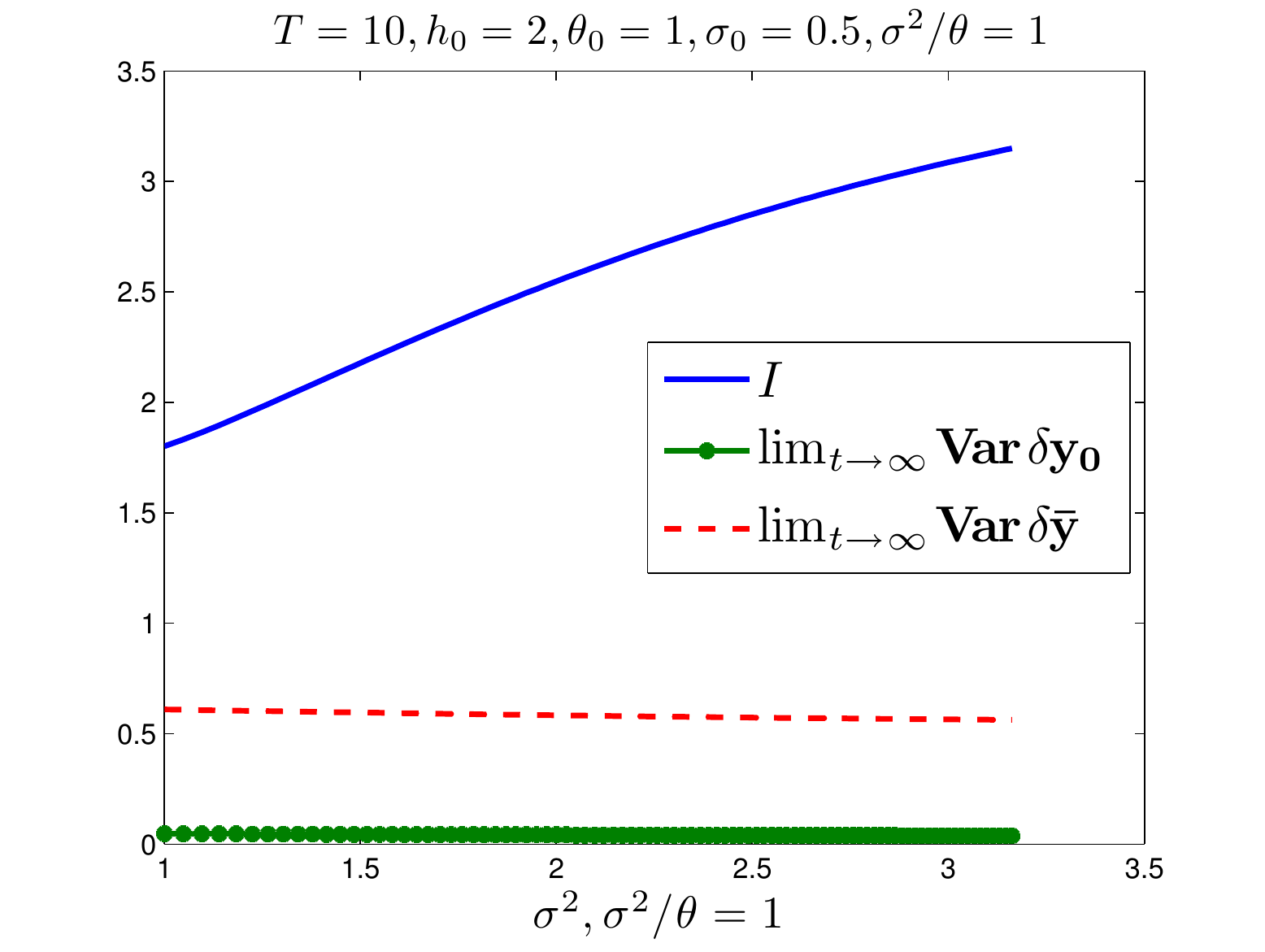}

	\caption{Plots of $\inf_{\bx\in\mathcal{A}}I(\bx)$, $\lim_{t\to\infty} \mathbf{Var}z_0(t)$ and 
	$\lim_{t\to\infty} \mathbf{Var}\bar{z}(t)$ for $\sigma^2$ from $1$ to $10$ with $\sigma^2/\theta=1$. We let $T=10$, 
	$\theta_0=1$. The left column is the case $\sigma_0=0$ and the right column is the case $\sigma_0=0.5$.}
	\label{fig:I to different sigma}
\end{figure}

Motivated by (\ref{eq:limit of Var delta y_0, large theta}) and (\ref{eq:limit of Var delta y_bar, large theta}), we know that 
$\lim_{t\to\infty} \mathbf{Var}z_0(t)$ and $\lim_{t\to\infty} \mathbf{Var}\bar{z}(t)$ are not significantly affected if we increase 
$\sigma$ and $\theta$ but keep the ratio $\sigma^2/\theta$ the same.
In Figure \ref{fig:I to different sigma}
we confirm this expectation and we also observe that $\inf_{\bx\in\mathcal{A}}I(\bx)$
increases as $\sigma$ increases, which means that systemic risk decreases.
This also means that, for a fixed level $\sigma^2/\theta$ of individual risk, the reduction of $\theta$, ie
the interaction of the local agent with the central agent, 
reduces the systemic risk.

\begin{figure}
	\centering
	\includegraphics[width=0.49\linewidth]{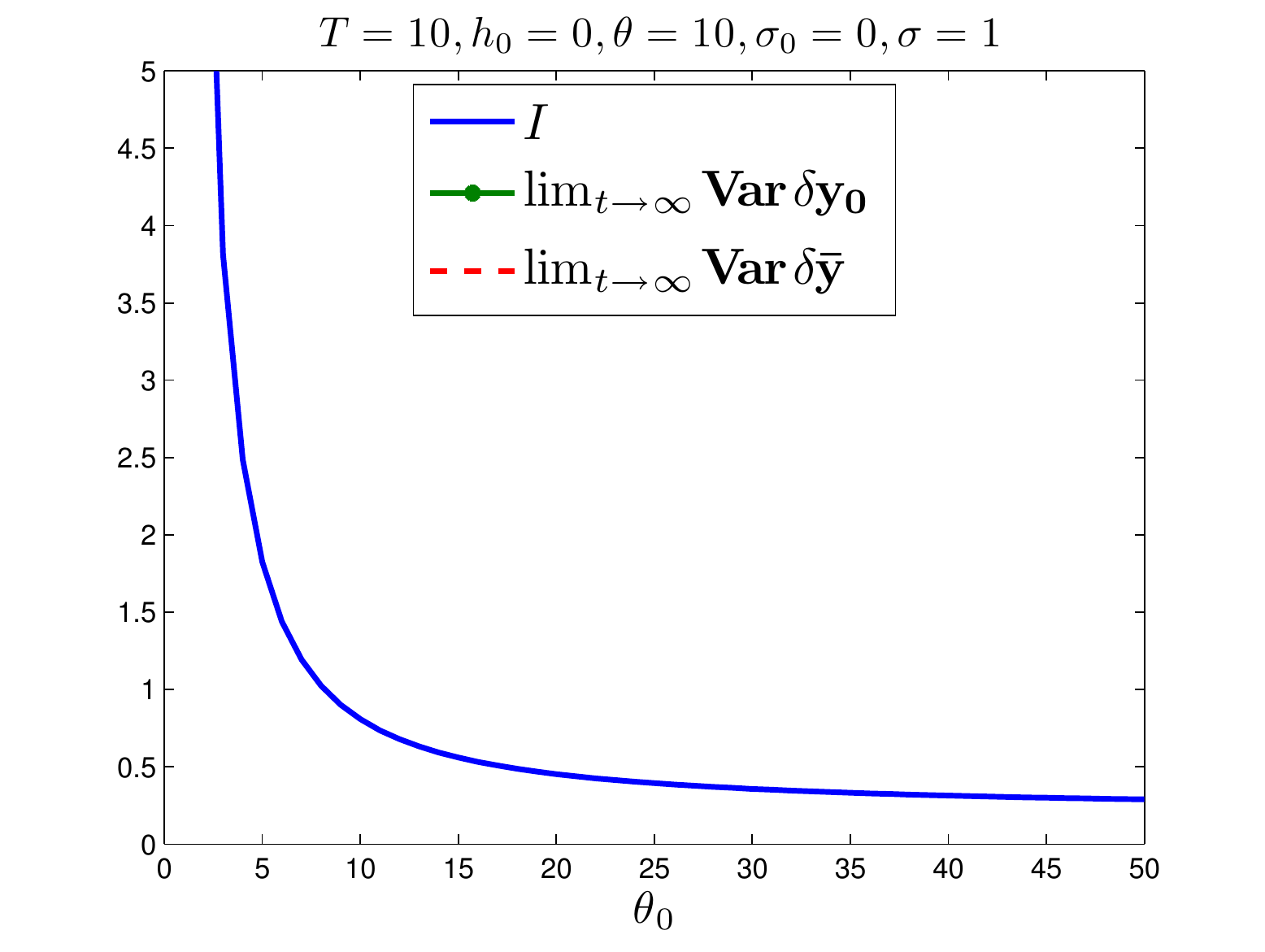}
	\includegraphics[width=0.49\linewidth]{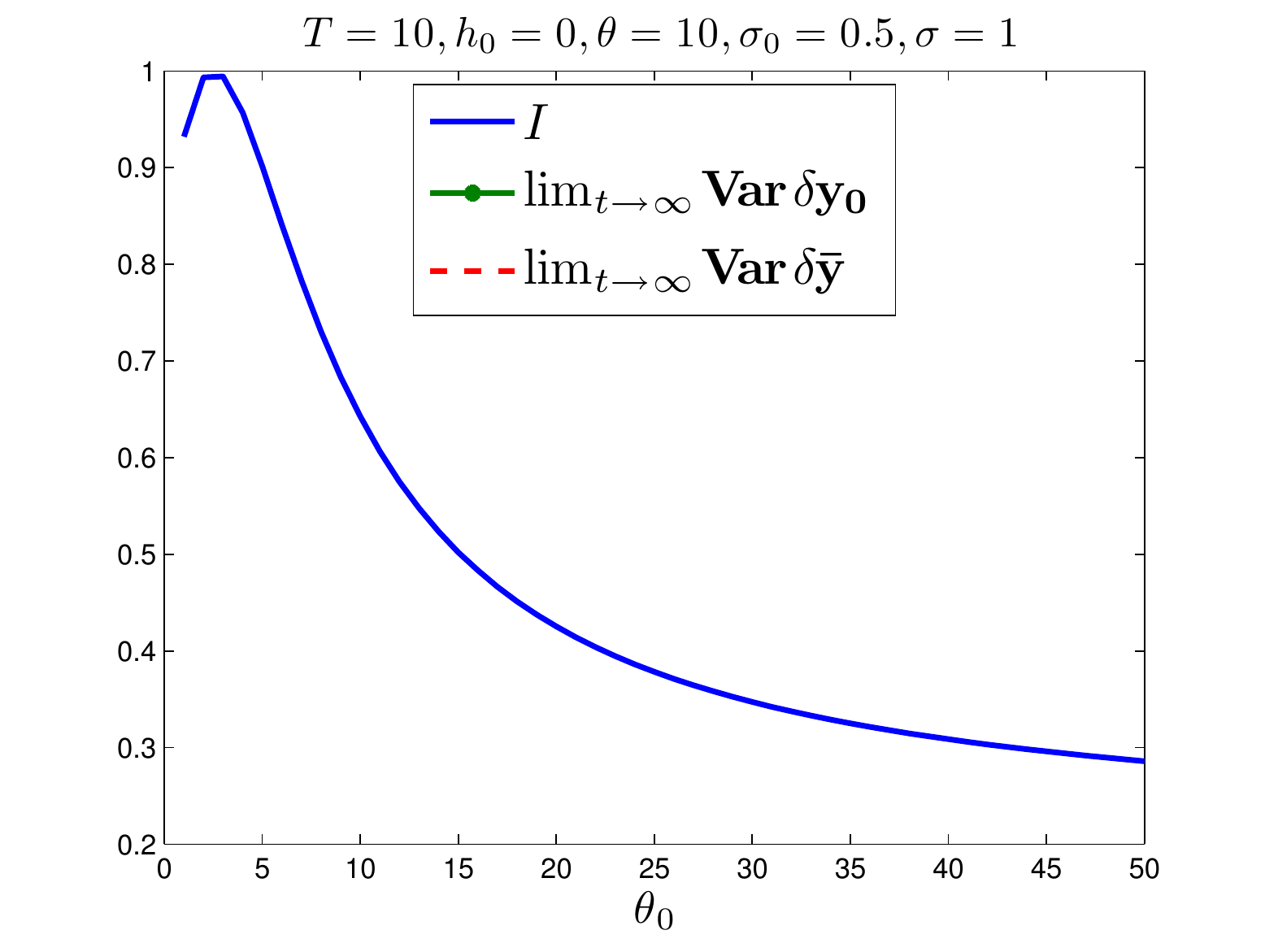}
	
	\includegraphics[width=0.49\linewidth]{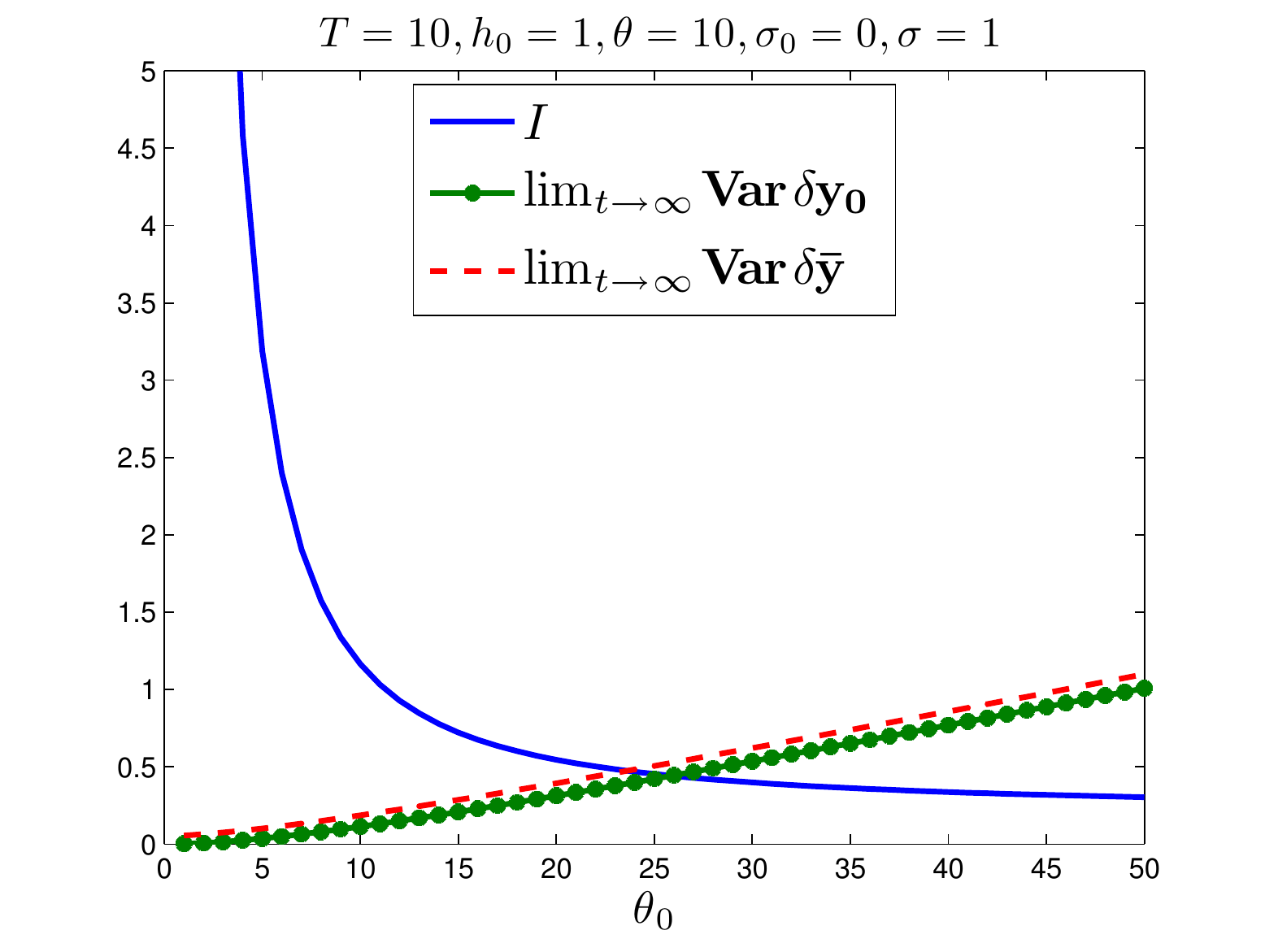}
	\includegraphics[width=0.49\linewidth]{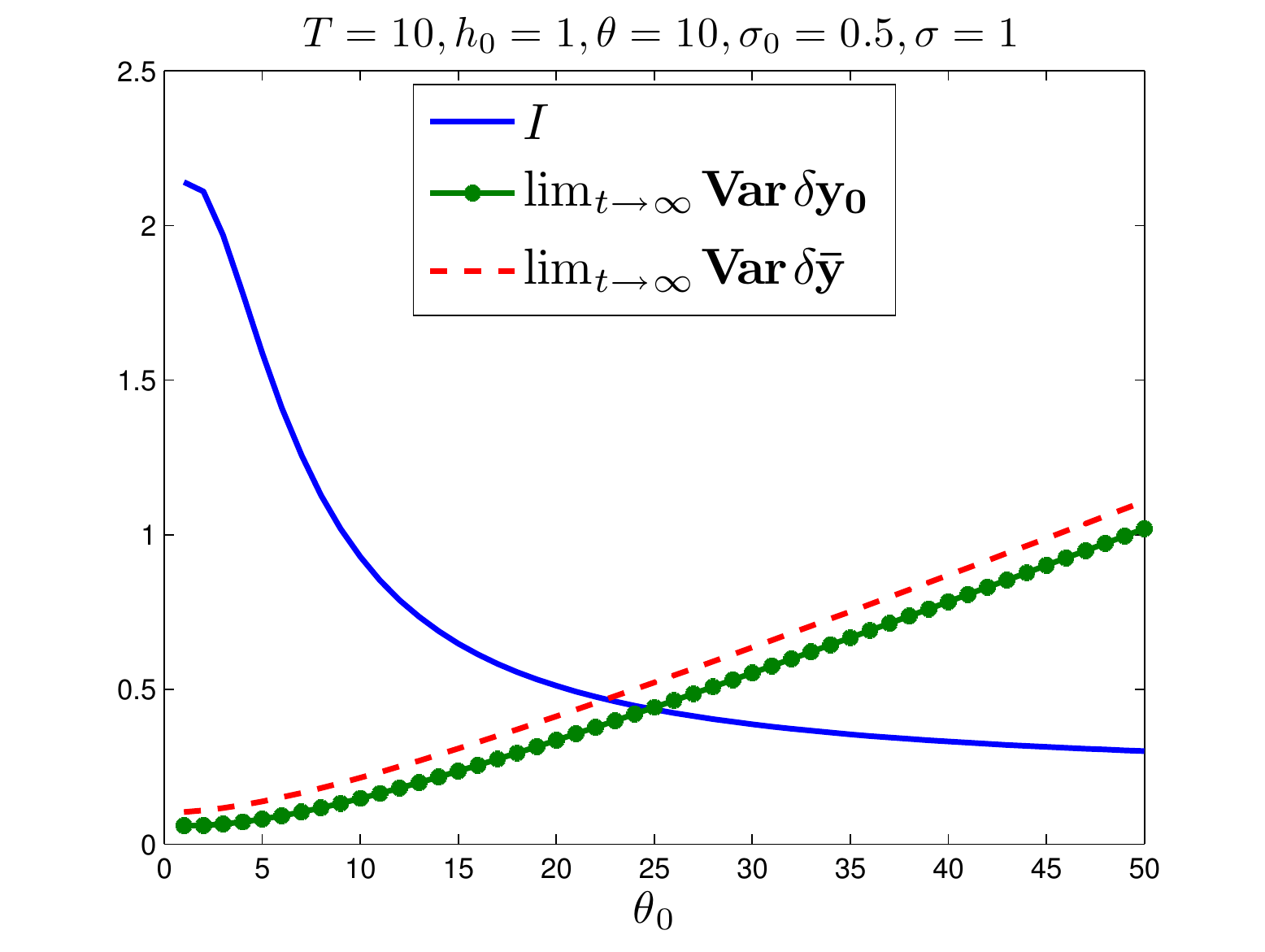}
	
	\includegraphics[width=0.49\linewidth]{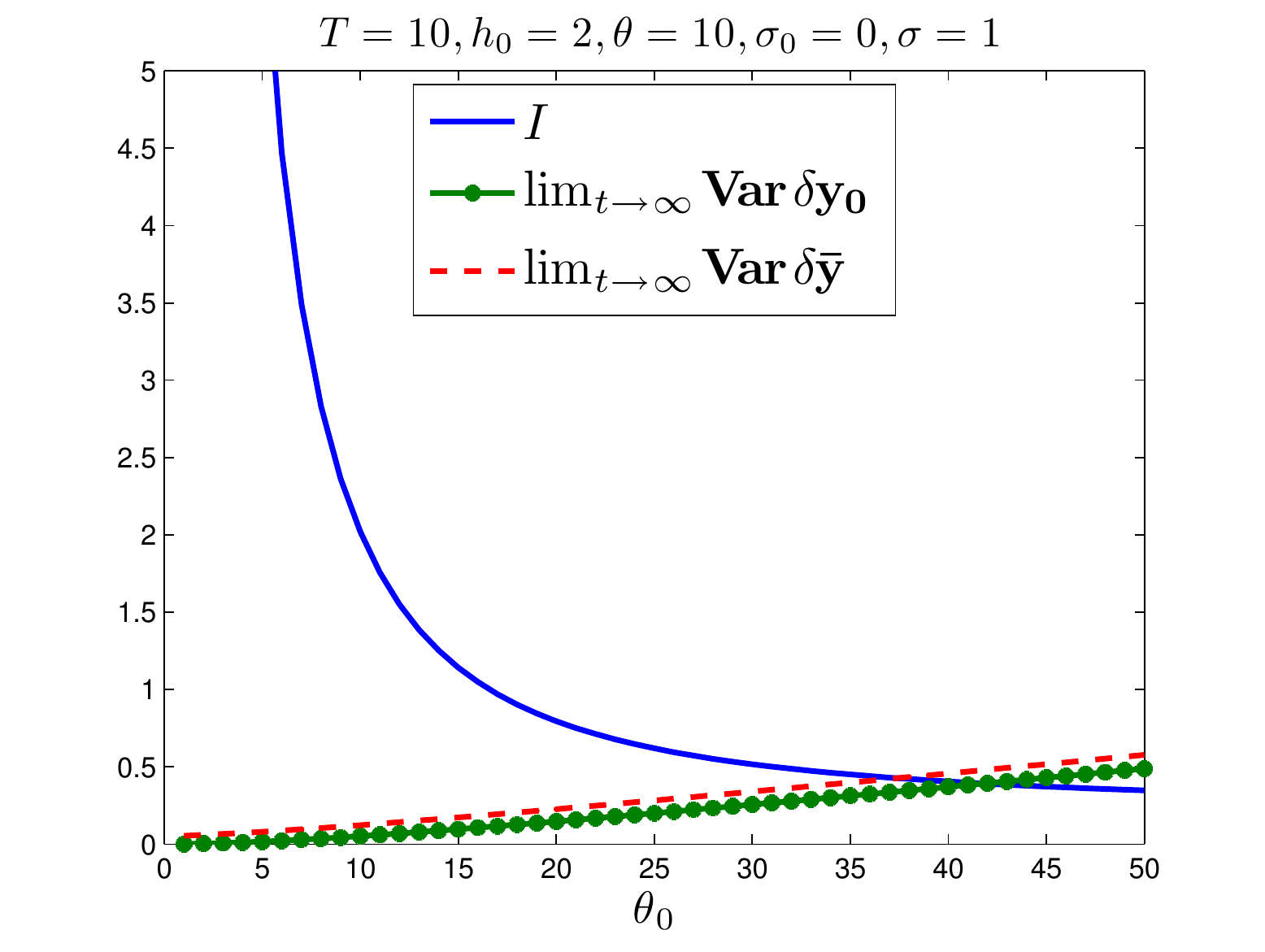}
	\includegraphics[width=0.49\linewidth]{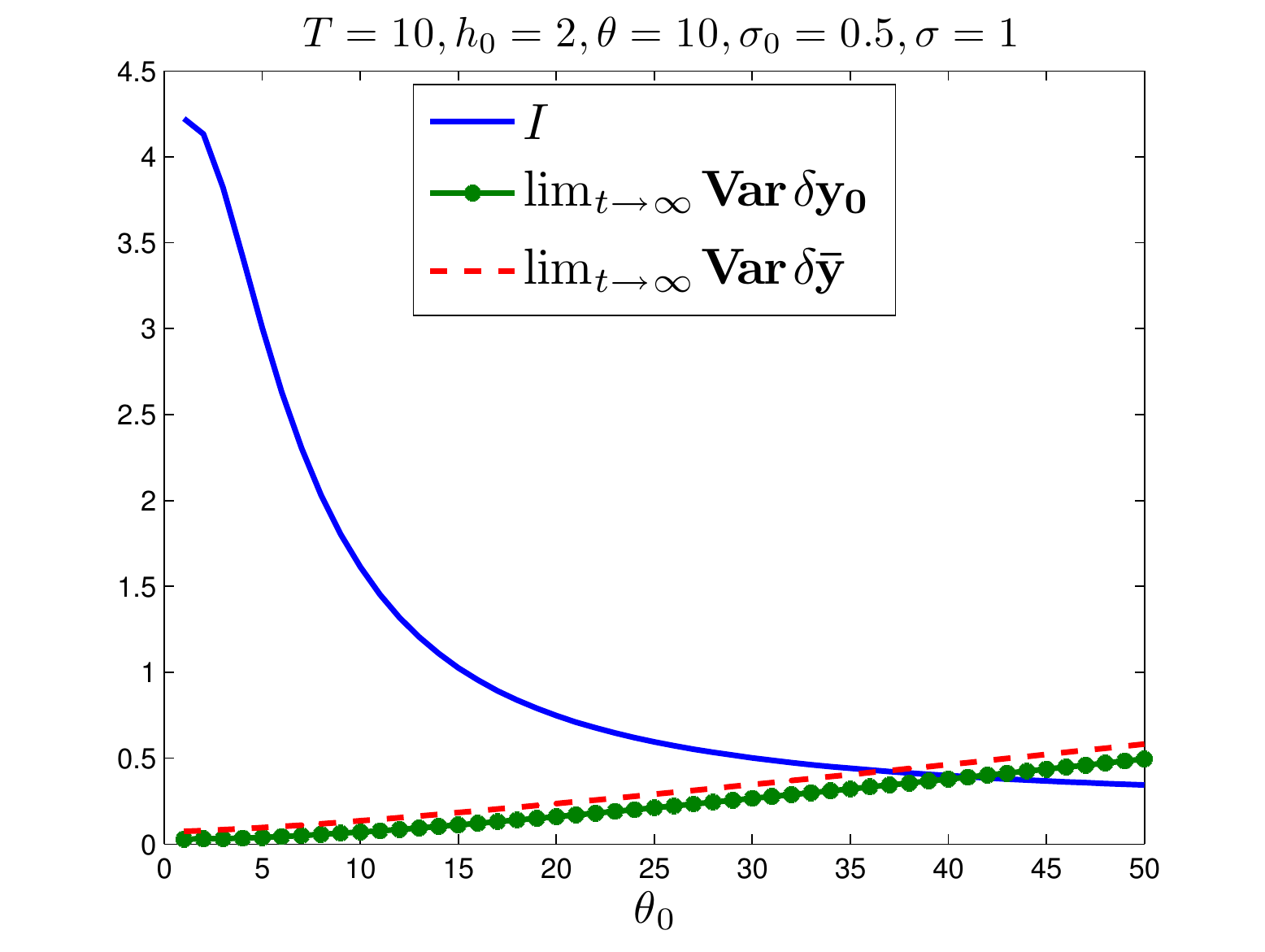}

	\caption{Plots of $\inf_{\bx\in\mathcal{A}}I(\bx)$, $\lim_{t\to\infty} \mathbf{Var}z_0(t)$ and 
	$\lim_{t\to\infty} \mathbf{Var}\bar{z}(t)$ for $\theta_0$ from $1$ to $50$. We let $T=10$, $\theta=10$, $\sigma=1$. The left 
	column is the case $\sigma_0=0$ and the right column is the case  $\sigma_0=0.5$.} 
	\label{fig:I to different theta0}
\end{figure}

One may also expect that $\theta_0$ does not greatly affect $\lim_{t\to\infty} \mathbf{Var}z_0(t)$ and 
$\lim_{t\to\infty} \mathbf{Var}\bar{z}(t)$; however, in Figure \ref{fig:I to different theta0} we see that 
the effect of $\theta_0$ on 
$\lim_{t\to\infty} \mathbf{Var}z_0(t)$ and $\lim_{t\to\infty} \mathbf{Var}\bar{z}(t)$ is not negligible. In other words, the 
independence of 
$\lim_{t\to\infty} \mathbf{Var}z_0(t)$ and $\lim_{t\to\infty} \mathbf{Var}\bar{z}(t)$
with respect to $\theta_0$ only holds in 
the limits (\ref{eq:limit of Var delta y_0, large theta}) and (\ref{eq:limit of Var delta y_bar, large theta}).

\subsection{Numerical results for optimal controls}
\label{sec:numerics of optimal control}

In this subsection, we use the Euler scheme to simulate (\ref{eq:dynamics of zero h}) with optimal controls:
\begin{align}
	\label{eq:Euler for optimal control}
	& x_0^{(N)}(n+1)=\frac{\sigma_0}{\sqrt{N}}\Delta W^0_{n+1}-h_0 V_0'(x_0^{(N)}(n))\Delta t - \theta_0 (x_0^{(N)}(n) - \bar{x}_N(n))\Delta t ,\\
	&\bar{x}_N(n+1) = \frac{\sigma}{\sqrt{N}} \Delta\bar{W}_{n+1} - \theta (\bar{x}_N(n) - x_0^{(N)}(n)) \Delta t + \alpha_j^\infty(n)\Delta t
	\notag
\end{align}
with $x_0^{(N)}(0)=\bar{x}_N(0)=-1$ and $\{\Delta W^0_{n+1}\}_n$, $\{\Delta \bar{W}_{n+1}\}_n$ i.i.d. Gaussian random variables with mean $0$ 
and variance $\Delta t$, where
\begin{equation}
	\label{eq:discrete stationary optimal control}
	\alpha_j^\infty(t)= -\theta_c \left(b_\infty (x_0^{(N)}(n)+1) + d_\infty (x_j(n)+1) + e_\infty (\bar{x}_N(n)+1)\right)
\end{equation}
and $(a_\infty, b_\infty, d_\infty, e_\infty)$ satisfies the algebraic Riccati equations (\ref{eq:algebraic Riccati equation}).

To obtain $(a_\infty, b_\infty, d_\infty, e_\infty)$, we numerically solve (\ref{eq:Riccati equation}) for large enough $T$ so that 
$(a(0), b(0), d(0), e(0))$ is essentially $(a_\infty, b_\infty, d_\infty, e_\infty)$. The values of the parameters used in 
(\ref{eq:Euler for optimal control}) are listed in Table \ref{tab:parameters of optimal control}.

\begin{figure}
	\centering
	\includegraphics[width=0.49\linewidth]{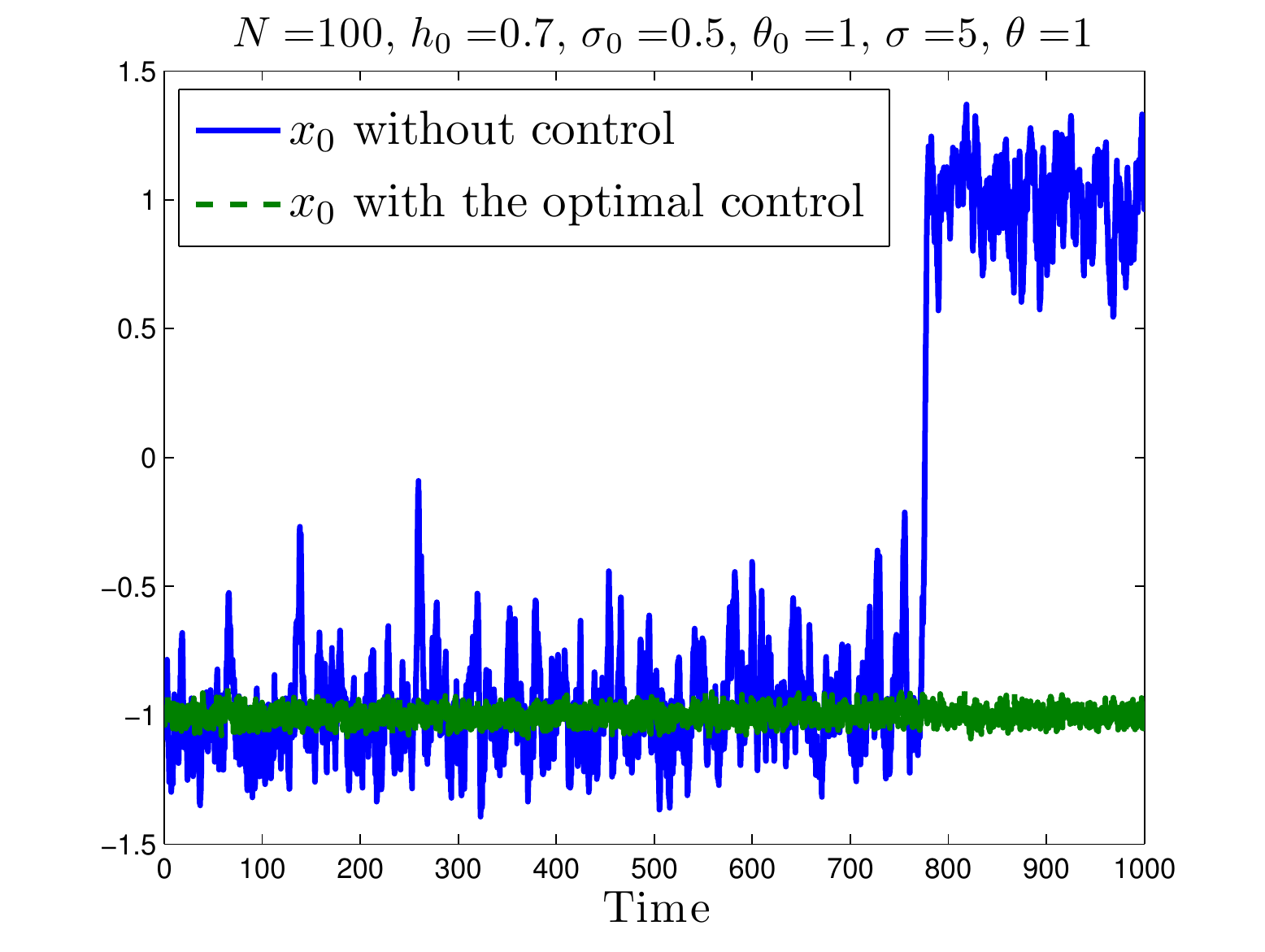}
	\includegraphics[width=0.49\linewidth]{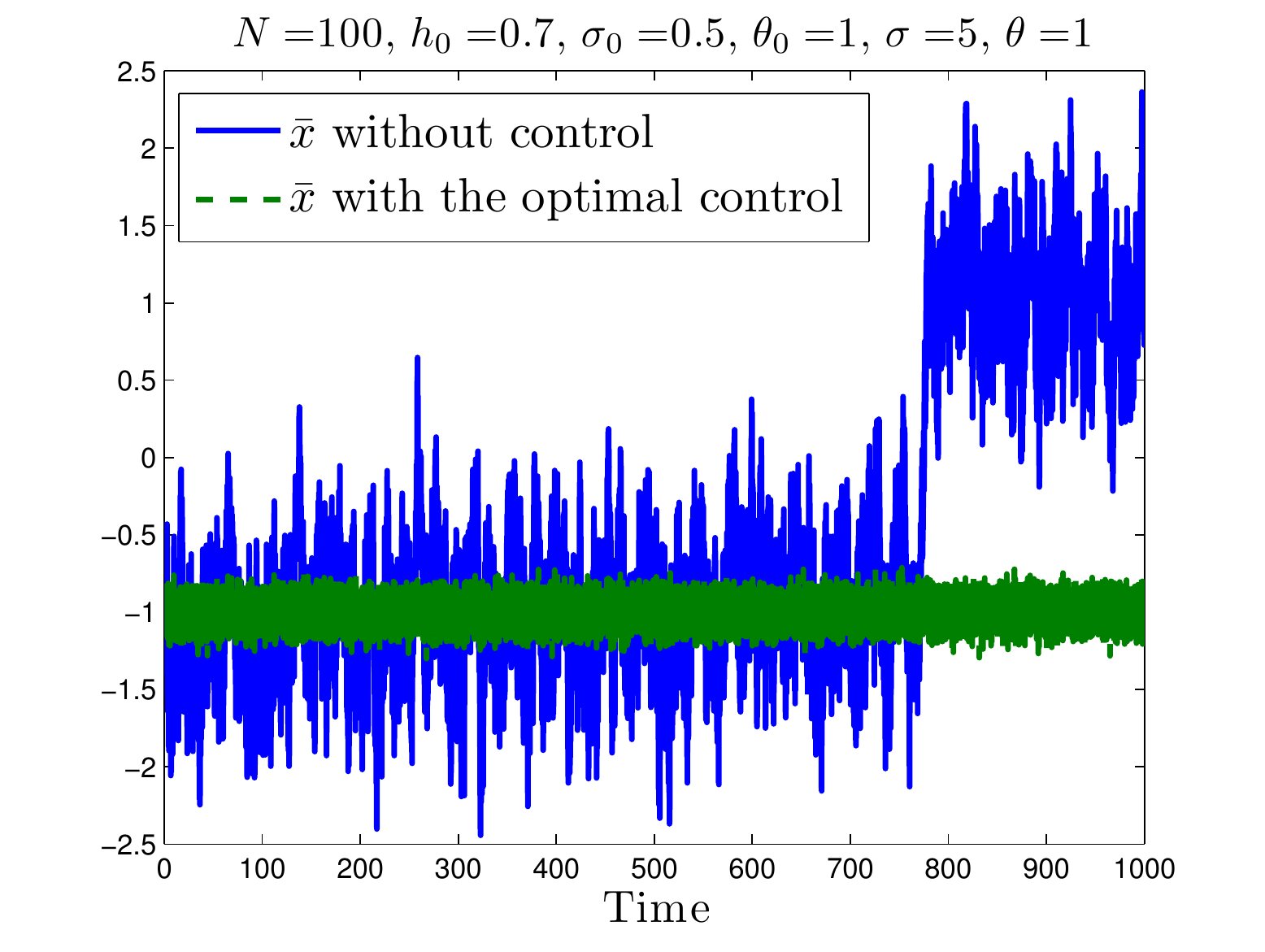}
	\caption{Sample paths of $x_0^{(N)}(t)$ and $\bar{x}_N(t)$ with and without the optimal control. With the optimal control, $x_0^{(N)}(t)$ and 
	$\bar{x}_N(t)$ are much more stable than the uncontrolled ones.}
	\label{fig:optimalControl}
\end{figure}

We see  from Figure \ref{fig:optimalControl} that the uncontrolled problem is very unstable in the sense that $x_0^{(N)}$ and $\bar{x}_N$ jump 
frequently between $-1$ and $+1$. On the other hand, under the same values of the parameters, the controlled $x_0^{(N)}$ and $\bar{x}_N$ are much 
more stable with no transition from $-1$ to $+1$.

\begin{table}
	\begin{center}
		\begin{tabular}{|c|c|c|c|c|c|c|c|}
			\hline 
			$N$&   $T$   & $\Delta t$& $h_0$& $\sigma_0$& $\theta_0$& $\sigma$& $\theta$\\ 
			\hline 
			$100$& $10^3$& $10^{-2}$ & $0.7$& $0.5$     & $1.0$     & $5.0$     & $1.0$\\ 
			\hline
		\end{tabular}
	\end{center}
	\caption{\label{tab:parameters of optimal control}
	The values of the parameters used in Sec.~\ref{sec:numerics of optimal control} for the controlled problem 
	(\ref{eq:Euler for optimal control}) and the uncontrolled problem (\ref{eq:Euler for zero h}).}
\end{table}

%% file: summary.tex
\section{Summary and Conclusions}
We have formulated and analyzed a multi-agent model for the evolution of individual and systemic risk 
when there is a central agent acting as a stabilizer in the system.
The local agents do not have an intrisinc stabilizing mechanism.
The main result of this paper can be visualized in Figures \ref{fig:I to different sigma} and \ref{fig:I to different theta0} and is briefly described as follows.
The systemic risk decreases when the rate of adherence of the local agents to the central agent
increases, but it increases when the rate of adherence of the central agent to the mean 
of the local agents increases. This is under the condition that the observed individual risk
is kept approximately constant.
We also show that the effect of drift controls on the local agents is to always
stabilize the systemic risk.

\section*{Acknowledgment}

This work is partly supported by the Department of Energy [National Nuclear Security Administration] under Award Number 
NA28614, and partly by AFOSR grant FA9550-11-1-0266. The authors thank the Institut des Hautes Etudes Scientifiques (IHES) for 
its hospitality while part of this work was carried out.

%% file: proof_zero_h.tex
\section{Proofs in Section \ref{sec:zero h}}

\subsection{Proof of Proposition \ref{prop:limits of fluctuations}}
\label{pf:limits of fluctuations}

We first consider the eigen-decomposition of ${\bf A}$: ${\bf A}={\bf Q}{\boldsymbol \Lambda} {\bf Q}^{-1}$, where
\begin{align*}
	&{\boldsymbol \Lambda} = \begin{pmatrix} \lambda_1 & 0 \\ 0 & \lambda_2 \end{pmatrix}, \quad
	{\bf Q} = \frac{\theta}{\lambda_1-\lambda_2}\begin{pmatrix} 1+\frac{\lambda_1}{\theta} 
	& 1+\frac{\lambda_2}{\theta} \\ 1 & 1 \end{pmatrix},\quad
	{\bf Q}^{-1} = \begin{pmatrix} 1 & -(1+\frac{\lambda_2}{\theta}) \\ 
	-1 & 1+\frac{\lambda_1}{\theta} \end{pmatrix},\\
	&\lambda_1 = \frac{1}{2}\left\{-[h_0V_0''(y_0^e)+\theta_0+\theta]
	+\sqrt{[h_0V_0''(y_0^e)+\theta_0+\theta]^2-4\theta h_0V_0''(y_0^e)} \right\},\\
	&\lambda_2 = \frac{1}{2}\left\{-[h_0V_0''(y_0^e)+\theta_0+\theta]
	-\sqrt{[h_0V_0''(y_0^e)+\theta_0+\theta]^2-4\theta h_0V_0''(y_0^e)} \right\}.
\end{align*}
We note that $\lambda_1$ and $\lambda_2$ are real and negative if $h_0$, $\theta_0$ and $\theta$ are positive. Then from 
(\ref{eq:expectations of fluctuations}), $\lim_{t\to\infty}z_0(t)=\lim_{t\to\infty}\bar{z}(t)=0$. In addition, from the 
eigen-decomposition we have
\begin{multline}
	\begin{pmatrix} 
		\mathbf{Var}z_0(t) & \mathbf{Cov}(z_0(t), \bar{z}(t)) \\ 
		\mathbf{Cov}(z_0(t), \bar{z}(t)) & \mathbf{Var}\bar{z}(t) 
	\end{pmatrix}\\
	= {\bf Q}\int_0^t e^{(t-s){\boldsymbol \Lambda}}{\bf Q}^{-1} \begin{pmatrix} \sigma_0^2 & 0 \\ 0 & \sigma^2 \end{pmatrix} 
		({\bf Q}^{-1})^{\mathbf{T}}e^{(t-s){\boldsymbol \Lambda}}ds {\bf Q}^\mathbf{T}.
\end{multline}
We observe that 
\begin{multline*}
	{\bf Q}^{-1} \begin{pmatrix} \sigma_0^2 & 0 \\ 0 & \sigma^2 \end{pmatrix} ({\bf Q}^{-1})^{\mathbf{T}}\\
	= \begin{pmatrix}
		\sigma_0^2+\sigma^2(1+\frac{\lambda_2}{\theta})^2 & 
		-\sigma_0^2-\sigma^2(1+\frac{\lambda_1}{\theta})(1+\frac{\lambda_2}{\theta}) \\
		-\sigma_0^2-\sigma^2(1+\frac{\lambda_1}{\theta})(1+\frac{\lambda_2}{\theta}) & 
		\sigma_0^2+\sigma^2(1+\frac{\lambda_1}{\theta})^2
	\end{pmatrix}.
\end{multline*}
Then 
\begin{multline*}
	\lim_{t\to\infty} \int_0^t e^{(t-s){\boldsymbol \Lambda}} {\bf Q}^{-1} \begin{pmatrix}\sigma_0^2 & 0\\
	0 & \sigma^2\end{pmatrix} ({\bf Q}^{-1})^{\mathbf{T}}e^{(t-s){\boldsymbol \Lambda}} ds\\
	= \begin{pmatrix}
		-\frac{1}{2\lambda_1}[\sigma_0^2+\sigma^2(1+\frac{\lambda_2}{\theta})^2] & 
		\frac{1}{\lambda_1+\lambda_2}[-\sigma_0^2-\sigma^2(1+\frac{\lambda_1}{\theta})(1+\frac{\lambda_2}{\theta})]\\
		\frac{1}{\lambda_1+\lambda_2}[-\sigma_0^2-\sigma^2(1+\frac{\lambda_1}{\theta})(1+\frac{\lambda_2}{\theta})] & 
		-\frac{1}{2\lambda_2}[\sigma_0^2+\sigma^2(1+\frac{\lambda_1}{\theta})^2]
	\end{pmatrix}.
\end{multline*}
So we obtain 
\begin{align}
	\label{eq:limit of Var delta y_0}
	&\lim_{t\to\infty} \mathbf{Var}z_0(t) 
	= \frac{\theta^2}{(\lambda_1-\lambda_2)^2}\left\{-\frac{1}{2\lambda_1}\left(1+\frac{\lambda_1}{\theta}\right)^2
	\left[\sigma_0^2+\sigma^2\left(1+\frac{\lambda_2}{\theta}\right)^2\right]\right. \\
	&\quad + \frac{2}{\lambda_1+\lambda_2}\left(1+\frac{\lambda_1}{\theta}\right)\left(1+\frac{\lambda_2}{\theta}\right)
	\left[\sigma_0^2+\sigma^2\left(1+\frac{\lambda_1}{\theta}\right)\left(1+\frac{\lambda_2}{\theta}\right)\right] \notag \\
	&\quad \left. -\frac{1}{2\lambda_2}\left(1+\frac{\lambda_2}{\theta}\right)^2
	\left[\sigma_0^2+\sigma^2\left(1+\frac{\lambda_1}{\theta}\right)^2\right]\right\},\notag
\end{align}
\begin{align}
	\label{eq:limit of Var delta y_bar}
	&\lim_{t\to\infty} \mathbf{Var}\bar{z}(t) 
	= \frac{\theta^2}{(\lambda_1-\lambda_2)^2}\left\{-\frac{1}{2\lambda_1}
	\left[\sigma_0^2+\sigma^2\left(1+\frac{\lambda_2}{\theta}\right)^2\right]\right.\\
	&\left. \quad + \frac{2}{\lambda_1+\lambda_2}
	\left[\sigma_0^2+\sigma^2\left(1+\frac{\lambda_1}{\theta}\right)\left(1+\frac{\lambda_2}{\theta}\right)\right]
	-\frac{1}{2\lambda_2}\left[\sigma_0^2+\sigma^2\left(1+\frac{\lambda_1}{\theta}\right)^2\right]\right\},\notag
\end{align}
\begin{align}
	\label{eq:limit of Cov(delta y_0, delta y_bar)}
	&\lim_{t\to\infty} \mathbf{Cov}(z_0(t), \bar{z}(t)) 
	= \frac{\theta^2}{(\lambda_1-\lambda_2)^2}\left\{-\frac{1}{2\lambda_1}\left(1+\frac{\lambda_1}{\theta}\right)
	\left[\sigma_0^2+\sigma^2\left(1+\frac{\lambda_2}{\theta}\right)^2\right]\right. \\
	&\quad + \frac{1}{\lambda_1+\lambda_2}\left(1+\frac{\lambda_1}{\theta}\right)
	\left[\sigma_0^2+\sigma^2\left(1+\frac{\lambda_1}{\theta}\right)\left(1+\frac{\lambda_2}{\theta}\right)\right] \notag \\
	&\quad + \frac{1}{\lambda_1+\lambda_2}\left(1+\frac{\lambda_2}{\theta}\right)
	\left[\sigma_0^2+\sigma^2\left(1+\frac{\lambda_1}{\theta}\right)\left(1+\frac{\lambda_2}{\theta}\right)\right] \notag \\
	&\quad \left. -\frac{1}{2\lambda_2}\left(1+\frac{\lambda_2}{\theta}\right)
	\left[\sigma_0^2+\sigma^2\left(1+\frac{\lambda_1}{\theta}\right)^2\right]\right\}.\notag
\end{align}

We are interested in the case that $\sigma$ and $\theta$ go to infinity while the ratio $\alpha=\sigma^2/\theta$ is fixed. For 
$\theta$ large and using the approximation $\sqrt{1+x}=1+\frac{1}{2}x+O(x^2)$, we have the following expansions:
\begin{align*}
	\frac{\lambda_1}{\theta}
	&= \frac{1}{2\theta} \left\{-[h_0V_0''(y_0^e)+\theta_0+\theta] 
	+ [h_0V_0''(y_0^e)+\theta_0+\theta]\sqrt{1-\frac{4\theta h_0V_0''(y_0^e)}{[h_0V_0''(y_0^e)+\theta_0+\theta]^2}}\right\}\\
	&= -\frac{h_0V_0''(y_0^e)}{h_0V_0''(y_0^e)+\theta_0+\theta} + O\left(\frac{1}{\theta^2}\right),
\end{align*}
\begin{align*}
	1+\frac{\lambda_2}{\theta}
	&= \frac{1}{2\theta} \left\{2\theta -[h_0V_0''(y_0^e)+\theta_0+\theta] 
	- [h_0V_0''(y_0^e)+\theta_0+\theta]\sqrt{1-\frac{4\theta h_0V_0''(y_0^e)}{[h_0V_0''(y_0^e)+\theta_0+\theta]^2}}\right\}\\
	&= -\frac{1}{\theta}[h_0V_0''(y_0^e)+\theta_0] + \frac{h_0V_0''(y_0^e)}{h_0V_0''(y_0^e)+\theta_0+\theta} 
	+ O\left(\frac{1}{\theta^2}\right).
\end{align*}
Thus $\lambda_1\to h_0V_0''(y_0^e)$ as $\theta\to\infty$ and $1+\frac{\lambda_2}{\theta}=O(\frac{1}{\theta})$ and finally we 
have the limits (\ref{eq:limit of Var delta y_0, large theta}), (\ref{eq:limit of Var delta y_bar, large theta}) and 
(\ref{eq:limit of Cov(delta y_0, delta y_bar), large theta}).

\subsection{Proof of Proposition \ref{prop:BVP for x0, degenerate case}}
\label{pf:BVP for x0, degenerate case}

If $x_0$ is the minimizer, then for any perturbation $\phi$ with $\phi(0)=\phi(T)=\dot{\phi}(0)=\dot{\phi}(T)=0$, the directional 
derivative of $I$ must be zero:
\begin{multline*}
	\left.\frac{d}{d\epsilon}\right|_{\epsilon=0}I(x_0+\epsilon \phi) 
	=\frac{1}{2\sigma^2}\int_0^T 2\left[\frac{1}{\theta_0}\ddot{x}_0+\frac{h_0}{\theta_0}V''_0(x_0)\dot{x}_0
	+\left(1+\frac{\theta}{\theta_0}\right)\dot{x}_0+\frac{\theta h_0}{\theta_0}V'_0(x_0)\right]\\
	\times\left[\frac{1}{\theta_0}\ddot{\phi}+\frac{h_0}{\theta_0}V'''_0(x_0)\phi\dot{x}_0+\frac{h_0}{\theta_0}V''_0(x_0)\dot{\phi}
	+\left(1+\frac{\theta}{\theta_0}\right)\dot{\phi}+\frac{\theta h_0}{\theta_0}V''_0(x_0)\phi\right]dt=0.
\end{multline*}
After integration by parts and using the fact that $\phi$ is arbitrary, the minimizer $x_0$ must satisfy the following equation:
\begin{align*}
	\frac{1}{\theta_0}\frac{d^2}{dt^2}\left[\frac{1}{\theta_0}\ddot{x}_0+\frac{h_0}{\theta_0}V''_0(x_0)\dot{x}_0
	+\left(1+\frac{\theta}{\theta_0}\right)\dot{x}_0+\frac{\theta h_0}{\theta_0}V'_0(x_0)\right]\\
	+\frac{h_0}{\theta_0}V'''_0(x_0)\dot{x}_0\left[\frac{1}{\theta_0}\ddot{x}_0+\frac{h_0}{\theta_0}V''_0(x_0)\dot{x}_0
	+\left(1+\frac{\theta}{\theta_0}\right)\dot{x}_0+\frac{\theta h_0}{\theta_0}V'_0(x_0)\right]\\
	-\frac{d}{dt}\left\{ \frac{h_0}{\theta_0}V''_0(x_0)\left[\frac{1}{\theta_0}\ddot{x}_0+\frac{h_0}{\theta_0}V''_0(x_0)\dot{x}_0
	+\left(1+\frac{\theta}{\theta_0}\right)\dot{x}_0+\frac{\theta h_0}{\theta_0}V'_0(x_0)\right]\right\} \\
	-\left(1+\frac{\theta}{\theta_0}\right)\frac{d}{dt}\left[\frac{1}{\theta_0}\ddot{x}_0+\frac{h_0}{\theta_0}V''_0(x_0)\dot{x}_0
	+\left(1+\frac{\theta}{\theta_0}\right)\dot{x}_0+\frac{\theta h_0}{\theta_0}V'_0(x_0)\right]\\
	+\frac{\theta h_0}{\theta_0}V''_0(x_0)\left[\frac{1}{\theta_0}\ddot{x}_0+\frac{h_0}{\theta_0}V''_0(x_0)\dot{x}_0
	+\left(1+\frac{\theta}{\theta_0}\right)\dot{x}_0+\frac{\theta h_0}{\theta_0}V'_0(x_0)\right] & =0.
\end{align*}
with the boundary conditions $x_0\left(0\right)=-1$, $x_0(t)=1$ and $\frac{d}{dt}x_0\left(0\right)=\frac{d}{dt}x_0(t)=0$. We then 
obtain (\ref{eq:BVP for x0, degenerate case}) after rearranging the above equation.

\subsection{Proof of Proposition \ref{prop:exponential rate of the transition probability, nondegenerate case}}
\label{pf:exponential rate of the transition probability, nondegenerate case}

If $h_0=0$, (\ref{eq:dynamics of zero h}) is a system of linear SDEs, and the explicit solution can be found:
\begin{equation*}
	\begin{pmatrix}x_0(T)\\ \bar{x}_{N}(T)\end{pmatrix}
	= e^{T{\bf A}_0}\begin{pmatrix}-1\\ -1 \end{pmatrix}
	+ \frac{1}{\sqrt{N}}\int_0^T e^{(T-s){\bf A}_0}\begin{pmatrix}\sigma_0 dW_s^0\\ \sigma d\bar{W}_s \end{pmatrix},\quad \quad
	{\bf A}_0 = \begin{pmatrix} - \theta_0 & \theta_0 \\ \theta & -\theta \end{pmatrix}.	 	
\end{equation*}
Since (\ref{eq:dynamics of zero h})  is linear, $(x_0(T), \bar{x}_N(T))$ is jointly Gaussian and can be completely characterized by its 
mean and covariance matrix. We note that $(-1,-1)^\mathbf{T}$ is in the null space of ${\bf A}_0$ and thus 
\begin{equation*}
	\EE\begin{pmatrix}x_0(T)\\ \bar{x}_{N}(T)\end{pmatrix}
	= e^{T{\bf A}_0}\begin{pmatrix}-1\\ -1 \end{pmatrix} = \begin{pmatrix}-1\\ -1 \end{pmatrix}.
\end{equation*}
In addition, ${\bf A}_0$ has the following eigen-decomposition: ${\bf A}_0={\bf Q}_0{\boldsymbol \Lambda}_0 {\bf Q}_0^{-1}$, where
\begin{equation*}
	{\boldsymbol \Lambda}_0 = \begin{pmatrix} 0 & 0 \\ 0 & -(\theta_0+\theta) \end{pmatrix}, \quad
	{\bf Q}_0 = \frac{\theta}{\theta_0+\theta}\begin{pmatrix} 1 & -\frac{\theta_0}{\theta} \\ 1 & 1 \end{pmatrix},\quad
	{\bf Q}^{-1}_0 = \begin{pmatrix} 1 & \frac{\theta_0}{\theta} \\ -1 & 1\end{pmatrix}.
\end{equation*}
Then the covariance matrix is 
\begin{align}
	\label{eq:covariance matrix of (x0(T),xbar(T)) for h0=0}
	\lefteqn{\begin{pmatrix} 
		\mathbf{Var}x_0(T) & \mathbf{Cov}(x_0(T), \bar{x}(T)) \\ 
		\mathbf{Cov}(x_0(T), \bar{x}(T)) & \mathbf{Var}\bar{x}(T) 
	\end{pmatrix}}\\
	&= \frac{1}{N} {\bf Q}_0\int_0^T e^{(T-s){\boldsymbol \Lambda}_0}{\bf Q}_0^{-1} \begin{pmatrix} \sigma_0^2 & 0 \\ 0 & \sigma^2 \end{pmatrix} 
	({\bf Q}_0^{-1})^{\mathbf{T}}e^{(T-s){\boldsymbol \Lambda}_0}ds {\bf Q}_0^\mathbf{T}\notag\\
	&= \frac{1}{N}{\bf Q}_0
	 {\boldsymbol{\Sigma}}
	{\bf Q}_0^\mathbf{T}, \notag   
\end{align}
with 
$$
 {\boldsymbol{\Sigma}} =\begin{pmatrix}
		T(\sigma_0^2+\theta_0^2\sigma^2/\theta^2)
		& \frac{1}{\theta_0+\theta}(-\sigma_0^2+\theta_0\sigma^2/\theta)[1-e^{-T(\theta_0+\theta)}]\\
		\frac{1}{\theta_0+\theta}(-\sigma_0^2+\theta_0\sigma^2/\theta)[1-e^{-T(\theta_0+\theta)}]
		& \frac{1}{2(\theta_0+\theta)}(\sigma_0^2+\sigma^2)[1-e^{-2T(\theta_0+\theta)}]
	\end{pmatrix} .
$$
When the terminal time $T$ is large, we can separate the middle matrix in (\ref{eq:covariance matrix of (x0(T),xbar(T)) for h0=0}) 
into the principle term and the correction term:
\begin{align*}
	 &{\boldsymbol{\Sigma}}=\begin{pmatrix} T(\sigma_0^2+\theta_0^2\sigma^2/\theta^2) & 0\\ 0 & 0 \end{pmatrix}\\
	& + \begin{pmatrix}
		0 & \frac{1}{\theta_0+\theta}(-\sigma_0^2+\theta_0\sigma^2/\theta)[1-e^{-T(\theta_0+\theta)}]\\
		\frac{1}{\theta_0+\theta}(-\sigma_0^2+\theta_0\sigma^2/\theta)[1-e^{-T(\theta_0+\theta)}]
		& \frac{1}{2(\theta_0+\theta)}(\sigma_0^2+\sigma^2)[1-e^{-2T(\theta_0+\theta)}]
	\end{pmatrix} .
\end{align*}
Then we have the approximation of the covariance matrix:
\begin{align}
	\label{eq:approximate covariance matrix of (x0(T),xbar(T)) for h0=0}
	\begin{pmatrix} 
		\mathbf{Var}x_0(T) & \mathbf{Cov}(x_0(T), \bar{x}(T)) \\ 
		\mathbf{Cov}(x_0(T), \bar{x}(T)) & \mathbf{Var}\bar{x}(T) 
	\end{pmatrix}
	&\approx \frac{1}{N}{\bf Q}_0 
	\begin{pmatrix} T(\sigma_0^2+\theta_0^2\sigma^2/\theta^2) & 0\\ 0 & 0 \end{pmatrix} 
	{\bf Q}_0^\mathbf{T}\\
	&= \frac{T}{N}\frac{\theta^2\sigma_0^2+\theta_0^2\sigma^2}{(\theta_0+\theta)^2}
	\begin{pmatrix}1 & 1\\ 1 & 1 \end{pmatrix}
	.\notag
\end{align}
From  (\ref{eq:approximate covariance matrix of (x0(T),xbar(T)) for h0=0}) we conclude that $x_0(T)$ and $\bar{x}(T)$ are 
approximately equal as $T$ becomes large and the probability in 
(\ref{eq:exponential rate of the transition probability, nondegenerate case}) is approximately $\PP(x_0(T)\in(1,1+dx))$, which 
gives the desired rate of decay by using the fact that $x_0(T)$ is Gaussian with mean $-1$ and approximate variance 
$\mathbf{Var}x_0(T)$ in (\ref{eq:approximate covariance matrix of (x0(T),xbar(T)) for h0=0}) for large $T$.

%% file: proof_empirical_measure.tex
\section{Proof of Proposition \ref{prop:gaussianpath}}
\label{pf:gaussianpath}

We prove it in three steps. The first step is to show that there exists a uniform lower bound for $\mathcal{J}$ over all feasible 
$\phi$.
\begin{lemma}
	If $h=0$, then for all $\phi(t,dx)$ such that $\langle \phi(t,dx), x\rangle =\bar{x}(t)$,
	\begin{eqnarray*}
		\mathcal{J}\big( (x_0(t),\phi(t,dx))_{t \in [0,T]} \big) 
		&\geq& \frac{1}{2\sigma_0^2} \int_0^T ( \dot{x}_0 + h_0 V_0'(x_0) + \theta_0(x_0 - \bar{x}) )^2 dt \\
		&&+ \frac{1}{2\sigma^2} \int_0^T ( \dot{\bar{x}} + \theta(\bar{x} -x_0))^2 dt,
	\end{eqnarray*}
	for $\sigma_0>0$ and for $\sigma_0=0$,
	\begin{equation*}
		\mathcal{J}\big( (x_0(t),\phi(t,dx))_{t \in [0,T]} \big)  \geq \frac{1}{2\sigma^2} \int_0^T ( \dot{\bar{x}} + \theta(\bar{x} -x_0))^2 dt,
	\end{equation*}
	if $\dot{x}_0 + h_0 V_0'(x_0) + \theta_0(x_0 - \bar{x})=0$ or $\mathcal{J}\big( (x_0(t),\phi(t,dx))_{t \in [0,T]} \big) =\infty$ otherwise.
\end{lemma}
\begin{proof}
	By taking $f(x)=x$, we have
	\begin{multline*}
		\int_0^T \sup_{f(x): \langle \phi, (f'(x))^2 \rangle \neq 0}
		\frac{\langle \phi_t - \frac{1}{2}\sigma^2 \phi_{xx} -\theta\frac{\partial}{\partial x}[(x-x_0(t))\phi], f(x)\rangle^2}
		{\langle \phi, (f'(x))^2 \rangle} dt\\
		\overset{f(x)=x}{\geq} \int_0^T \frac{\langle \phi_t - \frac{1}{2}\sigma^2 \phi_{xx} 
		-\theta\frac{\partial}{\partial x}[(x-x_0(t))\phi], x\rangle^2}{\langle \phi, 1\rangle}dt
		= \int_0^T ( \dot{\bar{x}} + \theta(\bar{x} -x_0))^2 dt.
	\end{multline*}
	Then we have the desired results.
\end{proof}

We then prove that $\mathcal{J}\big( (x_0(t),\bar{p}(t,dx))_{t \in [0,T]} \big)
=I\big( (x_0(t),\bar{x}(t))_{ \in [0,T]} \big)$ and consequently 
$\mathcal{I}(x_0,\bar{x})=I(x_0,\bar{x})$.
\begin{lemma}
	Let $\bar{p}$ defined in (\ref{eq:gaussianpath}) and $h=0$. Then $\mathcal{J}\big( (x_0(t),\bar{p}(t,dx))_{t \in [0,T]} \big)=I(x_0,\bar{x})$ in 
	(\ref{eq:rate function, degenerate case}) for $\sigma_0=0$ and $\mathcal{J}\big( (x_0(t),\bar{p}(t,dx))_{t \in [0,T]} \big)=I(x_0,\bar{x})$	in 
	(\ref{eq:rate function, nondegenerate case}) for $\sigma_0>0$. Consequently, $\bar{p}(t,dx)$ is a minimizer and 
	$\mathcal{I}(x_0,\bar{x})=I(x_0,\bar{x})$ for either $\sigma_0=0$ or $\sigma_0>0$.
\end{lemma}
\begin{proof}
	By using the same argument in \cite[Proposition 5.3]{Garnier2013}, if $\phi(t,dx)$ is absolutely continuous with respect to the 
	Lebesgue measure with the smooth density function $\phi(t,x)$, then  
	\begin{equation*}
		\int_0^T \sup_{f(x): \langle \phi, (f'(x))^2 \rangle \neq 0}
		\frac{\langle \phi_t - \frac{1}{2}\sigma^2 \phi_{xx} -\theta\frac{\partial}{\partial x}[(x-x_0(t))\phi], f(x)\rangle^2}
		{\langle \phi, (f'(x))^2 \rangle} dt = \int_0^T \langle \phi, (g(t,x))^2\rangle dt,
	\end{equation*}
	where $g(t,x)$ satisfies
	\begin{equation*}
		\phi_t(t,x) - \frac{1}{2}\sigma^2 \phi_{xx}(t,x) -\theta\frac{\partial}{\partial x}[(x-x_0(t))\phi(t,x)] 
		= \frac{\partial}{\partial x}(\phi(t,x) g(t,x)).
	\end{equation*}
	If $\phi(t,x)=\bar{p}(t,x)$, then by using the fact that $\bar{p}_t=-\dot{\bar{x}}(t)\bar{p}_x$ and 
	$\frac{1}{2}\sigma^2 \bar{p}_{xx} + \theta\frac{\partial}{\partial x}[(x-x_0(t))\bar{p}]=\theta[(\bar{x}(t)-x_0(t)) \bar{p}_x]$, the 
	corresponding $g(t,x)$ satisfies
	\begin{equation*}
			-\dot{\bar{x}}(t)\bar{p}_x - \theta[(\bar{x}(t)-x_0(t)) \bar{p}_x] = \frac{\partial}{\partial x}(g(t,x)\bar{p}).
	\end{equation*}
	Then $g(t,x)=-\dot{\bar{x}}(t)- \theta(\bar{x}(t)-x_0(t))$ and 
	$\int_0^T \langle \phi, (g(t,x))^2\rangle dt=\int_0^T(\dot{\bar{x}}(t) + \theta(\bar{x}(t)-x_0(t)))^2 dt$. We therefore obtain the 
	desired results.
\end{proof}

Finally we show that the minimizer $(\bar{p}(t,dx))_{t \in [0,T]}$ is unique.
\begin{lemma}
	The minimizer $(\bar{p}(t,dx))_{t \in [0,T]}$ of $\inf_{\phi(t,dx)}\mathcal{J} 
	\big( (x_0(t), \phi(t,dx))_{t \in [0,T]} \big)$ is unique for all $(\phi(t,dx))_{t \in [0,T]}$ such that 
	$\langle\phi(t,dx),x\rangle=\bar{x}(t)$ for all $t\in [0,T]$ and $\phi(0,dx)=\bar{p}(0,dx)$. 
\end{lemma}
\begin{proof}
	From the previous lemmas we conclude that if $(\phi(t,dx))_{t \in [0,T]}$ is a minimizer, then 
	\begin{equation*}
		x = \underset{f(x): \langle \phi, (f'(x))^2 \rangle \neq 0}{\arg\sup}
		\frac{\langle \phi_t - \frac{1}{2}\sigma^2 \phi_{xx} -\theta\frac{\partial}{\partial x}[(x-x_0(t))\phi], f(x)\rangle^2}
		{\langle \phi, (f'(x))^2 \rangle}.
	\end{equation*}
	Therefore for any perturbation $\hat{f}(x)$,
	\begin{equation*}
		\left. \frac{d}{d\epsilon} \right|_{\epsilon=0}
		\frac{\langle \phi_t - \frac{1}{2}\sigma^2 \phi_{xx} -\theta\frac{\partial}{\partial x}[(x-x_0(t))\phi], 
		x+\epsilon \hat{f}(x)\rangle^2}
		{\langle \phi, (1 + \epsilon\hat{f}'(x))^2 \rangle} = 0,
	\end{equation*}
	which leads to 
	\begin{align*}
		\phi_t - \frac{1}{2}\sigma^2 \phi_{xx} -\theta\frac{\partial}{\partial x}[(x-x_0(t))\phi]
		&= \langle \phi_t - \frac{1}{2}\sigma^2 \phi_{xx} -\theta\frac{\partial}{\partial x}[(x-x_0(t))\phi], x\rangle \phi\\
		&= [\dot{\bar{x}}(t) + \theta(\bar{x}(t)-x_0(t))]\phi.
	\end{align*}
	In other words, a minimizer $(\phi(t,dx))_{t \in [0,T]}$ must satisfy the above linear parabolic PDE that has a unique solution with the 
	given initial condition $\phi(0,dx)=\bar{p}(0,dx)$.
\end{proof}

%% file: proof_control.tex
\section{Proof of Proposition \ref{prop:optimal control}}
\label{pf:optimal control}

We can rewrite the problem in the matrix form:
\begin{equation*}
	\min_{(\balpha(t))_{t \in [0,T]}}\frac{1}{2}\EE\left[\int_0^T \balpha(t)^\TT \mathbf{R} \balpha(t) + \bX(t)^\TT  \mathbf{Q} \bX(t)dt\right],\quad
	d\bX = \mathbf{\Sigma} d\bW +  \mathbf{A} \bX + \mathbf{B} \balpha dt,
\end{equation*}
where
\begin{align*}
	&\mathbf{\Sigma} 
	=\begin{pmatrix}
		\frac{\sigma_0}{\sqrt{N}} & 0{\itbf u}^\TT \\
		0{\itbf u} & \sigma \mathbf{I}
	\end{pmatrix},\quad
	\mathbf{A} 
	=\begin{pmatrix} 
		- \theta_0 -H_0 & \frac{\theta_0}{N} {\itbf u}^\TT \\
		\theta {\itbf u} & -\theta\mathbf{I}
	\end{pmatrix},\quad 
	\mathbf{B}
	=\begin{pmatrix} 
		0 & 0{\itbf u}^\TT \\
		0 {\itbf u} & \mathbf{I}
	\end{pmatrix},\\
	&\mathbf{Q}
	=\theta_c\begin{pmatrix}
		N &- {\itbf u}^\TT \\
		-{\itbf u} & \mathbf{I}
	\end{pmatrix},\quad 
	\mathbf{R} 
	= \frac{1}{\theta_c} \begin{pmatrix}
		1 & 0{\itbf u}^\TT \\
		0 {\itbf u} & \mathbf{I}
	\end{pmatrix},\quad
	{\itbf u}=(1,\ldots,1)^\TT.
\end{align*}

We apply the standard theory \cite[Theorem 6.1]{Yong1999} and we find that the optimal control is
\begin{equation*}
	\balpha(t) = -\mathbf{R}^{-1} \mathbf{B}^\TT \mathbf{S}(t) \bX(t)
\end{equation*} 
where ${\bf S}(t)$ is solution of the matrix Riccati equation
\begin{equation*}
	-\frac{d}{dt}\mathbf{S} = \mathbf{A}^\TT \mathbf{S} + \mathbf{S}\mathbf{A} 
	- \mathbf{S}^\TT \mathbf{B} \mathbf{R}^{-1} \mathbf{B}^\TT \mathbf{S} + \mathbf{Q},
\end{equation*}
with the terminal condition $\mathbf{S}(T)=\mathbf{0}$. We find that
\begin{equation*}
	\mathbf{S}(t) 
	=\begin{pmatrix} 
		N a(t) & b(t) {\itbf u}^\TT \\
		b(t) {\itbf u} & d(t) \mathbf{I} + \frac{e(t)}{N} \mathbf{J}
	\end{pmatrix},
\end{equation*}
where $\mathbf{J}$ is the $N\times N$ matrix full of ones and $(a(t), b(t), d(t), e(t))_{t \in [0,T]}$ is the solution of
\begin{align*}
	\dot{a}(t) &= 2(\theta_0+H_0) a(t) - 2 \theta b(t) + \theta_c b^2(t) -\theta_c,\\
	\dot{b}(t) &= (\theta_0+H_0+\theta) b(t) -\theta d(t) -\theta_0 a(t) + \theta_c b(t)d(t) +\theta_c -\theta e(t) +\theta_c b(t)e(t),\\
	\dot{d}(t) &= 2 \theta d(t) +\theta_c d^2(t) -\theta_c,\\
	\dot{e}(t) &= -2\theta_0 b(t) +2 \theta e(t) + \theta_c (2d(t)e(t) +e^2(t)),
\end{align*}
with $(a(T), b(T), d(T), e(T))=(0, 0, 0, 0)$. Therefore the optimal control is
\begin{equation*}
	\alpha_j(t)= -\theta_c (b(t)X_0(t) + d(t)X_j(t) + e(t)\bar{X}_N(t)),\quad j=1,\ldots, N.
\end{equation*}